\documentclass[aps,prd,reprint,superscriptaddress,nofootinbib]{revtex4-2}

\usepackage{graphicx}
\graphicspath{{./figures/}}

\usepackage{multirow}
\usepackage{flafter}
\usepackage{placeins}
\usepackage{amsmath}
\usepackage{natbib}
\usepackage{hyperref}
\hypersetup{colorlinks=true,linkcolor=blue,citecolor=blue,urlcolor=blue}
\usepackage{etoolbox}
\usepackage{orcidlink}
\usepackage{soul}
\usepackage{threeparttable}
\usepackage{dcolumn}
\usepackage{float}

\newcommand{\nside}{\ensuremath{N_\mathrm{side}}}
\newcommand{\hii}{H{\textsc{ii}}}

\begin{document}

\title{SPT-3G D1: Maps of the millimeter-wave sky\texorpdfstring{\\}{ }from 2019 and 2020 observations of the SPT-3G Main field}

\affiliation{High-Energy Physics Division, Argonne National Laboratory, 9700 South Cass Avenue, Lemont, IL, 60439, USA}
\affiliation{Department of Physics, University of Chicago, 5640 South Ellis Avenue, Chicago, IL, 60637, USA}
\affiliation{Kavli Institute for Cosmological Physics, University of Chicago, 5640 South Ellis Avenue, Chicago, IL, 60637, USA}
\affiliation{Sorbonne Universit\'e, CNRS, UMR 7095, Institut d'Astrophysique de Paris, 98 bis bd Arago, 75014 Paris, France}
\affiliation{Universit\'e Paris-Saclay, Universit\'e Paris Cit\'e, CEA, CNRS, AIM, 91191, Gif-sur-Yvette, France}
\affiliation{Department of Astronomy, University of Illinois Urbana-Champaign, 1002 West Green Street, Urbana, IL, 61801, USA}
\affiliation{Department of Physics, University of California, Berkeley, CA, 94720, USA}
\affiliation{Department of Astronomy and Astrophysics, University of Chicago, 5640 South Ellis Avenue, Chicago, IL, 60637, USA}
\affiliation{Department of Physics \& Astronomy, University of California, One Shields Avenue, Davis, CA 95616, USA}
\affiliation{Department of Statistics, University of California, One Shields Avenue, Davis, CA 95616, USA}
\affiliation{Fermi National Accelerator Laboratory, MS209, P.O. Box 500, Batavia, IL, 60510, USA}
\affiliation{School of Physics, University of Melbourne, Parkville, VIC 3010, Australia}
\affiliation{Department of Physics and Astronomy, University of New Mexico, Albuquerque, NM, 87131, USA}
\affiliation{Kavli Institute for Particle Astrophysics and Cosmology, Stanford University, 452 Lomita Mall, Stanford, CA, 94305, USA}
\affiliation{Department of Physics, Stanford University, 382 Via Pueblo Mall, Stanford, CA, 94305, USA}
\affiliation{SLAC National Accelerator Laboratory, 2575 Sand Hill Road, Menlo Park, CA, 94025, USA}
\affiliation{University Observatory, Faculty of Physics, Ludwig-Maximilians-Universit{\"a}t, Scheinerstr.~1, 81679 Munich, Germany}
\affiliation{Enrico Fermi Institute, University of Chicago, 5640 South Ellis Avenue, Chicago, IL, 60637, USA}
\affiliation{Universit\'e de Gen\`eve, D\'epartement de Physique Th\'eorique, 24 Quai Ansermet, CH-1211 Gen\`eve 4, Switzerland}
\affiliation{Department of Physics \& Astronomy, University of Sussex, Brighton BN1 9QH, UK}
\affiliation{National Taiwan University, No. 1, Sec. 4, Roosevelt Road, Taipei 106319, Taiwan}
\affiliation{High Energy Accelerator Research Organization (KEK), Tsukuba, Ibaraki 305-0801, Japan}
\affiliation{Department of Physics and McGill Space Institute, McGill University, 3600 Rue University, Montreal, Quebec H3A 2T8, Canada}
\affiliation{Canadian Institute for Advanced Research, CIFAR Program in Gravity and the Extreme Universe, Toronto, ON, M5G 1Z8, Canada}
\affiliation{Joseph Henry Laboratories of Physics, Jadwin Hall, Princeton University, Princeton, NJ 08544, USA}
\affiliation{Department of Astronomy, University of Science and Technology of China, Hefei 230026, China}
\affiliation{School of Astronomy and Space Science, University of Science and Technology of China, Hefei 230026}
\affiliation{Department of Physics, University of Illinois Urbana-Champaign, 1110 West Green Street, Urbana, IL, 61801, USA}
\affiliation{Department of Physics and Astronomy, University of California, Los Angeles, CA, 90095, USA}
\affiliation{Department of Physics and Astronomy, Michigan State University, East Lansing, MI 48824, USA}
\affiliation{California Institute of Technology, 1200 East California Boulevard., Pasadena, CA, 91125, USA}
\affiliation{CASA, Department of Astrophysical and Planetary Sciences, University of Colorado, Boulder, CO, 80309, USA }
\affiliation{Department of Physics, University of Colorado, Boulder, CO, 80309, USA}
\affiliation{Department of Physics \& Astronomy, Box 41051, Texas Tech University, Lubbock TX 79409-1051, USA}
\affiliation{Center for AstroPhysical Surveys, National Center for Supercomputing Applications, Urbana, IL, 61801, USA}
\affiliation{NSF-Simons AI Institute for the Sky (SkAI), 172 E. Chestnut St., Chicago, IL 60611, USA}
\affiliation{Department of Physics, Case Western Reserve University, Cleveland, OH, 44106, USA}
\affiliation{Center for Astrophysics \textbar{} Harvard \& Smithsonian, 60 Garden Street, Cambridge, MA, 02138, USA}

\author{W.~Quan\,\orcidlink{0009-0002-2589-5501}}
\affiliation{High-Energy Physics Division, Argonne National Laboratory, 9700 South Cass Avenue, Lemont, IL, 60439, USA}
\affiliation{Department of Physics, University of Chicago, 5640 South Ellis Avenue, Chicago, IL, 60637, USA}
\affiliation{Kavli Institute for Cosmological Physics, University of Chicago, 5640 South Ellis Avenue, Chicago, IL, 60637, USA}
\author{E.~Camphuis\,\orcidlink{0000-0003-3483-8461}}
\affiliation{Sorbonne Universit\'e, CNRS, UMR 7095, Institut d'Astrophysique de Paris, 98 bis bd Arago, 75014 Paris, France}
\author{C.~Daley\,\orcidlink{0000-0002-3760-2086}}
\affiliation{Universit\'e Paris-Saclay, Universit\'e Paris Cit\'e, CEA, CNRS, AIM, 91191, Gif-sur-Yvette, France}
\affiliation{Department of Astronomy, University of Illinois Urbana-Champaign, 1002 West Green Street, Urbana, IL, 61801, USA}
\author{N.~Huang\,\orcidlink{0000-0003-3595-0359}}
\affiliation{Department of Physics, University of California, Berkeley, CA, 94720, USA}
\author{Y.~Omori}
\affiliation{Department of Astronomy and Astrophysics, University of Chicago, 5640 South Ellis Avenue, Chicago, IL, 60637, USA}
\affiliation{Kavli Institute for Cosmological Physics, University of Chicago, 5640 South Ellis Avenue, Chicago, IL, 60637, USA}
\author{F.~Guidi\,\orcidlink{0000-0001-7593-3962}}
\affiliation{Department of Physics \& Astronomy, University of California, One Shields Avenue, Davis, CA 95616, USA}
\affiliation{Sorbonne Universit\'e, CNRS, UMR 7095, Institut d'Astrophysique de Paris, 98 bis bd Arago, 75014 Paris, France}
\author{E.~Anderes}
\affiliation{Department of Statistics, University of California, One Shields Avenue, Davis, CA 95616, USA}
\author{A.~J.~Anderson\,\orcidlink{0000-0002-4435-4623}}
\affiliation{Fermi National Accelerator Laboratory, MS209, P.O. Box 500, Batavia, IL, 60510, USA}
\affiliation{Kavli Institute for Cosmological Physics, University of Chicago, 5640 South Ellis Avenue, Chicago, IL, 60637, USA}
\affiliation{Department of Astronomy and Astrophysics, University of Chicago, 5640 South Ellis Avenue, Chicago, IL, 60637, USA}
\author{B.~Ansarinejad}
\affiliation{School of Physics, University of Melbourne, Parkville, VIC 3010, Australia}
\author{M.~Archipley\,\orcidlink{0000-0002-0517-9842}}
\affiliation{Department of Astronomy and Astrophysics, University of Chicago, 5640 South Ellis Avenue, Chicago, IL, 60637, USA}
\affiliation{Kavli Institute for Cosmological Physics, University of Chicago, 5640 South Ellis Avenue, Chicago, IL, 60637, USA}
\author{L.~Balkenhol\,\orcidlink{0000-0001-6899-1873}}
\affiliation{Sorbonne Universit\'e, CNRS, UMR 7095, Institut d'Astrophysique de Paris, 98 bis bd Arago, 75014 Paris, France}
\author{D.~R.~Barron\,\orcidlink{0000-0002-1623-5651}}
\affiliation{Department of Physics and Astronomy, University of New Mexico, Albuquerque, NM, 87131, USA}
\author{K.~Benabed}
\affiliation{Sorbonne Universit\'e, CNRS, UMR 7095, Institut d'Astrophysique de Paris, 98 bis bd Arago, 75014 Paris, France}
\author{A.~N.~Bender\,\orcidlink{0000-0001-5868-0748}}
\affiliation{High-Energy Physics Division, Argonne National Laboratory, 9700 South Cass Avenue, Lemont, IL, 60439, USA}
\affiliation{Kavli Institute for Cosmological Physics, University of Chicago, 5640 South Ellis Avenue, Chicago, IL, 60637, USA}
\affiliation{Department of Astronomy and Astrophysics, University of Chicago, 5640 South Ellis Avenue, Chicago, IL, 60637, USA}
\author{B.~A.~Benson\,\orcidlink{0000-0002-5108-6823}}
\affiliation{Fermi National Accelerator Laboratory, MS209, P.O. Box 500, Batavia, IL, 60510, USA}
\affiliation{Kavli Institute for Cosmological Physics, University of Chicago, 5640 South Ellis Avenue, Chicago, IL, 60637, USA}
\affiliation{Department of Astronomy and Astrophysics, University of Chicago, 5640 South Ellis Avenue, Chicago, IL, 60637, USA}
\author{F.~Bianchini\,\orcidlink{0000-0003-4847-3483}}
\affiliation{Kavli Institute for Particle Astrophysics and Cosmology, Stanford University, 452 Lomita Mall, Stanford, CA, 94305, USA}
\affiliation{Department of Physics, Stanford University, 382 Via Pueblo Mall, Stanford, CA, 94305, USA}
\affiliation{SLAC National Accelerator Laboratory, 2575 Sand Hill Road, Menlo Park, CA, 94025, USA}
\author{L.~E.~Bleem\,\orcidlink{0000-0001-7665-5079}}
\affiliation{High-Energy Physics Division, Argonne National Laboratory, 9700 South Cass Avenue, Lemont, IL, 60439, USA}
\affiliation{Kavli Institute for Cosmological Physics, University of Chicago, 5640 South Ellis Avenue, Chicago, IL, 60637, USA}
\affiliation{Department of Astronomy and Astrophysics, University of Chicago, 5640 South Ellis Avenue, Chicago, IL, 60637, USA}
\author{S.~Bocquet\,\orcidlink{0000-0002-4900-805X}}
\affiliation{University Observatory, Faculty of Physics, Ludwig-Maximilians-Universit{\"a}t, Scheinerstr.~1, 81679 Munich, Germany}
\author{F.~R.~Bouchet\,\orcidlink{0000-0002-8051-2924}}
\affiliation{Sorbonne Universit\'e, CNRS, UMR 7095, Institut d'Astrophysique de Paris, 98 bis bd Arago, 75014 Paris, France}
\author{M.~G.~Campitiello}
\affiliation{High-Energy Physics Division, Argonne National Laboratory, 9700 South Cass Avenue, Lemont, IL, 60439, USA}
\author{J.~E.~Carlstrom\,\orcidlink{0000-0002-2044-7665}}
\affiliation{Kavli Institute for Cosmological Physics, University of Chicago, 5640 South Ellis Avenue, Chicago, IL, 60637, USA}
\affiliation{Enrico Fermi Institute, University of Chicago, 5640 South Ellis Avenue, Chicago, IL, 60637, USA}
\affiliation{Department of Physics, University of Chicago, 5640 South Ellis Avenue, Chicago, IL, 60637, USA}
\affiliation{High-Energy Physics Division, Argonne National Laboratory, 9700 South Cass Avenue, Lemont, IL, 60439, USA}
\affiliation{Department of Astronomy and Astrophysics, University of Chicago, 5640 South Ellis Avenue, Chicago, IL, 60637, USA}
\author{J.~Carron\,\orcidlink{0000-0002-5751-1392}}
\affiliation{Universit\'e de Gen\`eve, D\'epartement de Physique Th\'eorique, 24 Quai Ansermet, CH-1211 Gen\`eve 4, Switzerland}
\affiliation{Department of Physics \& Astronomy, University of Sussex, Brighton BN1 9QH, UK}
\author{C.~L.~Chang}
\affiliation{High-Energy Physics Division, Argonne National Laboratory, 9700 South Cass Avenue, Lemont, IL, 60439, USA}
\affiliation{Kavli Institute for Cosmological Physics, University of Chicago, 5640 South Ellis Avenue, Chicago, IL, 60637, USA}
\affiliation{Department of Astronomy and Astrophysics, University of Chicago, 5640 South Ellis Avenue, Chicago, IL, 60637, USA}
\author{P.~M.~Chichura\,\orcidlink{0000-0002-5397-9035}}
\affiliation{Department of Physics, University of Chicago, 5640 South Ellis Avenue, Chicago, IL, 60637, USA}
\affiliation{Kavli Institute for Cosmological Physics, University of Chicago, 5640 South Ellis Avenue, Chicago, IL, 60637, USA}
\author{A.~Chokshi}
\affiliation{Department of Astronomy and Astrophysics, University of Chicago, 5640 South Ellis Avenue, Chicago, IL, 60637, USA}
\author{T.-L.~Chou\,\orcidlink{0000-0002-3091-8790}}
\affiliation{Department of Astronomy and Astrophysics, University of Chicago, 5640 South Ellis Avenue, Chicago, IL, 60637, USA}
\affiliation{Kavli Institute for Cosmological Physics, University of Chicago, 5640 South Ellis Avenue, Chicago, IL, 60637, USA}
\affiliation{National Taiwan University, No. 1, Sec. 4, Roosevelt Road, Taipei 106319, Taiwan}
\author{A.~Coerver\,\orcidlink{0000-0002-2707-1672}}
\affiliation{Department of Physics, University of California, Berkeley, CA, 94720, USA}
\author{T.~M.~Crawford\,\orcidlink{0000-0001-9000-5013}}
\affiliation{Department of Astronomy and Astrophysics, University of Chicago, 5640 South Ellis Avenue, Chicago, IL, 60637, USA}
\affiliation{Kavli Institute for Cosmological Physics, University of Chicago, 5640 South Ellis Avenue, Chicago, IL, 60637, USA}
\author{T.~de~Haan\,\orcidlink{0000-0001-5105-9473}}
\affiliation{High Energy Accelerator Research Organization (KEK), Tsukuba, Ibaraki 305-0801, Japan}
\author{K.~R.~Dibert}
\affiliation{Department of Astronomy and Astrophysics, University of Chicago, 5640 South Ellis Avenue, Chicago, IL, 60637, USA}
\affiliation{Kavli Institute for Cosmological Physics, University of Chicago, 5640 South Ellis Avenue, Chicago, IL, 60637, USA}
\author{M.~A.~Dobbs}
\affiliation{Department of Physics and McGill Space Institute, McGill University, 3600 Rue University, Montreal, Quebec H3A 2T8, Canada}
\affiliation{Canadian Institute for Advanced Research, CIFAR Program in Gravity and the Extreme Universe, Toronto, ON, M5G 1Z8, Canada}
\author{M.~Doohan}
\affiliation{School of Physics, University of Melbourne, Parkville, VIC 3010, Australia}
\author{D.~Dutcher\,\orcidlink{0000-0002-9962-2058}}
\affiliation{Joseph Henry Laboratories of Physics, Jadwin Hall, Princeton University, Princeton, NJ 08544, USA}
\author{C.~Feng}
\affiliation{Department of Astronomy, University of Science and Technology of China, Hefei 230026, China}
\affiliation{School of Astronomy and Space Science, University of Science and Technology of China, Hefei 230026}
\affiliation{Department of Physics, University of Illinois Urbana-Champaign, 1110 West Green Street, Urbana, IL, 61801, USA}
\author{K.~R.~Ferguson\,\orcidlink{0000-0002-4928-8813}}
\affiliation{Department of Physics and Astronomy, University of California, Los Angeles, CA, 90095, USA}
\affiliation{Department of Physics and Astronomy, Michigan State University, East Lansing, MI 48824, USA}
\author{N.~C.~Ferree\,\orcidlink{0000-0002-7130-7099}}
\affiliation{California Institute of Technology, 1200 East California Boulevard., Pasadena, CA, 91125, USA}
\affiliation{Kavli Institute for Particle Astrophysics and Cosmology, Stanford University, 452 Lomita Mall, Stanford, CA, 94305, USA}
\affiliation{Department of Physics, Stanford University, 382 Via Pueblo Mall, Stanford, CA, 94305, USA}
\author{K.~Fichman}
\affiliation{Department of Physics, University of Chicago, 5640 South Ellis Avenue, Chicago, IL, 60637, USA}
\affiliation{Kavli Institute for Cosmological Physics, University of Chicago, 5640 South Ellis Avenue, Chicago, IL, 60637, USA}
\author{A.~Foster\,\orcidlink{0000-0002-7145-1824}}
\affiliation{Joseph Henry Laboratories of Physics, Jadwin Hall, Princeton University, Princeton, NJ 08544, USA}
\author{S.~Galli}
\affiliation{Sorbonne Universit\'e, CNRS, UMR 7095, Institut d'Astrophysique de Paris, 98 bis bd Arago, 75014 Paris, France}
\author{A.~E.~Gambrel}
\affiliation{Kavli Institute for Cosmological Physics, University of Chicago, 5640 South Ellis Avenue, Chicago, IL, 60637, USA}
\author{A.~K.~Gao}
\affiliation{Department of Physics, University of Illinois Urbana-Champaign, 1110 West Green Street, Urbana, IL, 61801, USA}
\author{F.~Ge}
\affiliation{California Institute of Technology, 1200 East California Boulevard., Pasadena, CA, 91125, USA}
\affiliation{Kavli Institute for Particle Astrophysics and Cosmology, Stanford University, 452 Lomita Mall, Stanford, CA, 94305, USA}
\affiliation{Department of Physics, Stanford University, 382 Via Pueblo Mall, Stanford, CA, 94305, USA}
\affiliation{Department of Physics \& Astronomy, University of California, One Shields Avenue, Davis, CA 95616, USA}
\author{S.~Guns}
\affiliation{Department of Physics, University of California, Berkeley, CA, 94720, USA}
\author{N.~W.~Halverson}
\affiliation{CASA, Department of Astrophysical and Planetary Sciences, University of Colorado, Boulder, CO, 80309, USA }
\affiliation{Department of Physics, University of Colorado, Boulder, CO, 80309, USA}
\author{E.~Hivon\,\orcidlink{0000-0003-1880-2733}}
\affiliation{Sorbonne Universit\'e, CNRS, UMR 7095, Institut d'Astrophysique de Paris, 98 bis bd Arago, 75014 Paris, France}
\author{G.~P.~Holder\,\orcidlink{0000-0002-0463-6394}}
\affiliation{Department of Physics, University of Illinois Urbana-Champaign, 1110 West Green Street, Urbana, IL, 61801, USA}
\author{W.~L.~Holzapfel}
\affiliation{Department of Physics, University of California, Berkeley, CA, 94720, USA}
\author{J.~C.~Hood}
\affiliation{Kavli Institute for Cosmological Physics, University of Chicago, 5640 South Ellis Avenue, Chicago, IL, 60637, USA}
\author{A.~Hryciuk}
\affiliation{Department of Physics, University of Chicago, 5640 South Ellis Avenue, Chicago, IL, 60637, USA}
\affiliation{Kavli Institute for Cosmological Physics, University of Chicago, 5640 South Ellis Avenue, Chicago, IL, 60637, USA}
\author{T.~Jhaveri}
\affiliation{Department of Astronomy and Astrophysics, University of Chicago, 5640 South Ellis Avenue, Chicago, IL, 60637, USA}
\affiliation{Kavli Institute for Cosmological Physics, University of Chicago, 5640 South Ellis Avenue, Chicago, IL, 60637, USA}
\author{F.~K\'eruzor\'e}
\affiliation{High-Energy Physics Division, Argonne National Laboratory, 9700 South Cass Avenue, Lemont, IL, 60439, USA}
\author{A.~R.~Khalife\,\orcidlink{0000-0002-8388-4950}}
\affiliation{Sorbonne Universit\'e, CNRS, UMR 7095, Institut d'Astrophysique de Paris, 98 bis bd Arago, 75014 Paris, France}
\author{L.~Knox}
\affiliation{Department of Physics \& Astronomy, University of California, One Shields Avenue, Davis, CA 95616, USA}
\author{K.~Kornoelje}
\affiliation{Department of Astronomy and Astrophysics, University of Chicago, 5640 South Ellis Avenue, Chicago, IL, 60637, USA}
\affiliation{Kavli Institute for Cosmological Physics, University of Chicago, 5640 South Ellis Avenue, Chicago, IL, 60637, USA}
\affiliation{High-Energy Physics Division, Argonne National Laboratory, 9700 South Cass Avenue, Lemont, IL, 60439, USA}
\author{C.-L.~Kuo}
\affiliation{Kavli Institute for Particle Astrophysics and Cosmology, Stanford University, 452 Lomita Mall, Stanford, CA, 94305, USA}
\affiliation{Department of Physics, Stanford University, 382 Via Pueblo Mall, Stanford, CA, 94305, USA}
\affiliation{SLAC National Accelerator Laboratory, 2575 Sand Hill Road, Menlo Park, CA, 94025, USA}
\author{K.~Levy}
\affiliation{School of Physics, University of Melbourne, Parkville, VIC 3010, Australia}
\author{Y.~Li\,\orcidlink{0000-0002-4820-1122}}
\affiliation{Kavli Institute for Cosmological Physics, University of Chicago, 5640 South Ellis Avenue, Chicago, IL, 60637, USA}
\author{A.~E.~Lowitz\,\orcidlink{0000-0002-4747-4276}}
\affiliation{Kavli Institute for Cosmological Physics, University of Chicago, 5640 South Ellis Avenue, Chicago, IL, 60637, USA}
\author{C.~Lu}
\affiliation{Department of Physics, University of Illinois Urbana-Champaign, 1110 West Green Street, Urbana, IL, 61801, USA}
\author{G.~P.~Lynch\,\orcidlink{0009-0004-3143-1708}}
\affiliation{Department of Physics \& Astronomy, University of California, One Shields Avenue, Davis, CA 95616, USA}
\author{T.~J.~Maccarone\,\orcidlink{0000-0003-0976-4755}}
\affiliation{Department of Physics \& Astronomy, Box 41051, Texas Tech University, Lubbock TX 79409-1051, USA}
\author{A.~S.~Maniyar\,\orcidlink{0000-0002-4617-9320}}
\affiliation{Kavli Institute for Particle Astrophysics and Cosmology, Stanford University, 452 Lomita Mall, Stanford, CA, 94305, USA}
\affiliation{Department of Physics, Stanford University, 382 Via Pueblo Mall, Stanford, CA, 94305, USA}
\affiliation{SLAC National Accelerator Laboratory, 2575 Sand Hill Road, Menlo Park, CA, 94025, USA}
\author{E.~S.~Martsen}
\affiliation{Department of Astronomy and Astrophysics, University of Chicago, 5640 South Ellis Avenue, Chicago, IL, 60637, USA}
\affiliation{Kavli Institute for Cosmological Physics, University of Chicago, 5640 South Ellis Avenue, Chicago, IL, 60637, USA}
\author{F.~Menanteau}
\affiliation{Department of Astronomy, University of Illinois Urbana-Champaign, 1002 West Green Street, Urbana, IL, 61801, USA}
\affiliation{Center for AstroPhysical Surveys, National Center for Supercomputing Applications, Urbana, IL, 61801, USA}
\author{M.~Millea\,\orcidlink{0000-0001-7317-0551}}
\affiliation{Department of Physics, University of California, Berkeley, CA, 94720, USA}
\author{J.~Montgomery}
\affiliation{Department of Physics and McGill Space Institute, McGill University, 3600 Rue University, Montreal, Quebec H3A 2T8, Canada}
\author{Y.~Nakato}
\affiliation{Department of Physics, Stanford University, 382 Via Pueblo Mall, Stanford, CA, 94305, USA}
\author{T.~Natoli}
\affiliation{Kavli Institute for Cosmological Physics, University of Chicago, 5640 South Ellis Avenue, Chicago, IL, 60637, USA}
\author{A.~Ouellette\,\orcidlink{0000-0003-0170-5638}}
\affiliation{Department of Physics, University of Illinois Urbana-Champaign, 1110 West Green Street, Urbana, IL, 61801, USA}
\author{Z.~Pan\,\orcidlink{0000-0002-6164-9861}}
\affiliation{High-Energy Physics Division, Argonne National Laboratory, 9700 South Cass Avenue, Lemont, IL, 60439, USA}
\affiliation{Kavli Institute for Cosmological Physics, University of Chicago, 5640 South Ellis Avenue, Chicago, IL, 60637, USA}
\affiliation{Department of Physics, University of Chicago, 5640 South Ellis Avenue, Chicago, IL, 60637, USA}
\author{P.~Paschos}
\affiliation{Enrico Fermi Institute, University of Chicago, 5640 South Ellis Avenue, Chicago, IL, 60637, USA}
\author{K.~A.~Phadke\,\orcidlink{0000-0001-7946-557X}}
\affiliation{Department of Astronomy, University of Illinois Urbana-Champaign, 1002 West Green Street, Urbana, IL, 61801, USA}
\affiliation{Center for AstroPhysical Surveys, National Center for Supercomputing Applications, Urbana, IL, 61801, USA}
\affiliation{NSF-Simons AI Institute for the Sky (SkAI), 172 E. Chestnut St., Chicago, IL 60611, USA}
\author{A.~W.~Pollak}
\affiliation{Department of Astronomy and Astrophysics, University of Chicago, 5640 South Ellis Avenue, Chicago, IL, 60637, USA}
\author{K.~Prabhu}
\affiliation{Department of Physics \& Astronomy, University of California, One Shields Avenue, Davis, CA 95616, USA}
\author{S.~Raghunathan\,\orcidlink{0000-0003-1405-378X}}
\affiliation{Department of Physics \& Astronomy, University of California, One Shields Avenue, Davis, CA 95616, USA}
\affiliation{Center for AstroPhysical Surveys, National Center for Supercomputing Applications, Urbana, IL, 61801, USA}
\author{M.~Rahimi}
\affiliation{School of Physics, University of Melbourne, Parkville, VIC 3010, Australia}
\author{A.~Rahlin\,\orcidlink{0000-0003-3953-1776}}
\affiliation{Department of Astronomy and Astrophysics, University of Chicago, 5640 South Ellis Avenue, Chicago, IL, 60637, USA}
\affiliation{Kavli Institute for Cosmological Physics, University of Chicago, 5640 South Ellis Avenue, Chicago, IL, 60637, USA}
\author{C.~L.~Reichardt\,\orcidlink{0000-0003-2226-9169}}
\affiliation{School of Physics, University of Melbourne, Parkville, VIC 3010, Australia}
\author{M.~Rouble}
\affiliation{Department of Physics and McGill Space Institute, McGill University, 3600 Rue University, Montreal, Quebec H3A 2T8, Canada}
\author{J.~E.~Ruhl}
\affiliation{Department of Physics, Case Western Reserve University, Cleveland, OH, 44106, USA}
\author{A.~C.~Silva~Oliveira\,\orcidlink{0000-0001-5755-5865}}
\affiliation{California Institute of Technology, 1200 East California Boulevard., Pasadena, CA, 91125, USA}
\affiliation{Kavli Institute for Particle Astrophysics and Cosmology, Stanford University, 452 Lomita Mall, Stanford, CA, 94305, USA}
\affiliation{Department of Physics, Stanford University, 382 Via Pueblo Mall, Stanford, CA, 94305, USA}
\author{A.~Simpson}
\affiliation{Department of Astronomy and Astrophysics, University of Chicago, 5640 South Ellis Avenue, Chicago, IL, 60637, USA}
\affiliation{Kavli Institute for Cosmological Physics, University of Chicago, 5640 South Ellis Avenue, Chicago, IL, 60637, USA}
\author{J.~A.~Sobrin\,\orcidlink{0000-0001-6155-5315}}
\affiliation{Fermi National Accelerator Laboratory, MS209, P.O. Box 500, Batavia, IL, 60510, USA}
\affiliation{Kavli Institute for Cosmological Physics, University of Chicago, 5640 South Ellis Avenue, Chicago, IL, 60637, USA}
\author{A.~A.~Stark}
\affiliation{Center for Astrophysics \textbar{} Harvard \& Smithsonian, 60 Garden Street, Cambridge, MA, 02138, USA}
\author{J.~Stephen}
\affiliation{Enrico Fermi Institute, University of Chicago, 5640 South Ellis Avenue, Chicago, IL, 60637, USA}
\author{C.~Tandoi}
\affiliation{Department of Astronomy, University of Illinois Urbana-Champaign, 1002 West Green Street, Urbana, IL, 61801, USA}
\author{C.~Trendafilova}
\affiliation{Center for AstroPhysical Surveys, National Center for Supercomputing Applications, Urbana, IL, 61801, USA}
\author{J.~D.~Vieira\,\orcidlink{0000-0001-7192-3871}}
\affiliation{Department of Astronomy, University of Illinois Urbana-Champaign, 1002 West Green Street, Urbana, IL, 61801, USA}
\affiliation{Department of Physics, University of Illinois Urbana-Champaign, 1110 West Green Street, Urbana, IL, 61801, USA}
\affiliation{Center for AstroPhysical Surveys, National Center for Supercomputing Applications, Urbana, IL, 61801, USA}
\author{A.~G.~Vieregg\,\orcidlink{0000-0002-4528-9886}}
\affiliation{Kavli Institute for Cosmological Physics, University of Chicago, 5640 South Ellis Avenue, Chicago, IL, 60637, USA}
\affiliation{Department of Astronomy and Astrophysics, University of Chicago, 5640 South Ellis Avenue, Chicago, IL, 60637, USA}
\affiliation{Enrico Fermi Institute, University of Chicago, 5640 South Ellis Avenue, Chicago, IL, 60637, USA}
\affiliation{Department of Physics, University of Chicago, 5640 South Ellis Avenue, Chicago, IL, 60637, USA}
\author{A.~Vitrier\,\orcidlink{0009-0009-3168-092X}}
\affiliation{Sorbonne Universit\'e, CNRS, UMR 7095, Institut d'Astrophysique de Paris, 98 bis bd Arago, 75014 Paris, France}
\author{Y.~Wan}
\affiliation{Department of Astronomy, University of Illinois Urbana-Champaign, 1002 West Green Street, Urbana, IL, 61801, USA}
\affiliation{Center for AstroPhysical Surveys, National Center for Supercomputing Applications, Urbana, IL, 61801, USA}
\author{N.~Whitehorn\,\orcidlink{0000-0002-3157-0407}}
\affiliation{Department of Physics and Astronomy, Michigan State University, East Lansing, MI 48824, USA}
\author{W.~L.~K.~Wu\,\orcidlink{0000-0001-5411-6920}}
\affiliation{California Institute of Technology, 1200 East California Boulevard., Pasadena, CA, 91125, USA}
\affiliation{Kavli Institute for Particle Astrophysics and Cosmology, Stanford University, 452 Lomita Mall, Stanford, CA, 94305, USA}
\affiliation{SLAC National Accelerator Laboratory, 2575 Sand Hill Road, Menlo Park, CA, 94025, USA}
\author{M.~R.~Young}
\affiliation{Fermi National Accelerator Laboratory, MS209, P.O. Box 500, Batavia, IL, 60510, USA}
\affiliation{Kavli Institute for Cosmological Physics, University of Chicago, 5640 South Ellis Avenue, Chicago, IL, 60637, USA}
\author{J.~A.~Zebrowski}
\affiliation{Kavli Institute for Cosmological Physics, University of Chicago, 5640 South Ellis Avenue, Chicago, IL, 60637, USA}
\affiliation{Department of Astronomy and Astrophysics, University of Chicago, 5640 South Ellis Avenue, Chicago, IL, 60637, USA}
\affiliation{Fermi National Accelerator Laboratory, MS209, P.O. Box 500, Batavia, IL, 60510, USA}

\collaboration{SPT-3G Collaboration}
\noaffiliation

\date{\today}

\begin{abstract}
\fontsize{10pt}{11pt}\selectfont
  Maps of the sky in millimeter wavelengths contain rich information on cosmology through anisotropies of the cosmic microwave background (CMB).
  Creating multifrequency sky maps of anisotropies in the $I$, $Q$, and $U$ Stokes parameters is one of the first steps of CMB cosmology analyses.
  In this work, we describe the production and validation of a set of sky maps from the South Pole Telescope's third-generation camera, SPT-3G.
  The maps are from data taken in frequency bands centered at 95, 150, and 220\;GHz and taken during the first two years, 2019 and 2020, of the SPT-3G Main survey, which covers 4\% of the sky.
  We applied high-pass filters to time series of individual detectors and binned the filtered time series samples into map pixels.
  After that, we calibrated and cleaned the maps to reduce known systematic errors.
  In addition, we searched for other systematic errors through null tests and mitigated a significant systematic error detected therein.
  The white noise levels of the full-depth maps of the $I$ Stokes parameter are 5.4, 4.4, and 16.2 $\mathrm{\mu}$K--arcmin in the 95, 150, and 220\;GHz bands, respectively, and 8.4, 6.6, and 25.8 $\mathrm{\mu}$K--arcmin for $Q/U$.
  These maps are the deepest to date used for measurements of mid-to-high-$\ell$ primary temperature and $E$-mode polarization CMB anisotropies, and reconstructions of the CMB gravitational lensing potential.
  We make these maps and supporting data products publicly accessible.
\end{abstract}

\maketitle
\newpage
\tableofcontents

\section{\label{sec:introduction}Introduction}
  Observations of the cosmic microwave background (CMB), especially the primary temperature and polarization anisotropies and the effects of gravitational lensing on the CMB, are the cornerstone of our current cosmological model.
  For the last decade, the strongest constraints on cosmology from the CMB---or indeed from any observable---have come from the satellite experiment \textit{Planck}.
  Data from \textit{Planck} have been used to place subpercent-precision constraints on most parameters of the standard cosmological model, the $\Lambda$CDM model \citep[e.g.,][]{planck18-1}.
  The constraints from \textit{Planck} are dominated by measurements of medium-scale ($\ell \lesssim 1000$) primary temperature anisotropy of the CMB.
  More recently, ground-based experiments such as the Atacama Cosmology Telescope (ACT, \citealt{henderson16}) and the South Pole Telescope (SPT, \citealt{carlstrom11}) have taken data that have the potential to enable yet more powerful tests of $\Lambda$CDM and its extensions through high-precision measurements of small angular-scale primary CMB polarization anisotropy and the effects of gravitational lensing on the CMB \citep[e.g.,][]{louis25, ge25, prabhu24, camphuis25}.

  SPT is a 10-meter telescope located at the geographical South Pole and optimized for observing faint, diffuse millimeter-wave emission such as anisotropies of the CMB.
  The telescope has been in operation since 2007, and there have been three generations of cameras on the telescope.
  The third-generation camera, SPT-3G \citep{sobrin22}, has $\sim$16\,000 detectors, which are evenly split among three frequency bands centered at 95, 150, and 220\;GHz.
  SPT-3G is a significant upgrade from the previous cameras, increasing the detector count (and thus the mapping speed) by an order of magnitude.

  To date, we have used two datasets from SPT-3G to measure the power spectra of the primary temperature and $E$-mode polarization anisotropies ($TT/TE/EE$ spectra) and the power spectrum of the projected gravitational potential between us and the CMB last scattering surface ($\phi\phi$ spectrum).
  The first dataset comprises observations of the SPT-3G Main field \citep[Figure~1]{prabhu24} conducted in the 2018 austral winter observing season.
  In that year, we acquired data during only one half of the full observing season and with only $\sim$6600 detectors passing cuts \citep[Section~III\,D]{dutcher21}.
  Measurements of the $TT/TE/EE$ and $\phi\phi$ spectra were presented in a combination of three works: \citet{dutcher21}, \citet{balkenhol23}, and \citet{pan23}.

  The second dataset, which we call SPT-3G D1, comprises observations of the SPT-3G Main field conducted in the 2019 and 2020 austral winter\footnote{
  We use ``austral winter'' as shorthand for the period from Polar sunset until approximately December 1; this period in fact comprises the entirety of the Polar night plus approximately two months with the Sun above the horizon.}
observing seasons, the first two observing seasons of the SPT-3G Main survey \citep[Table~1]{prabhu24}.
  In each of the two years, we acquired data during the full observing season and with $\sim$11\,000 detectors passing cuts.
  As a result, the 2019--2020 dataset is significantly larger than the 2018 dataset.
  Measurements of the $TT/TE/EE$ and $\phi\phi$ spectra using SPT-3G D1 have been presented in a combination of two works: \citet{ge25} (hereafter G25) and \citet{camphuis25} (hereafter C25).
  A quadratic-estimator-based measurement of the $\phi\phi$ spectrum will be presented in Omori et al., in preparation (hereafter O26).
  The 2019--2020 set of measurements represents significant improvements over the 2018 set and contributes to state-of-the-art high-precision measurements at small angular scales.

  In this work, we describe the production and validation of the primary data products used for the SPT-3G D1 measurements: sky maps of anisotropies in three Stokes parameters, total intensity ($I$ or $T$) and two linear polarization intensities ($Q$ and $U$), in each of the three frequency bands.
  We produced the maps by processing time series (hereafter timestreams) of individual detectors into map pixel values (Sections~\ref{sec:observations} and \ref{sec:mapmaking}) and then by calibrating and cleaning the resultant maps (Sections~\ref{sec:map_calibration} and \ref{sec:residual_calibration_biases}).
  We validated the maps by searching for systematic errors through consistency tests between two halves of the dataset, which we call null tests, and mitigating a significant systematic error detected therein (Section~\ref{sec:map_level_null_tests}).
  We also briefly describe ancillary products needed to make unbiased power spectrum measurements from the maps (Section~\ref{sec:beam_and_filter_transfer_function}); more details can be found in C25 and two future works: Huang et al., in preparation, and Hivon et al., in preparation.
  Finally, we summarize the main results of this work (Section~\ref{sec:conclusion}) and share information on how to access the maps (Appendix~\ref{app:data_availability}).

\section{\label{sec:observations}Observations}
  We started the SPT-3G Main survey in 2019 and have thus far spent six austral winter observing seasons (between late March and late November in 2019--2023 and in 2025) on the SPT-3G Main field.
  We call the set of observations conducted during the first two observing seasons SPT-3G D1.
  In this section, we first describe the field observations and calibration observations that we routinely conduct in the survey (Sections~\ref{sec:field_observations} and \ref{sec:calibration_observations}) and then describe SPT-3G D1 and its timestreams that passed data quality cuts and were used to produce maps (Section~\ref{sec:spt3g_d1}).

\subsection{\label{sec:field_observations}Field Observations}
\begin{figure}
  \includegraphics[scale=1.00]{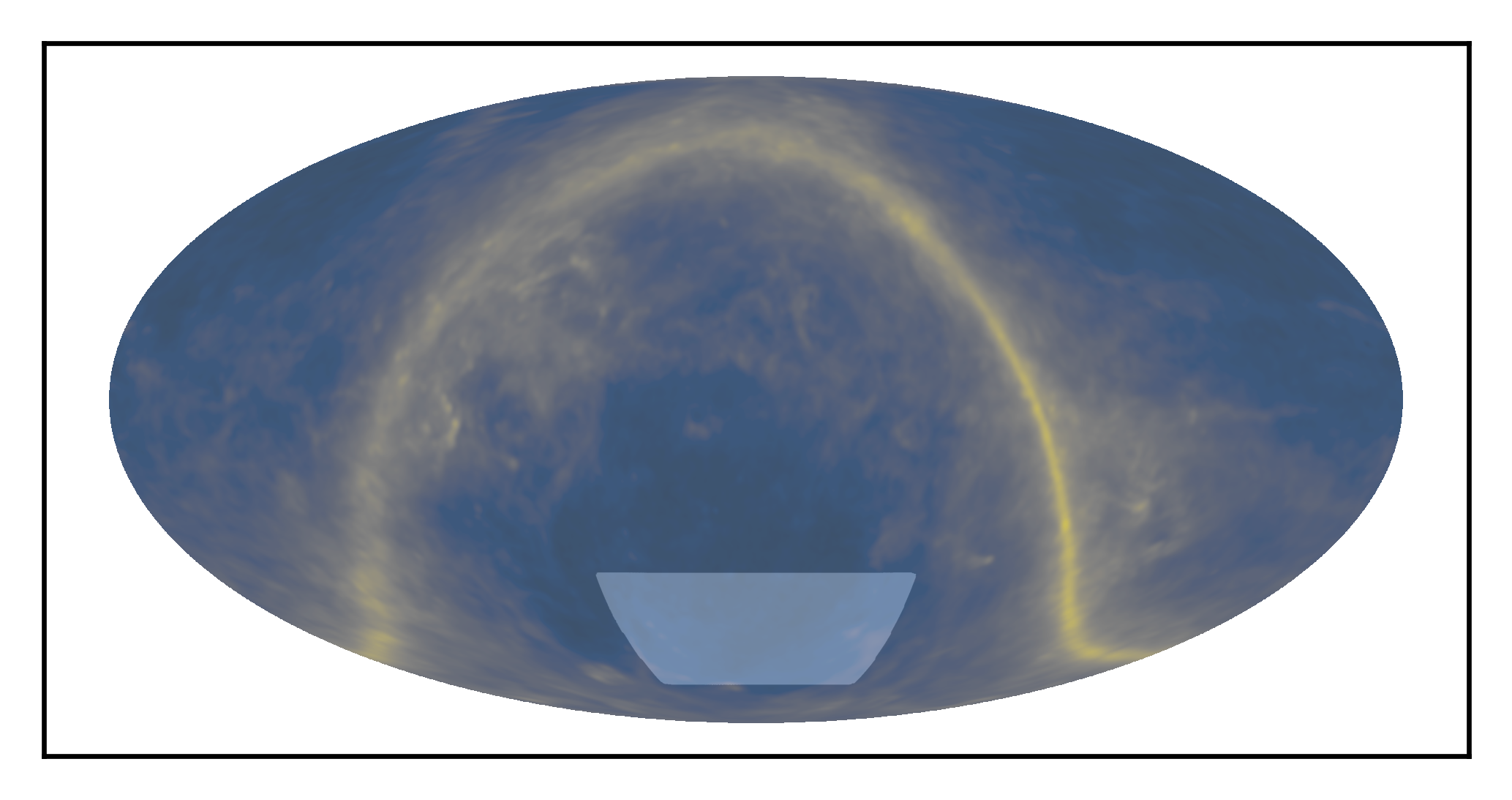}
  \caption{\label{fig:survey_field_full}%
  The target uniform coverage region of the SPT-3G Main field (the light blue, approximately trapezoidal region) overlaid on top of the full-sky galactic dust map from \textit{Planck} Public Data Release 3 (the background image) in the equatorial coordinate system and the Mollweide projection.
  The dust map is the Stokes $I$ map in the file \texttt{COM\_CompMap\_IQU-thermaldust-gnilc-unires\_2048\_R3.00.\\fits} downloaded from Planck Legacy Archive (\url{https://pla.esac.esa.int/pla/}).}
\end{figure}

  The SPT-3G Main field is defined as the region spanning right ascension from $20^{\mathrm{h}}40^{\mathrm{m}}0^{\mathrm{s}}$ to $3^{\mathrm{h}}30^{\mathrm{m}}0^{\mathrm{s}}$ (from $-50^\circ$ to $50^\circ$) and declination from $-42^\circ$ to $-70^\circ$.
  The region bounded by these limits has an area of $1550\ \mathrm{deg}^2$ and is the target uniform coverage region.
  As shown in Figure~\ref{fig:survey_field_full}, this region is a small area that has low emission from galactic dust.
  We scan the center of the full detector array ($2.8\ \mathrm{deg}^2$ field of view) past these limits so that every detector observes the target uniform-coverage region.
  As a result, surrounding regions are also observed but by fewer detectors.
  The total observed area is close to $1870\ \mathrm{deg}^2$.
  Individual science analyses may use some of the non-uniform coverage regions if the advantages of having more area outweigh the complexities associated with spatially varying noise.

\begin{figure}
  \includegraphics[scale=1.00]{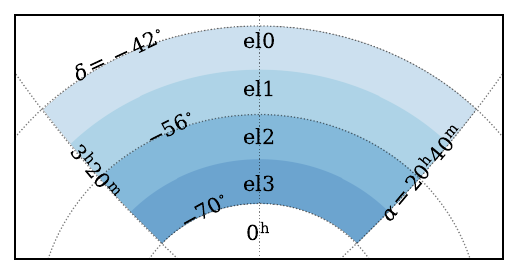}
  \caption{\label{fig:survey_field_subdivision}%
  The target uniform coverage region of the SPT-3G Main field and the footprint of each of the four subfields in a Lambert azimuthal equal-area projection (ZEA, a flat-sky projection, \citet{calabretta02}).
  The subfields are the four shaded regions colored by the different hues.
  Three contours of constant right ascension ($\alpha$) and three contours of constant declination ($\delta$) are also shown.}
\end{figure}

  The SPT-3G Main field is divided into four subfields.
  Each subfield covers the entire range of right ascension spanned by the full field but covers only one-fourth of the range of declination spanned by the full field.
  The four subfields are all centered at right ascension $0^\mathrm{h}$, and the declination centers are $-44.75^\circ$, $-52.25^\circ$, $-59.75^\circ$, and $-67.25^\circ$.
  We call these subfields \texttt{el0}, \texttt{el1}, \texttt{el2}, and \texttt{el3}, respectively; here \texttt{el} denotes elevation of the subfield, which is approximately constant and equal to the negative of declination given the unique location of the South Pole.
  Figure~\ref{fig:survey_field_subdivision} shows this subdivision.

  This subfield division is defined to ensure stable operation of detectors.
  As the telescope scans across the field, changing radiation from the sky is absorbed by each detector and causes its resistance to change.
  Each detector is a transition-edge sensor tuned (biased by a constant voltage) to be in its normal-to-superconducting transition, such that its resistance depends strongly on its temperature.
  Details of the detectors and other aspects of the SPT-3G camera can be found in \citet{sobrin22} (hereafter S22).
  If the full elevation range of the SPT-3G Main field were observed without retuning the detectors (adjusting the bias voltage), the change in radiative load on the detectors---caused by the changing atmospheric column density as a function of elevation---would cause at least some of the detectors to move out of their transition at some point during the observation, rendering them unusable.
  To avoid this issue, we observe only one subfield at a time and retune the detectors before observing another subfield.
  This places each detector in its transition for the typical loading in the elevation range of that subfield.

  An observation of a subfield proceeds through a raster pattern of telescope scans.
  It lasts for 128 minutes and comprises 36 pairs of increasing-azimuth (``right-going'') and decreasing-azimuth (``left-going'') scans at constant elevation.
  The constant-speed portion of each scan covers between 102.6 and 103.4 degrees of azimuth (depending on subfield) at a scan speed of 1.0 deg/s (modulo the sidereal rate as described later).
  The scan speed and duration are the same regardless of the subfield observed or the elevation of a scan within a subfield.
  To ensure that right-going and left-going scans have the same scan speed of 1.0 deg/s on sky (in right ascension), we use slightly faster speed in azimuth for left-going scans than for right-going scans so that the speed difference compensates for the rotation of the Earth.
  After each pair of right/left scans, the telescope elevation is increased by ${12^\prime}.5$, and another scan pair is executed at the new elevation.
  For each subfield observation, the scans start at the bottom elevation of the subfield and reach the top elevation of the subfield at the end.
  For each subfield, we offset the starting elevation of different observations to achieve uniform coverage.
  We use 25 different values of starting elevation offset, or ``dither steps," separated by ${0^\prime}.5$.

  During each scan, we record a timestream from each detector.
  In SPT-3G, detector timestreams represent the changing electrical power dissipated on a detector as a function of time caused by changing incoming radiation from the sky during a scan.
  The sampling rate of every timestream is 152.6 Hz.
  Telescope attitude (pointing) information, the most important being the value of the azimuth and elevation encoders, is also recorded for all the scans, initially at 100 Hz.

  In an ``observing day,'' which is a $\sim$15-hour period of observing time bounded by pauses in observation to recycle the cryogenic system used to cool the detectors and to tune the detectors, we typically conduct six observations of two subfields.
  We conduct three or four observations of \texttt{el0} first and then three or two observations of \texttt{el1} (or \texttt{el2} first and then \texttt{el3}).

\subsection{\label{sec:calibration_observations}Calibration Observations}
  In addition to the subfield observations, we conduct multiple calibration observations in each observing day.
  One purpose is to estimate the gain of each detector that converts the units of the timestreams of that detector from electrical power to CMB fluctuation temperature, and  the other purpose is to monitor detector properties, such as the time constant.
  Different types of calibration observations have different lengths and are conducted at different intervals.

  To estimate the gain of each detector, we conduct weekly, long (80 minutes) observations of two \hii{} regions in our Galaxy, RCW38 and Mat5a (NGC 3576).
  These observations are designed so that a map of an \hii{} region in units of electrical power can be made from every detector individually and compared with a stored map\footnote{
  More information on this stored map can be found in Section~\ref{sec:gain_calibration}.}
in units of CMB fluctuation temperature with absolute calibration reference to \textit{Planck} data.
  This comparison provides the primary estimate of the gain of every detector.
  RCW38 is used to calibrate the timestreams from observations of \texttt{el0} and \texttt{el1} because the declination of RCW38 ($-48^{\circ}$) is within the range of declination spanned by the two subfields.
  For the same reason, Mat5a (declination $-61^{\circ}$) is used for \texttt{el2} and \texttt{el3}.
  While in principle these gains could be measured once, and corrected for any subsequent variations using the procedures described later, in practice we re-measure and re-apply the fundamental \hii{} region gains after every weekly observation.

  We apply two corrections to the primary estimate of the gain before using it for the unit conversion.
  These corrections reflect the changes in the sky transmission and detector gains between a long \hii{} region observation and a later subfield observation.
  To estimate the correction factors, before every subfield observation, we conduct a short (15 minutes) observation of the corresponding \hii{} region and an observation of a thermal source installed on the telescope.
  We call the latter observation a ``calibrator stare,'' in which every detector is exposed to the radiation from the source chopped at 4 Hz for one minute.
  Details of the method that we use to estimate the corrected gain can be found in S22 (Section~6.2).

  To monitor detector properties, one type of observation that we conduct is an ``elevation nod'' (more commonly known as a ``sky dip'' in radio astronomy).
  In an elevation nod, the telescope scans up and down over two degrees in elevation in one minute.
  The amount of change in the electrical power dissipated on a detector caused by the changing atmospheric column density can be used to assess the detector's responsiveness.
  We conduct one elevation nod before every subfield observation.

  Other observations for monitoring detector properties include additional calibrator stares conducted at different elevation or with different chop frequencies.

  While the calibrator stare conducted before each subfield observation takes place when the telescope is at the bottom elevation of a subfield, we also conduct a calibrator stare at the top elevation of the subfield after the subfield observation ends.
  A comparison between the detector responses from the two calibrator stares allows us to measure the elevation-dependent gain change of each detector over the course of the subfield observation.

  We call the additional calibrator stares conducted at different chop frequencies ``calibrator sweeps.''
  A calibrator sweep comprises a series of calibrator stares with the chop frequencies spanning the range between 4 and 64 Hz.
  This allows us to measure the time constant of each detector.
  Calibrator sweeps take place during the transition from one subfield to the other in an observing day.
  We measure the time constants at the center elevation of the first subfield, retune the detectors as described in Section~\ref{sec:field_observations}, and measure the time constants again at the center elevation of the second subfield.

\subsection{\label{sec:spt3g_d1}SPT-3G D1}
  SPT-3G D1 comprises 1047, 908, 788, and 591 observations of \texttt{el0}, \texttt{el1}, \texttt{el2}, and \texttt{el3}, respectively, and the associated calibration observations.
  We conducted these observations during the 2019 and 2020 observing seasons, from March 22 to November 29 in 2019 and from March 23 to November 25 in 2020.
  Because the duration of one subfield observation is the same regardless of which subfield is observed, and different subfields have different areas, the observation time per unit area is longer for a smaller subfield in one observation.
  To have similar total observation time per unit area in all four subfields, we conducted more observations of the larger subfields.

  A vast majority of the subfield observations had good data quality.
  Out of the 3334 total subfield observations, we had to discard 48 because of occasional issues with the data acquisition system.
  There are 56 subfield observations that ended early and did not have the complete set of 72 scans, a common reason being that the cryogenic system used to cool the detectors ran out of its cooling power and had to be recycled.
  For these subfield observations, we still retained the data from the scans that did get executed.

  For each subfield observation and each scan, we assessed the quality of the timestream of each detector against several criteria.
  We cut detector timestreams over full observations by checking whether a detector was operating within its normal-to-superconducting transition and whether the detector showed good optical response during a calibrator stare that preceded the subfield observation, and we cut detector timestreams on a single scan by checking whether the timestream of the detector was free from large discontinuities.
  Over the two observing seasons, we operated 11\,316 detectors on average, and 10\,697 detectors passed these data quality cuts, with an even distribution of 3621, 3688, and 3388 good detectors in the 95, 150, and 220 GHz bands, respectively.\footnote{
  Details of the detector and readout yield can be found in S22 (Section~5.2).}
  We calibrated all the timestreams from these good detectors to have the units of CMB fluctuation temperature using a combination of the calibration observations described in Section~\ref{sec:calibration_observations}.
  We also produced a downsampled version of these calibrated timestreams with the sampling rate reduced from 152.6 to 76.3 Hz for analyses that do not require small angular-scale information.
  Both the full-rate and downsampled timestreams have been processed into $T/Q/U$ maps, a process that we call ``mapmaking'' throughout this work and describe in Section~\ref{sec:mapmaking}.

\section{\label{sec:mapmaking}Mapmaking}
  The SPT-3G D1 timestreams have been used for multiple mapmaking runs, which were optimized for measurements of different signals.
  We call the maps produced for the measurements of the $TT/TE/EE$ and $\phi\phi$ spectra (G25, C25, and O26) the ``mid-$\ell$ maps'' to distinguish them from the ``low-$\ell$ maps,'' which were produced for the measurement of the $B$-mode polarization anisotropy at low $\ell$ as reported in \citet{zebrowski25} (hereafter Z25), and the ``high-$\ell$ $TT$ maps,'' which were produced for the measurement of the secondary temperature anisotropies at high $\ell$ as reported in \citet{chaubal26}.
  All three sets of maps were produced from the downsampled timestreams.
  In addition, we have produced maps using the full-rate timestreams from SPT-3G D1 and observations conducted in additional austral winter observing seasons of the SPT-3G Main survey for detection of discrete objects such as active galactic nuclei and dusty, star-forming galaxies, galaxy clusters, and transient events.
  We call these maps ``discrete-source maps'' in this work.
  While the focus of this work is on the mid-$\ell$ maps, we compare different sets of maps in some places.

  All the mapmaking runs are based on the same general method, in which we apply high-pass filters to the timestreams and bin the filtered timestreams into map pixels.
  In this section, we first describe this method (Section~\ref{sec:filter_bin_method}), and then we describe important parameters of the mid-$\ell$ mapmaking pipeline (Section~\ref{sec:filter_bin_details}).
  We used this pipeline to produce $T/Q/U$ maps from the individual subfield observations and combined the maps to form various coadded maps (Section~\ref{sec:coadds}).
  We also describe the simulations run to characterize the mid-$\ell$ timestream filtering (Section~\ref{sec:simulations}).

  We make maps described in Sections~\ref{sec:coadds} and \ref{sec:simulations} publicly accessible.
  The access information can be found in Appendix~\ref{app:data_availability}.

\subsection{\label{sec:filter_bin_method}Filter-and-bin Method}
  The mapmaking for SPT data has been based on the filter-and-bin method introduced in \citet{hivon02}.
  In this method, a high-pass filter is applied to the timestreams of every detector to reduce correlated low-frequency noise, and each timestream sample is then binned into a corresponding pixel with a weight.
  We now review this method.

  When an ideal polarization-sensitive detector is pointed in a particular direction in the sky, a measurement of the electric field intensity made by the detector is equal to a linear combination of the $T$, $Q$, and $U$ Stokes parameters of the incoming radiation from that direction.
  Here we define an ideal polarization-sensitive detector as one that has no noise and absorbs all the intensity along one polarization direction and rejects the intensity along the orthogonal direction.
  For detector $\alpha$, the linear combination of the Stokes parameters is given by the following (e.g., \citet{jones07}):
\begin{equation}
\label{eqn:detector_signal_model_simple}
  I_\alpha(\hat{n}) = \frac{1}{2}
    \left[\,T(\hat{n}) + 
          Q(\hat{n})\,\cos\,2\psi_\alpha + U(\hat{n})\,\sin\,2\psi_\alpha \,\right],
\end{equation}
where $I_{\alpha}$ is the intensity absorbed by the detector, and $\psi_{\alpha}$ is the polarization direction (angle) of the detector.\footnote{Throughout this work, we define the polarization angle using the IAU convention.}

  When multiple polarization-sensitive detectors that have noise but are otherwise ideal and that span a range of polarization angles register the incoming radiation in the same frequency band from a region of the sky, a maximum-likelihood solution can be constructed to estimate the Stokes parameter values from that region.
  Combining the Stokes parameter values as a function of sky position into a single vector $\mathbf{m}$, it can be shown that for detector noise covariance $\mathbf{N}$, where $\mathbf{N}$ encodes both the covariance between samples from a given detector and between different detectors, the maximum-likelihood estimate for $\mathbf{m}$ is a version of the typical linear least-squares solution (e.g., \citet{couchot99}):
\begin{equation}
\label{eqn:lls_solution}
  \mathbf{\bar{m}} =
    (\mathbf{A}^T \mathbf{N}^{-1} \mathbf{A})^{-1} \
     \mathbf{A}^T \mathbf{N}^{-1} \mathbf{d},
\end{equation}
where $\mathbf{\bar{m}}$ contains the best-fit values for $T(\hat{n})$, $Q(\hat{n})$, and $U(\hat{n})$, $\mathbf{A}$ is the ``pointing matrix'' that encodes the location on the sky each detector was pointed at each time sample and the polarization angle of the detector, and $\mathbf{d}$ contains the measurement for every detector at every time sample and is equal to $I_{\alpha}(\hat{n}_i)$ in Equation~\ref{eqn:detector_signal_model_simple} plus noise for detector $\alpha$ at time sample $i$.

  When applied to modern, high-resolution CMB data, the maximum-likelihood solution becomes computationally expensive (though not intractable---for example, conjugate-gradient approximations to the maximum-likelihood solution are used for most ACT mapmaking, see, e.g., \citet{naess25}).
  The covariance matrix $\mathbf{N}$ is high-dimensional (number of detectors times number of time samples in each dimension), and correlations between time samples and detectors---typically from atmospheric emission or instrumental effects---makes the matrix dense and not trivially invertible.

  Applying a high-pass filter to timestreams can suppress the off-diagonal elements of  $\mathbf{N}$ and justify the approximation that $\mathbf{N}$ is diagonal, and, in this approximation, the maximum-likelihood solution reduces to a binning operation.
  For $T$-only mapmaking, and now writing the summation over the pixel and time sample indices explicitly, Equation~\ref{eqn:lls_solution} reduces to the following:
\begin{equation}
\label{eqn:t_solution}
  \bar{T}(\hat{n}) =
    \left( \sum\limits_{i,\alpha \in \hat{n}} w_{i\alpha} \right)^{-1}
    \sum_{i,\alpha \in \hat{n}} w_{i\alpha} d_{i\alpha},
\end{equation}
where $\bar{T}(\hat{n})$ is the best-fit value of the $T$ Stokes parameter in direction $\hat{n}$, the sum runs over all the timestream samples at which a detector was pointed in direction $\hat{n}$, $d_{i\alpha}$ is the value of the timestream sample from detector $\alpha$ at time sample $i$, and $w_{i\alpha}$ is the weight assigned to that sample.
  In the diagonal-covariance maximum-likelihood case, $w_{i\alpha} = \sigma^{-2}_{i\alpha}$, where $\sigma_{i\alpha}$ is the noise uncertainty on that timestream sample.
  Including $Q$ and $U$, the best-fit $T$, $Q$, and $U$ Stokes parameter values in direction $\hat{n}$ are the following:
\begin{equation}
\label{eqn:tqu_solution}
  \begin{pmatrix}
    \bar{T}(\hat{n}) \\ \bar{Q}(\hat{n}) \\ \bar{U}(\hat{n}) \\
  \end{pmatrix}
  = {W(\hat{n})}^{-1}
  \begin{pmatrix}
     \bar{T}^W(\hat{n}) \\ \bar{Q}^W(\hat{n}) \\ \bar{U}^W(\hat{n}) \\
  \end{pmatrix},
\end{equation}
where
\begin{equation}
\label{eqn:weighted_tqu}
  \begin{pmatrix}
    \bar{T}^W(\hat{n}) \\ \bar{Q}^W(\hat{n}) \\ \bar{U}^W(\hat{n}) \\
  \end{pmatrix} = \sum\limits_{i,\alpha \in \hat{n}} w_{i\alpha} d_{i\alpha}
  \begin{pmatrix}
      1 \\
      \cos{2 \psi_\alpha} \\
      \sin{2 \psi_\alpha} \\
  \end{pmatrix},
\end{equation}
and
\begin{equation}
\label{eqn:weight_matrix}
  \begin{split}
    W&(\hat{n}) = \sum\limits_{i,\alpha \in \hat{n}} w_{i\alpha} \times \\
    &\begin{pmatrix}
      1 & \cos{2\psi_\alpha} & \sin{2\psi_\alpha} \\
      \cos{2\psi_\alpha} & {\cos}^2{2\psi_\alpha} & \frac{1}{4}\sin{4\psi_\alpha} \\
      \sin{2\psi_\alpha} & \frac{1}{4}\sin{4\psi_\alpha} & {\sin}^2{2\psi_\alpha}
    \end{pmatrix}
  \end{split}
\end{equation}
(e.g., \citet{couchot99}).

  While the filter-and-bin method simplifies the linear least-squares solution, the filtering removes not only low-frequency noise but also signals of interest contained in those frequency components.
  As a result, the resultant maps are a biased representation of the signals.
  To characterize the bias, which we call the ``filter transfer function'' (Section~\ref{sec:filter_transfer_function}), many simulations are needed.

\subsection{\label{sec:filter_bin_details}Filter-and-bin Parameters}
  We now describe details of the implementation of the filter-and-bin method for the production of the mid-$\ell$ maps.
  We describe the filters applied to the timestreams, the procedure used to determine the pointing information of the timestream samples, and the weights and pixelization schemes used to bin the timestream samples.
  We performed all these operations on a scan-by-scan basis for each subfield observation and frequency band.

\subsubsection{\label{sec:high_pass_filter_cutoff}High-pass Filter Cutoff}
  To reduce correlated low-frequency noise within and between detector timestreams, we used a high-pass filter that deprojects slowly varying signal templates from every detector timestream.
  The templates consist of Legendre polynomials up to 30th order and sines and cosines up to a cutoff frequency that varies from one scan to another, depending on declination.
  We estimated the best-fit coefficients of each polynomial and sinusoid using linear least-squares fitting and subtracted the fitting result.
  As described in Section~\ref{sec:masking}, the need to mask certain timestream samples when applying the high-pass filter is the reason why we performed what is effectively a Fourier high-pass filter in real space using linear least-squares fitting.

  The varying cutoff frequency of the high-pass filter corresponds to a fixed cutoff in scan-direction multipole number of $\ell_{x,\,c}$ = 300.
  Similar to the relation $\ell \sim 360/\theta$, we define the relation $\ell_x = 360/\theta_x$.
  In the former relation, $\theta$ is the characteristic wavelength (expressed as an angular distance in degrees) of a periodic fluctuation on the celestial sphere, and $\ell$ is the corresponding spherical harmonic multipole number.
  In the latter relation, $\theta_x$ is the wavelength (also expressed as an angular distance in degrees) of a periodic fluctuation along the arc on the celestial sphere traced by a scan, and $\ell_x$ is the corresponding scan-direction multipole number.
  The cutoff frequency, $f_c$, was recalculated for every scan as follows: $f_c = (\ell_{x,\,c}/360)\,v_{\alpha}\,\cos\,\delta$, where $\delta$ is the declination at which the center of the full detector array is pointed, $v_{\alpha}$ is the scan speed in right ascension, which is 1.0 deg/s for every scan.
  Across the declination range of the SPT-3G Main field, $f_c$ ranges from 0.29 to 0.62 Hz.
  The shape of the resulting filter transfer function is dominated by the sinusoid removal, but the low-order polynomials were included to remove long-timescale fluctuations, such as a slope, that are more concisely represented as polynomials than as sinusoids.

  In the filter-and-bin method, the timestreams should have white noise after the filtering for the diagonal-covariance assumption to hold.
  However, the cutoff frequencies reported in the preceding paragraph are in fact not high enough to make the filtered timestreams have white noise.
  We now describe our considerations behind the choice of those cutoff frequencies.

  The timestream from one detector in a scan predominantly contains noise, and the main sources of the $1/f$ part of the noise include the atmosphere and the detector readout system.
  While the $1/f$ noise from the atmosphere is correlated among detectors, the $1/f$ noise from the readout is uncorrelated among detectors but causes correlation among samples within the timestream from each detector.
  As described in \citet[Section~3.3]{bender19}, the $1/f$ part of the readout noise contributes negligibly to the total readout noise beyond $\sim$0.1 Hz, which is below the cutoff frequencies reported earlier.
  However, as described in S22 (Section~7.5), because of the atmospheric $1/f$ noise, the power spectral density of a timestream does not become flat until $~\sim$1 Hz.
  Thus, the cutoff frequencies reported earlier are not high enough to reduce the atmospheric $1/f$ noise to a negligible level.

  Even if the atmospheric $1/f$ noise is not reduced to a negligible level, in which case the covariance matrix still has non-negligible off-diagonal elements, binning the timestream samples by ignoring the off-diagonal elements is not necessarily suboptimal.
  If the scan speed is fast enough that the atmospheric fluctuation patterns change little over the course of a scan (the atmosphere appears ``frozen''), they can be treated as a signal component like the static astrophysical signals, such as the CMB, over the course of that scan.
  In that case, we expect that only the readout $1/f$ noise needs to be removed for the diagonal-covariance assumption to hold.

  Although in principle the cutoff frequencies only need to be high enough to remove the readout $1/f$ noise (and incidentally the atmospheric $1/f$ noise up to those frequencies) under the ideal frozen-atmosphere assumption, in practice we have found that sampling and windowing effects make it important to increase the cutoff frequencies.
  These effects result in low-frequency power in timestreams from the atmosphere affecting not only low-$\ell$ but also high-$\ell$ noise in the resultant maps.
  A model for these effects is described in Appendix \ref{app:high_pass_filter_cutoff}.

  To make a decision on the $\ell_{x,\,c}$ to use to produce the mid-$\ell$ maps, we conducted four test mapmaking runs, in which we used a small subset (5\%) of all the subfield observations and increased $\ell_{x,\,c}$ from 100 to 400.
  In each mapmaking run, we produced two sets of $T/Q/U$ maps in the HEALPix pixelization scheme\footnote{\url{https://healpix.sourceforge.net/}} with a typical pixel size of ${1^\prime}.7$ (the resolution parameter $N_{\mathrm{side}}$ = 2048) for each subfield observation and each frequency band.
  We used the timestreams from only the left-going scans to produce one set of maps and the timestreams from only the right-going scans to produce the other set.
  Then, for each scan direction, we added the maps from all the subfield observations together.
  After that, we subtracted the ``left-going'' coadded $T/Q/U$ maps from the ``right-going'' ones to remove the common astrophysical signals and obtain noise maps.
  Finally, we calculated the 2D $TT$ and $EE$ noise spectra (${|a_{{\ell}m,T}|}^2$ and ${|a_{{\ell}m,E}|}^2$, or the square of the amplitudes of the spherical harmonic coefficients of the temperature and $E$-mode polarization anisotropies, of the noise maps) using HEALPix tools and compared the four cases.

\begin{figure*}
  \includegraphics[scale=1.00]{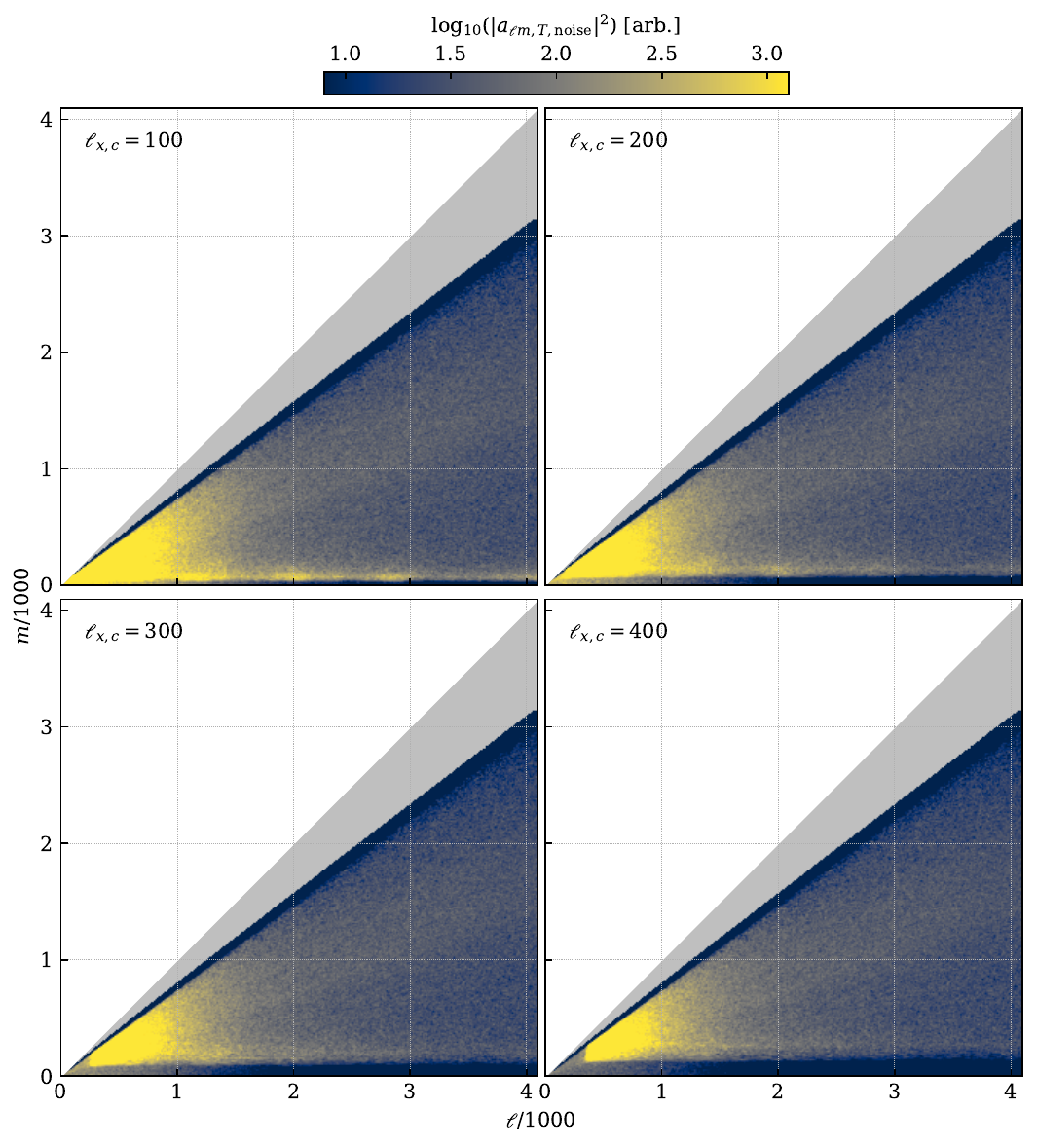}
  \caption{\label{fig:noise_tt_alm_triangle}%
  The 2D $TT$ noise spectra in the 150\;GHz band obtained from the four test mapmaking runs.
  The top left corner of each panel indicates the $\ell_{x,\,c}$ used in a particular mapmaking run.
  The color represents the logarithm of the noise power with an arbitrary normalization and is saturated in a large fraction of the bright region near the bottom left corner.
  The spherical harmonic coefficients are 0.0 in the gray region in each panel.
  This region is above the line $m \sim \ell\,\cos\,\delta_{\mathrm{low}}$, where $\delta_{\mathrm{low}}$ is the lowest declination covered by the SPT-3G Main field ($-40$ degrees).
  The high-$m$ modes in the gray region are concentrated near the equator of the spherical coordinates and have little overlap with the SPT-3G Main field.
  As a result, their spherical harmonic coefficients are practically zero.}
\end{figure*}

  One outcome of increasing $\ell_{x,\,c}$ is to remove noisy spherical harmonic modes that have low $m$.
  This effect can be seen in Figure~\ref{fig:noise_tt_alm_triangle}, which shows the 2D $TT$ noise spectra in the 150\;GHz band in the four cases.
  Because the spherical coordinates are defined such that the azimuthal angle is equal to right ascension and because a scan has constant declination, power in timestreams at low frequencies generally contributes to power in the resultant maps at spherical harmonic modes with low $m$.
  Given the definition of $\ell_x$ provided earlier, $m = \ell_x\,\cos\,\delta$ (when the quantities on the right side are equal to an integer).
  As $\ell_{x,\,c}$ is increased from 100 to 400, the high-pass filter suppresses more low-$m$ modes, and the low-$m$ bright horizontal band visible in the $\ell_{x,\,c}$ = 100 case, which contributes to both low-$\ell$ and high-$\ell$ noise, are absent in the $\ell_{x,\,c}$ = 400 case.
  The removal of the low-$m$ noisy modes in fact does not have to be achieved by increasing $\ell_{x,\,c}$.
  It can be achieved by filtering the timestreams with a low $\ell_{x,\,c}$, calculating the spherical harmonic coefficients of the resultant maps, and downweighting the noisy low-$m$ coefficients in further analysis.

\begin{figure}
  \includegraphics[scale=1.00]{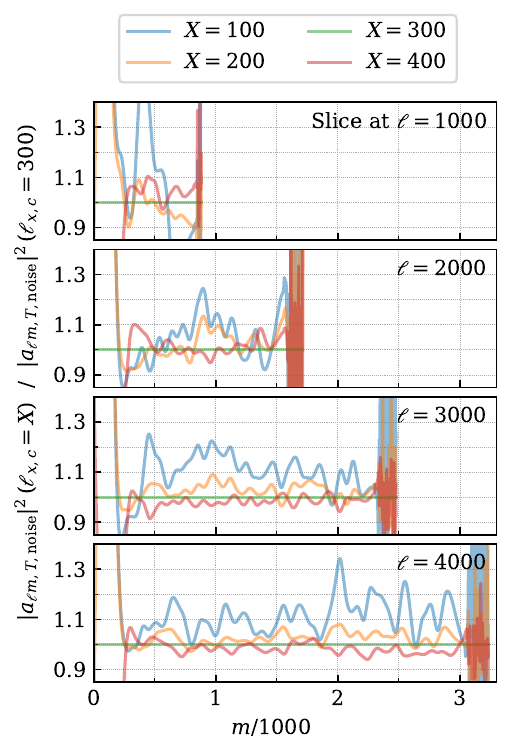}
  \caption{\label{fig:noise_tt_alm_xsec_ratio}%
  The ratio of each 2D power spectrum shown in Figure~\ref{fig:noise_tt_alm_triangle} to the power spectrum in the $\ell_{x,\,c}$ = 300 case along four constant-$\ell$ slices.
  Each panel shows the four ratios at a particular $\ell$, which is indicated in the upper right corner.
  The color of each ratio represents the spectrum used in the numerator.
  The denominator is the spectrum in the $\ell_{x,\,c}$ = 300 case for every ratio.}
\end{figure}

  The outcome that we mainly wanted to achieve by increasing $\ell_{x,\,c}$ is in fact to generally improve noise in spherical harmonic modes that have high $\ell$ and $m$ and whose $m$ are well above what the filter removes.
  This effect is not large enough to be easily seen in Figure~\ref{fig:noise_tt_alm_triangle}, and we use Figure~\ref{fig:noise_tt_alm_xsec_ratio} to show the effect.
  Figure~\ref{fig:noise_tt_alm_xsec_ratio} shows ratios of the spectra shown in Figure~\ref{fig:noise_tt_alm_triangle}.
  From each of the four triangles, we extracted the vertical slice at each of four $\ell$: 1000, 2000, 3000, and 4000.
  Then, we smoothed the values in each slice with a Gaussian kernel ($\sigma$ = 30) and divided them by the corresponding smoothed values from the $\ell_{x,\,c}$ = 300 case.
  The figure shows that increasing $\ell_{x,\,c}$ from 100 to 400 generally improves noise in all the modes by $\sim$10\%, and the amount of improvement becomes smaller for each increment in $\ell_{x,\,c}$.
  The effect in the 2D $EE$ noise spectra in the 150\;GHz band is similar and can be seen in the figures in Appendix \ref{app:high_pass_filter_cutoff}.
  In the 220\;GHz band, the improvement achieved by increasing $\ell_{x,\,c}$ from 100 to 400 is also $\sim$10\% in both $TT$ and $EE$.
  In the 95\;GHz band, the improvement is only a few percent.
  Using these comparison results, we set $\ell_{x,\,c}$ to 300 as a good compromise between the noise improvement and mode loss.

  A comparison of real noise spectra from these tests and a model for the leakage of atmospheric noise from low to high $\ell$ is shown in Appendix~\ref{app:high_pass_filter_cutoff}.
  The agreement between model and data indicates that the only departure from the frozen-screen atmosphere assumption is caused by the finite sampling of the full detector array, and that, with sufficient high-pass filtering, the departure from the frozen-screen model is not detectable.
  This implies that, for the modes preserved by the filtering, the filter-and-bin mapmaking procedure is close to formally optimal even in the case in which residual atmospheric noise creates a non-white detector noise spectrum.
  Furthermore, as shown in C25 (Figure~10), we obtained nearly flat 1D $EE$ noise spectra by using $\ell_{x,\,c}$ = 300, so the earlier argument is only relevant for $T$ mapmaking; the $Q/U$ maps described here are nearly optimal regardless of the correctness of the model described in Appendix~\ref{app:high_pass_filter_cutoff}.

\subsubsection{\label{sec:masking}Masking in High-pass Filter}
  When applying the high-pass filter to each timestream, we masked timestream samples near and on high-S/N objects such as bright active galactic nuclei (AGN) to prevent an undesirable feature in the resultant maps.\footnote{
  This process is called ``mapmaking masking'' in C25.}
  If the linear least-squares fitting is performed on a timestream that includes a bright AGN, the source dominates the estimates of the best-fit coefficients of the templates to be subtracted, and the resulting image of the source in the map is high-pass filtered in the scan direction, producing a scan-direction feature that can extend for many degrees across the map (see, e.g., \citep[Figure~1]{vieira10}).
  Following other SPT publications, including C25, we call this feature ``filtering wings'' in this work.

  Having filtering wings would not be an issue if we were willing to leave all the sources in a map for post-map analysis.
  The filtering wings represent missing low-frequency power of the sources, which would be properly corrected for, together with the missing low-frequency power of the other astrophysical signals, when a spectrum measured from the map is divided by the filter transfer function.
  However, map pixels near and on high-S/N objects are often inpainted or masked in power spectrum and lensing reconstruction analyses, and filtering wings would complicate these steps.
  For example, they would necessitate significant extra pixels that need to be inpainted or masked.

  Masking the timestream samples within a certain radius of each known high-S/N object (simply not using those timestream samples) in the linear least-squares fitting prevents the sources from contributing to the best-fit coefficients of a template.
  When the best-fit template is then subtracted from the entire timestream (both unmasked and masked samples), the sources in the filtered timestream have no low-frequency power missing, and hence no filtering wings in the resultant map.

  Although the masking preserves low-frequency power of the sources, it also prevents clean removal of low-frequency power of the other components in the timestreams, such as the CMB, not only inside but also outside the masked regions.
  Given a simulated timestream that does not contain any high-S/N object, if some timestream samples are masked regardless when the best-fit coefficients of templates are calculated, these coefficients differ from the coefficients calculated by using all the samples.
  Using the coefficients estimated with the masking then causes less clean subtraction of the templates and leaves a small residual across the entire timestream.
  As shown in C25 (Appendix~A\,1), the difference between maps made from simulated timestreams filtered with and without the masking has oscillatory structures, and the amplitudes are larger in pixels closer to the masked regions.
  In spherical harmonic space, the oscillatory structures exist as low-$m$ spherical harmonic modes that should ideally be removed by the high-pass filter.
  These structures are called ``filtering artifacts'' in C25.
  We treated the filtering artifacts demonstrated in the simulations as an additive bias present in the $TT/TE/EE$ spectra measured from the mid-$\ell$ maps and subtracted the bias from the measurements.
  Details of this method are described in C25 (Section~IV\,A\,1).

  Across the SPT-3G Main field, we masked 2118 emissive sources and 537 galaxy clusters when applying the high-pass filter to the timestreams.
  These objects were detected in maps produced in a different mapmaking run (the discrete-source maps) and filtered to enhance the detection significance of the objects.
  The emissive sources are the ones detected at greater than 16 $\sigma$, or equivalently a flux of 6 mJy, in 150\;GHz maps made using observations taken in the 2019, 2020, and 2021 observing seasons of the SPT-3G Main survey.
  The galaxy clusters are the decrements detected at greater than 10 $\sigma$ (from the thermal Sunyaev-Zel'dovich effect) in the 95 and 150\;GHz maps made using observations taken in the 2019 and 2020 observing seasons.

  We masked each object with a radius determined by the detection significance.
  For each emissive source, we chose the radius at which the product of the detection significance and the radial profile of the point spread function of the telescope falls below unity.
  Sources at the low-flux threshold were typically masked to 2$^\prime$ or 3$^\prime$, while the brightest source (PKS 0208-512) was masked to 15$^\prime$.
  For each galaxy cluster, we used the same criterion except that the point spread function was replaced with the convolution of the point spread function and an approximate radial profile of the galaxy cluster determined as part of the detection.

\subsubsection{\label{sec:low_pass_filter}Low-pass Filter}
  We also applied an anti-aliasing filter to reduce high-frequency power in each timestream before we binned its samples into pixels.
  The filter was applied in Fourier space and has the following form:
\begin{equation}
  F^\mathrm{AA}(f) = e^{-\left( f/f_\mathrm{c} \right)^6}.
\end{equation}
  As is the case with the high-pass filter, we adjusted the cutoff frequency for every scan to have a fixed cutoff of $\ell_{x,\,c}$ = 13\,000.
  Across the declination range of the SPT-3G Main field, $f_c$ ranges from 12 to 27 Hz, which is lower than the Nyquist frequency associated with the sampling rate of the timestreams by at least 10 Hz.\footnote{
  As described in the beginning Section~\ref{sec:mapmaking}, we used the downsampled timestreams to produce the mid-$\ell$ maps, and the Nyquist frequency is 38.2 Hz.}
  Removing this much bandwidth using the masked linear least-squares fitting adopted for the high-pass filter would have required too many sinusoidal templates to fit and would have been too computationally expensive.
  As a result, we applied the anti-aliasing filter using fast Fourier transforms.
  Because the timestream samples near the high-S/N objects were not masked in this filtering step, some amount of Fourier ringing is expected around them.
  However, the chosen $\ell_{x,\,c}$ is high enough that there is not any noticeable ringing in the maps even around the brightest source.

\subsubsection{\label{sec:pointing}Pointing}
  Another step before binning the timestream samples was to calculate the pointing information of each sample.
  For this, we used a model of the telescope pointing and corrections derived by observing astrophysical sources.
  We briefly summarize this approach here, and a full description can be found in \citet{chichura25}.
  The pointing model incorporates non-idealities in the telescope structure such as flexure of structural components and tilts of the azimuth and elevation bearings relative to local gravity.
  The best-fit parameters of the model are determined through dedicated observations of many bright point-like sources across the sky.

  The model parameters are assumed to be constant in time, but for SPT certain parameters are empirically determined to vary on several-hour timescales.
  As a result, even though the relative pointing within a subfield over the course of one observation is calculated accurately using the model, there are errors in the absolute pointing from observation to observation.
  To correct for these errors, we compared the positions of bright AGN in the 95\;GHz $T$ maps of the individual observations to known positions (from the AT20G catalog, \citet{murphy10}).
  The measured positional offsets were used to calculate corrections to the two pointing parameters known to vary on these timescales, the pointing model was recalculated with these parameter corrections, and maps were re-made.
  In a typical map, at least five sources are detectable at greater than 10 $\sigma$, and, with the corrected parameters, the observation-to-observation pointing error is reduced to the few-arcsecond level.

\subsubsection{\label{sec:binning}Binning}
  After filtering each timestream and calculating the pointing information of each timestream sample, we binned the sample with an inverse-variance weight into the corresponding pixel in two pixelization schemes.
  We calculated the weight using the following equation and used the same weight for every sample of a timestream:
\begin{equation}
  w_{i\alpha} = \left( \frac{1}{f_h - f_l} \int_{f_l}^{f_h} \mathrm{PSD}_{\alpha}(f)\;df \right)^{-1},
\end{equation}
where $w_{i\alpha}$ is the weight used in Equations~\ref{eqn:t_solution}, \ref{eqn:weighted_tqu}, and \ref{eqn:weight_matrix}, and $\mathrm{PSD}_{\alpha}(f)$ is the power spectral density of the timestream from detector $\alpha$ to which the timestream sample $i\alpha$ belongs.
  We adjusted the limits of the integration, $f_l$ and $f_h$, for every scan so that they correspond to fixed limits of $\ell_x$ at 320 and 4000, respectively.\footnote{
  The lower limit is slightly higher than the high-pass cutoff to avoid including frequency components removed by the filter.}
  We chose this $\ell_x$ range to optimize the 1D $TT$ and $EE$ noise spectra of the resultant maps in the $\ell$ range [320, 4000], which encompasses the range used in C25.

  We binned the weighted samples to build two sets of weighted $T/Q/U$ maps (Equation~\ref{eqn:weighted_tqu}) and weight matrices (Equation~\ref{eqn:weight_matrix}) for each subfield observation.
  One set uses the HEALPix pixelization scheme with a typical pixel size of ${0^\prime}.43$ (the resolution parameter $N_{\mathrm{side}}$ = 8192), while the other set uses the Lambert azimuthal equal-area projection (ZEA, a flat-sky projection, \citet{calabretta02}) pixelized with ${0^\prime}.5625$ square pixels.
  We used the HEALPix maps in C25 and O26 and the flat-sky maps in G25.
  In the latter, the flat-sky maps were downsampled by a factor of four to have ${2^\prime}.25$ pixels.

\subsubsection{\label{sec:other_pipelines}Other Pipelines}
  As described in the beginning of Section~\ref{sec:mapmaking}, in addition to the mid-$\ell$ maps, other sets of maps have been produced using the SPT-3G D1 timestreams to enable different analyses.
  Those maps were produced using filter-and-bin pipelines that have different high-pass and low-pass filter cutoffs, numbers of masked high-S/N objects, weighting methods, and pixelization schemes.

  For example, the low-$\ell$ maps used in Z25 are a set of HEALPix \nside\ 2048 maps created with a minimal high-pass filter (only a 10th-order polynomial and no sinusoids deprojected) and weights calculated from ``pair-differenced'' timestreams instead of timestreams of individual detectors.\footnote{
  A pair-differenced timestream is the difference between the timestreams from a pair of detectors sensitive to orthogonal polarization directions.}
  These settings were chosen to optimize the sensitivity to degree-scale polarization anisotropy.
  At the other extreme are the discrete-source maps.
  These maps were made in the ZEA projection with ${0^\prime}.25$ pixels and made with more aggressive high-pass filters to reduce atmospheric $1/f$ noise.
  The high-pass filters include an FFT-based high-pass filter with $\ell_{x,\,c}$ = 500 and a common-mode filter that removes the mean timestream over all the detectors in the same frequency band and on the same detector wafer.
  These maps also have high-S/N objects interpolated over before (instead of masked during) filtering, which allows the high-pass filter to be applied in Fourier space and faster.

  As described in Section~\ref{sec:introduction}, two datasets from SPT-3G have been used to measure the $TT/TE/EE$ and $\phi\phi$ spectra: the 2018 dataset and the 2019--2020 dataset (SPT-3G D1).
  In Appendix~\ref{app:changes_from_2018}, we highlight important changes between the SPT-3G D1 mid-$\ell$ maps and the 2018 counterparts.

\subsection{\label{sec:coadds}Coadded Maps}
  After producing the weighted $T/Q/U$ maps and weight map matrix for each subfield observation and frequency band (hereafter individual-observation maps) using the filter-and-bin method and parameters described in Sections~\ref{sec:filter_bin_method} and \ref{sec:filter_bin_details}, we combined the individual-observation maps to produce coadded $T/Q/U$ maps.
  We describe fives types of coadded $T/Q/U$ maps that we produced for different analysis tasks involved in the measurements of the $TT/TE/EE$ and $\phi\phi$ spectra and show full-depth coadds in a series of figures.

\subsubsection{\label{sec:coadding_procedure}Coadding Procedure}
  In general, we produce a set of coadded $T/Q/U$ maps by combining the individual-observation maps from many observations through a weighted average as follows:
\begin{equation}
\label{eqn:coadd}
  {\begin{pmatrix}
    \bar{T}(\hat{n}) \\ \bar{Q}(\hat{n}) \\ \bar{U}(\hat{n}) \\
  \end{pmatrix}}_{c}
  = \left[ \sum\limits_{j} W_j (\hat{n}) \right]^{-1}
  \sum\limits_{j}
  \begin{pmatrix}
    \bar{T}_j^W(\hat{n}) \\ \bar{Q}_j^W(\hat{n}) \\ \bar{U}_j^W(\hat{n}) \\
  \end{pmatrix},
\end{equation}
where the subscript $c$ on the left-hand side denotes the coadded maps, $\bar{T}^W(\hat{n})$, $\bar{Q}^W(\hat{n})$, and $\bar{U}^W(\hat{n})$ are defined in Equation~\ref{eqn:weighted_tqu}, $W(\hat{n})$ is defined in Equation~\ref{eqn:weight_matrix}, and the index $j$ on the right-hand side runs over individual observations.
  This coadding process is equivalent to binning the timestream samples from multiple observations at once.
  In Z25 there is another step in the coadding process, which is to measure noise spectra of each individual-observation $Q/U$ maps and rescale the weight map matrix for that observation accordingly to optimize low-$\ell$ noise.\footnote{
  This additional weighting step was introduced to mitigate correlated low frequency noise between detectors from polarized atmosphere, which is not detectable in individual detector or pair-differenced timestreams.}
  Hereafter, we call a group of three weight-removed coadded maps in one frequency band produced by Equation~\ref{eqn:coadd} ($\bar{T}(\hat{n})$, $\bar{Q}(\hat{n})$, and $\bar{U}(\hat{n})$) a ``coadd.''

  To create the final mid-$\ell$ coadds, we first combined all the individual-observation maps to produce five types of coadds for each subfield, which we call the following: \texttt{full}, \texttt{half}, \texttt{one-thirtieth}, \texttt{pre-null}, and \texttt{signflip-noise}.

  All except the \texttt{signflip-noise} coadds were created using Equation~\ref{eqn:coadd}, with $j$ running over different sets of observations.
  To produce the \texttt{full} coadd, which is the full-depth signal coadd, we simply coadded all the individual-observation maps.
  To produce the \texttt{half} coadd, which is a pair of half-depth signal coadds, we randomly divided all the individual-observation maps into two parts of equal map weight and coadded the maps within each part.
  We used the \texttt{full} and \texttt{half} coadds mainly to estimate coefficients needed to calibrate and clean all the coadds as described in Section~\ref{sec:map_calibration}.
  The \texttt{one-thirtieth} coadds, which form a set of 30 partial-depth coadds, were produced in the same way as the \texttt{half} coadds except that 30 parts were used instead of two parts.
  Hereafter, we use the term ``bundle'' to mean a partial-depth coadd.
  We used the \texttt{one-thirtieth} coadds to measure the $TT/TE/EE$ spectra through a cross-spectrum approach (C25, Section~IV\,D).
  To produce the \texttt{pre-null} coadds, which are six sets of 50 bundles, we divided all the individual-observation maps not randomly but in ways that allow us to probe potential systematic errors in the \texttt{full}, \texttt{half}, and \texttt{one-thirtieth} coadds using the null tests described in Section~\ref{sec:map_level_null_tests}.

  The \texttt{signflip-noise} coadds retain the same noise properties as the \texttt{full} coadds and were constructed with two extra steps applied to the individual-observation maps before coadding.
  In one step, to remove the astrophysical signals in the individual-observation maps, we subtracted the \texttt{full} coadd from them.
  In the other step, to create many (500) full-depth noise coadds that have different noise patterns, we multiplied the individual-observation maps by a random sequence of $+1$s and $-1$s when constructing each \texttt{signflip-noise} coadd.\footnote{
  The way in which the weight map matrices associated with the individual observations were used in these two steps is described in Appendix~\ref{app:weight_map_matrix}.}
  The assignment of $+1$s and $-1$s used for each coadd was determined by randomly permuting all the observations, splitting the list of observations into two halves at a point where the total weights of the two halves are nearly equal, and assigning $+1$ to the first half and $-1$ to the second half.\footnote{
  The total weight of each half is defined as the sum of all the pixel values in the first diagonal element of the coadded weight map matrix from each half.}
  For a given individual observation, the same sign was used for all three frequency bands to retain the interfrequency noise correlation structure present in the \texttt{full} coadds.

  After producing the five types of coadds for each subfield, we combined these subfield coadds to produce the full-field coadds.
  For example, to produce the full-field coadd for the fifth bundle of the \texttt{one-thirtieth} coadds, we combined the fifth bundle from each subfield.
  Because two adjacent subfields have an overlap in their non-uniform coverage regions, the overlapped region has the weighted average of the pixel values from the two subfields.
  For the \texttt{full}, \texttt{half}, \texttt{one-thirtieth}, and \texttt{signflip-noise} coadds, prior to combining the four subfields, we calibrated and cleaned the subfield coadds as described in Section~\ref{sec:map_calibration}.
  For the \texttt{pre-null} coadds, we combined the subfields without the calibration and cleaning steps because we did not expect the steps to significantly affect the null tests.
  We make the five types of full-field coadds publicly accessible.
  The access information can be found in Appendix~\ref{app:data_availability}.

\subsubsection{\label{sec:full_depth_coadds}Full-depth Coadds}
  The full-field \texttt{full} and \texttt{signflip-noise} coadds serve as a summary of SPT-3G D1 as they were produced by adding all the individual-observation maps.
  Here we show some of the coadds and describe their characteristics using a series of figures.

\begin{figure}
  \includegraphics[scale=1.00]{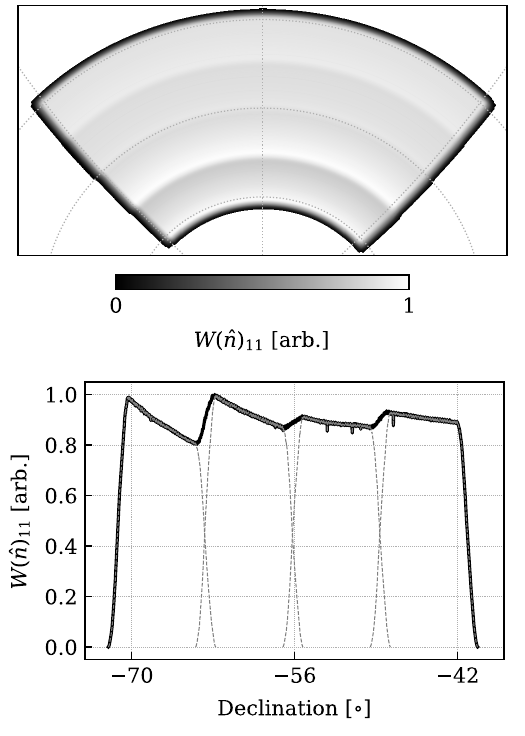}
  \caption{\label{fig:coadds_spt_only_weight_tt}%
  The first diagonal element of the weight map matrix associated with the 95\;GHz full-field \texttt{full} coadd (top) and the cross section of the weight map along the contour of constant right ascension at $0^{\mathrm{h}}$ (bottom).
  The weight map is normalized to its maximum value.
  The contours of constant declination and those of constant right ascension in the top panel are the same as the ones shown in Figure~\ref{fig:survey_field_subdivision}.
  In the bottom panel, each gray dashed cross section is the weight map associated with one subfield.
  The sum of the four gray cross sections is equal to the black cross section.}
\end{figure}

  The \texttt{full} coadds cover the target uniform coverage region of the SPT-3G Main field with good uniformity.
  Figure~\ref{fig:coadds_spt_only_weight_tt} shows a normalized version of the first diagonal element of the weight map matrix associated with the 95\;GHz full-field \texttt{full} coadd and a cross section of this weight map.
  The weight map is a representation of the noise variance in the pixels across the full field.
  All the low-weight regions are outside the target uniform coverage region as expected, and the median weight of a subfield varies by less than 10\% across the full field.

  Two features of the weight map are worth noting.
  First, within each subfield, the weight has a gradient along declination.
  This is caused by the fact that the telescope scan speed in right ascension is the same at different declination.
  As declination decreases, each scan covers a smaller angular extent (${\sim}100\, \cos\,\delta$, where 100 is the angle covered in degrees in right ascension, and $\delta$ is the declination of a scan), but the scan speed remains the same, so the observing time per unit area increases, which causes the weight per pixel to increase.
  Second, the several small notches revealed in the cross section are caused by high-S/N emissive sources.
  These sources are bright enough to make non-negligible contributions to the variance of the timestreams of the detectors that scan directly over the sources, which causes a decrease in the weights.
  While this is formally suboptimal and will be fixed in future mapmaking runs by masking sources when calculating the variance on which the weights are based, the notches have no practical effect on any downstream analyses.

\begin{figure*}
  \includegraphics[scale=1.00]{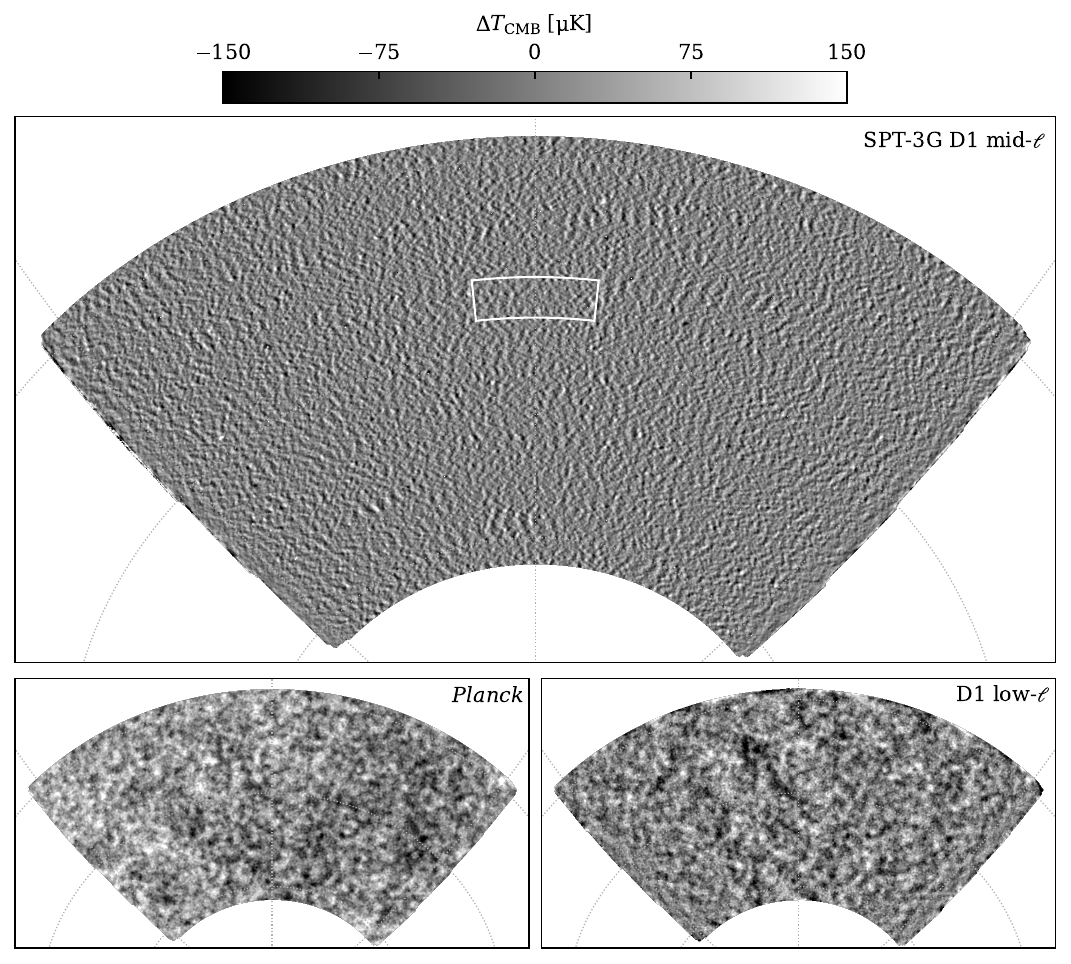}
  \caption{\label{fig:coadds_spt_planck_comparison_t}%
  The $T$ map of the 150\;GHz SPT-3G D1 mid-$\ell$ \texttt{full} coadd (top) and, for comparison, the same region of the \textit{Planck} PR3 full-mission 143\;GHz $T$ map (bottom left) and the $T$ map of the 150\;GHz SPT-3G D1 low-$\ell$ full-depth coadd.
  All the maps are shown in the ZEA projection used in Figure~\ref{fig:survey_field_subdivision}, and the contours of constant declination and those of constant right ascension shown here are the same as the ones shown in that figure.
  For the top panel, the gray scale ranges from $-150$ to $150$ $\mathrm{{\mu}K}$ (approximately ${\pm}4$ $\sigma$ of the distribution of the pixel values of the mid-$\ell$ map).
  For the bottom two panels, the range is from $-300$ to $300$ $\mathrm{{\mu}K}$ (approximately ${\pm}4$ $\sigma$ of the pixel values of the \textit{Planck} map).
  The small box with white edges in the top panel represents the small region shown in Figures~\ref{fig:coadds_spt_only_signal_tqu_strip} and \ref{fig:coadds_spt_only_noise_tqu_strip}.}
\end{figure*}

  Because of the scan-direction timestream high-pass filter, the CMB anisotropies in the \texttt{full} coadds are suppressed along contours of constant declination.
  This can be seen in Figure~\ref{fig:coadds_spt_planck_comparison_t}, which shows the $T$ map of the 150\;GHz mid-$\ell$ \texttt{full} coadd and compares it with the same region of the \textit{Planck} PR3 full-mission 143\;GHz $T$ map.\footnote{
  The \textit{Planck} map is the Stokes $I$ map in the file \texttt{HFI\_SkyMap\_\\143-field-IQU\_2048\_R3.00\_full.fits} downloaded from Planck Legacy Archive (\url{https://pla.esac.esa.int/pla/}).}
  The figure also compares the mid-$\ell$ map with its counterpart from the low-$\ell$ mapmaking run (Section~\ref{sec:other_pipelines}) conducted for Z25.
  While the low-$\ell$ map shows that we have access to large angular-scale CMB fluctuations, we chose to remove them in the mid-$\ell$ mapmaking run to improve noise  at small angular scales (Section~\ref{sec:high_pass_filter_cutoff}).

\begin{figure*}
  \includegraphics[scale=1.00]{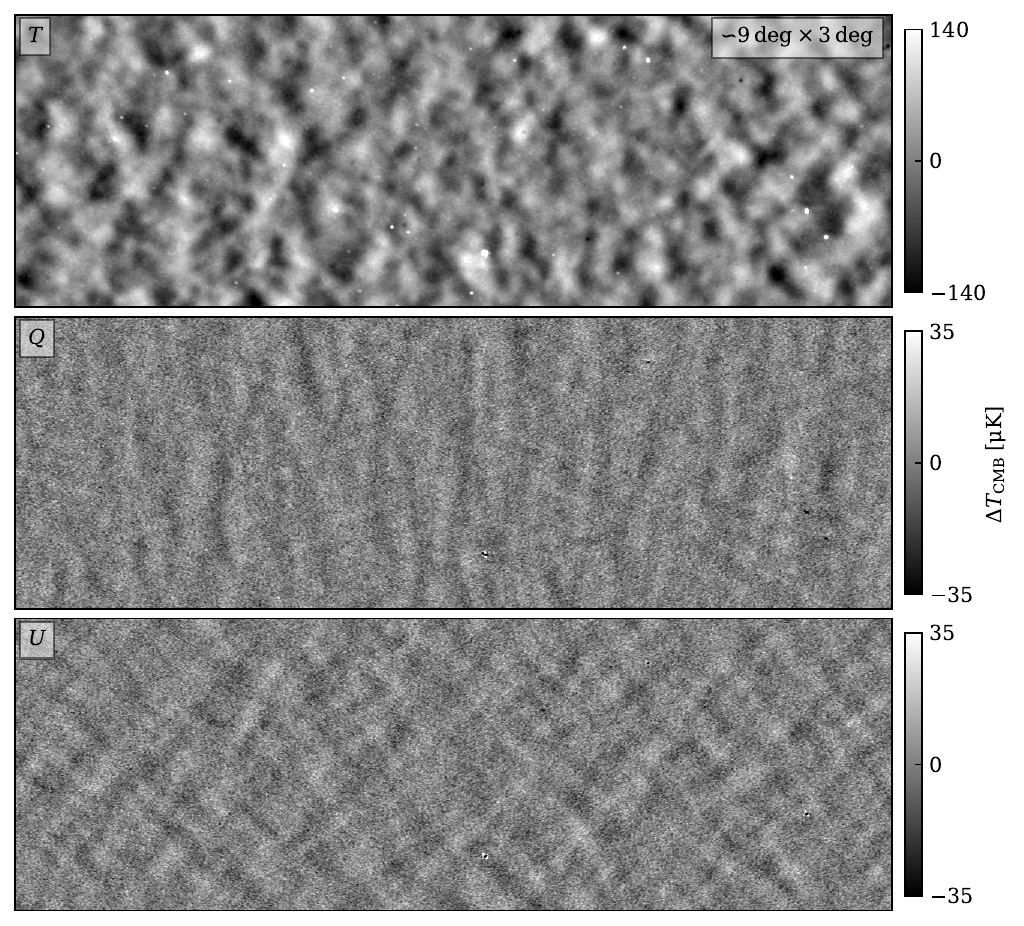}
  \caption{\label{fig:coadds_spt_only_signal_tqu_strip}%
  The 150\;GHz \texttt{full} coadd in a small region around the center of \texttt{el1} in the Gnomonic projection.
  This region is approximately 9-degree wide and 3-degree tall (the small box with white edges shown in Figure~\ref{fig:coadds_spt_planck_comparison_t}).
  Each panel shows one of the three Stokes parameters as indicated in the top left corner.
  The gray scale for each panel represents approximately ${\pm}4$ $\sigma$ of the distribution of the pixel values in that region.}
\end{figure*}

\begin{figure*}
  \includegraphics[scale=1.00]{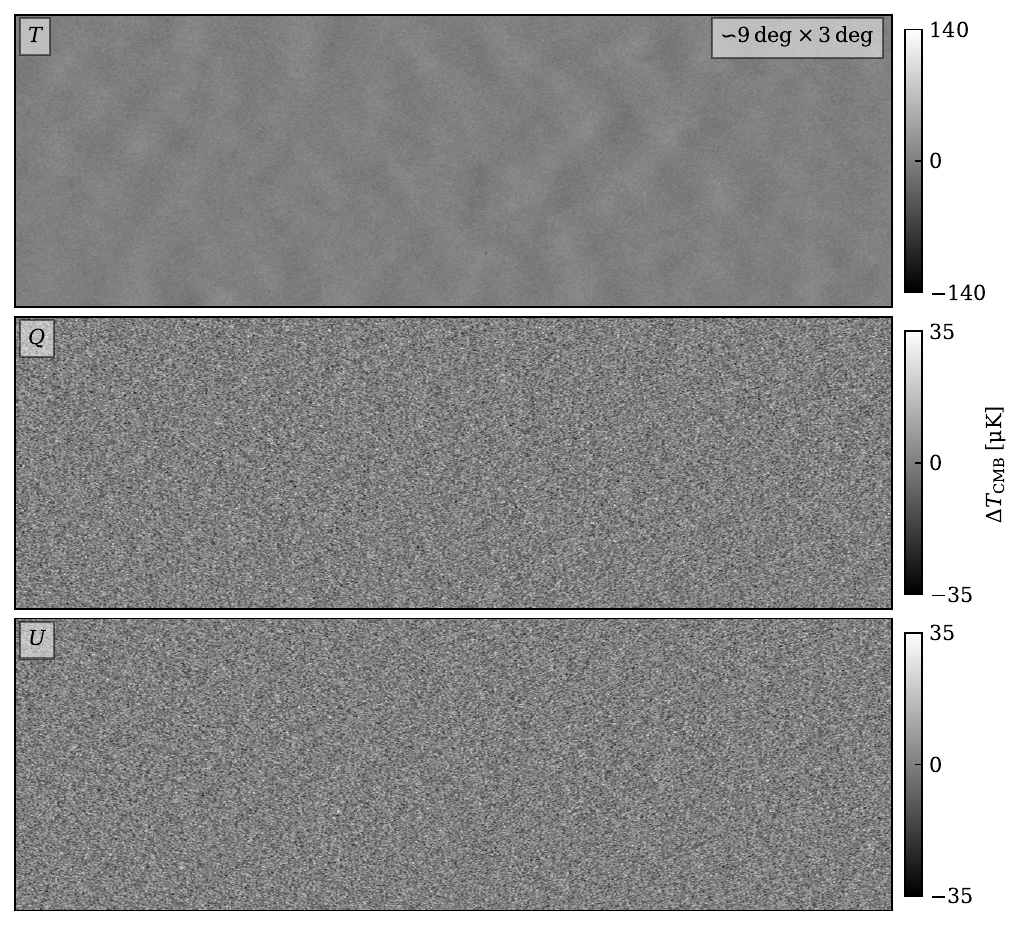}
  \caption{\label{fig:coadds_spt_only_noise_tqu_strip}%
  A 150\;GHz \texttt{signflip-noise} coadd in the same region and projection as those used in Figure~\ref{fig:coadds_spt_only_signal_tqu_strip}.
  The gray scales represent the same ranges of values as before so that the signal and noise are compared on the same color scales.}
\end{figure*}

  Another property of the high-pass filter is that, for a sky in which the polarized signal is dominated by $E$ modes, the filter suppresses CMB $Q$ anisotropy more than $U$.
  Figure~\ref{fig:coadds_spt_only_noise_tqu_strip} shows a small region of the 150\;GHz \texttt{full} coadd in a Gnomonic projection.
  The fact that the CMB polarization is dominated by the $E$-mode polarization means that the $Q$ map in Figure~\ref{fig:coadds_spt_only_signal_tqu_strip} should have crisscross patterns along the horizontal and vertical directions and that the $U$ map should have diagonal crisscross patterns.
  While the $U$ map shows clear crisscross patterns along the diagonal directions, the patterns in the $Q$ map are less obvious; horizontal stripes are not seen as clearly as vertical stripes.
  This is again because of the high-pass filter, which suppress anisotropies along the horizontal direction.

  To estimate the power spectra of the noise fluctuations in the \texttt{full} coadd for each frequency band, we used the \texttt{signflip-noise} coadds.
  Figure~\ref{fig:coadds_spt_only_noise_tqu_strip} shows one 150\;GHz \texttt{signflip-noise} coadd in the same small region as the one used in Figure~\ref{fig:coadds_spt_only_signal_tqu_strip}.
  While the dominant feature in the $T$ map is 1/$\ell$ noise from the atmosphere, noise in the $Q/U$ maps resembles white noise.
  Average $TT$ and $EE$ noise spectra over all the 500 \texttt{signflip-noise} coadds (after applying the calibration and cleaning steps described in Section~\ref{sec:map_calibration}) are shown in C25 (Figure~10).
  From the spectra, we calculated the white noise levels of the $T$ maps of the \texttt{full} coadds to be 5.4, 4.4, and 16.2 $\mathrm{\mu}$K--arcmin in the 95, 150, and 220\;GHz bands, respectively, and calculated the white noise levels of the $Q/U$ maps to be 8.4, 6.6, and 25.8 $\mathrm{\mu}$K--arcmin.

  SPT-3G D1 is from a small but deep survey that complements the shallower but wider surveys from \textit{Planck} and ACT.
  Table~\ref{tab:survey_statistics_spt_planck_act} compares the depth and observed sky fraction of three datasets: SPT-3G D1, ACT DR6 \citep{naess25}, and \textit{Planck} PR3 \citep{planck18-1}.
  While SPT-3G D1 covers only 4\% of the sky, it is three and eight times deeper than ACT DR6 and \textit{Planck} PR3, respectively.

\begin{figure*}
  \includegraphics[scale=1.00]{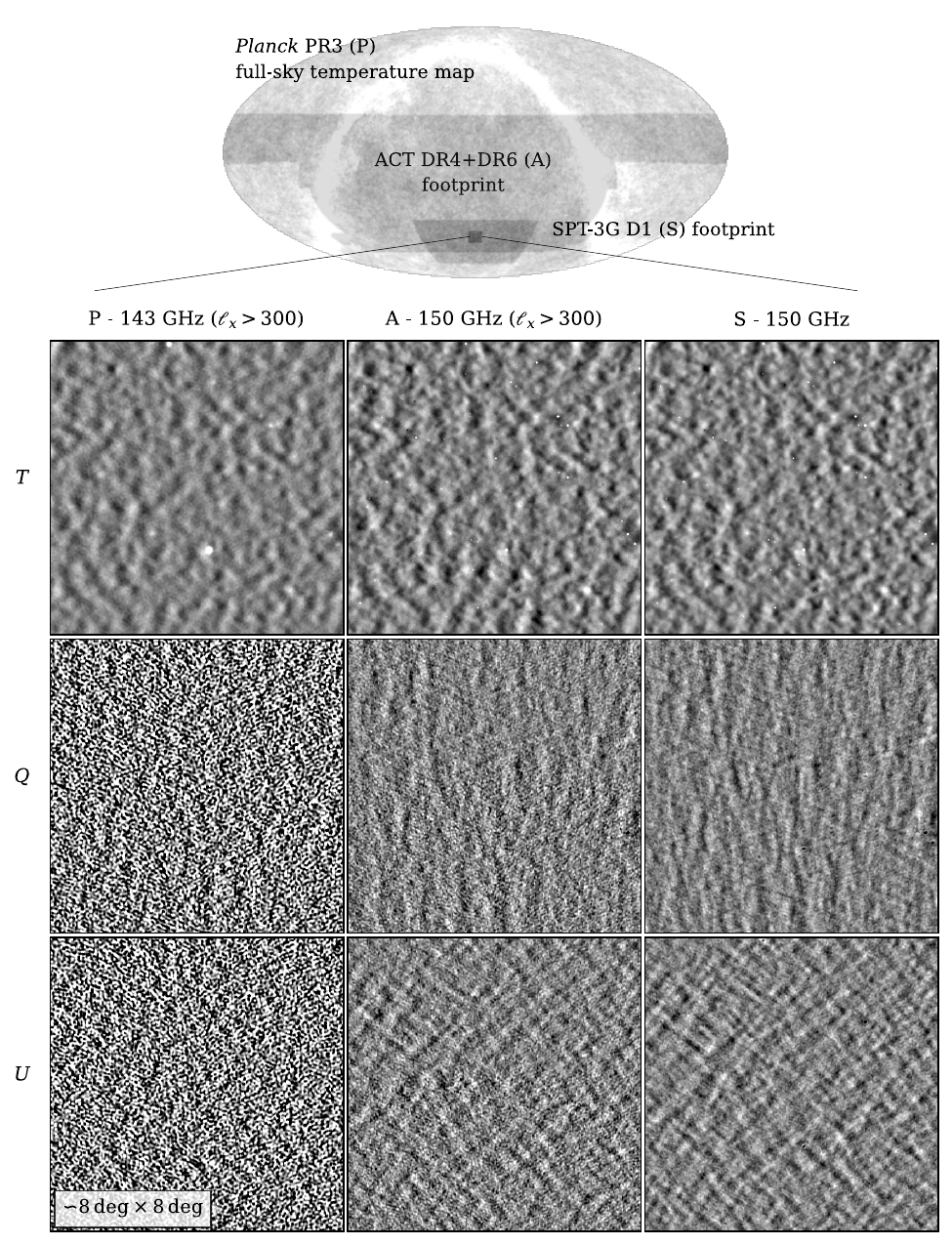}
  \caption{\label{fig:coadds_spt_act_planck_comparison_tqu_filtered}%
  A comparison of $T/Q/U$ maps from SPT-3G D1, ACT DR4+DR6, and \textit{Planck} PR3.
  The limits of the gray scales for the $T$ ($Q/U$) rows are ${\pm}150$ (${\pm}15$) $\mathrm{{\mu}K}$, approximately ${\pm}4$ $\sigma$ of the SPT map pixel values.
  The ACT and \textit{Planck} maps were high-pass filtered to remove the large angular-scale information that the mid-$\ell$ mapmaking removes.}
\end{figure*}

  Figure~\ref{fig:coadds_spt_act_planck_comparison_tqu_filtered} shows the observed sky areas of the three datasets and $T/Q/U$ maps of a small region of the sky (around the center of \texttt{el1} in the Gnomonic projection, approximately 8-degree wide and tall) from each dataset in the frequency band near 150\;GHz.\footnote{
  The \textit{Planck} maps were downloaded from Planck Legacy Archive (\url{https://pla.esac.esa.int/pla/}). The maps are the 143\;GHz full-mission maps stored in the file \texttt{HFI\_SkyMap\_143-field-IQU\_\\2048\_R3.00\_full.fits}. The ACT maps were downloaded from NASA LAMBDA (\url{https://lambda.gsfc.nasa.gov/product/act/act_dr6.02/}). The maps contain data from both DR6 and their previous, smaller dataset DR4 and are stored in the file \texttt{act\_dr4dr6\_coadd\_AA\_daynight\_f150\_map\_healpix.fits}.}
  Because the SPT-3G D1 mid-$\ell$ maps do not retain large angular-scale CMB anisotropies, to compare only the common information present in all the maps, we high-pass filtered the ACT and \textit{Planck} maps using the mock-observation pipeline described in Section~\ref{sec:simulations}.
  In addition, we applied an isotropic low-pass filter to the high-resolution (\nside\ = 8192) SPT and ACT $Q/U$ maps to remove the information at $\ell$ above 6143, which    is predominantly noise, and enhance the CMB.
  The lower-resolution (\nside\ = 2048) \textit{Planck} $Q/U$ maps do not contain information above this $\ell$ in any case.
  We did not filter the $T$ maps because doing so creates visible Fourier ringing around the high-S/N emissive sources.
  Although the noise level of the \textit{Planck} $T$ map is low enough that the CMB temperature anisotropy is imaged cleanly, that is not the case for the polarization anisotropy.
  However, significantly deeper maps of the polarization anisotropy have been produced by SPT and ACT.
  The $Q/U$ maps from the two ground-based experiments show good consistency visually, and the SPT maps show the $E$-mode crisscross patterns more clearly.

\begin{table}
\begin{threeparttable}
  \caption{\label{tab:survey_statistics_spt_planck_act}%
  The depth and observed sky fraction of each of the three datasets: SPT-3G D1, ACT DR6, and \textit{Planck} PR3.}
  \begin{ruledtabular}
  \begin{tabular}{c|ccc}
      & Sky fraction [\%] & Depth [$\mathrm{{\mu}K}$-arcmin]\tnote{1} \\
    \hline
    SPT-3G D1 & 4 & 3.3 \\
    ACT DR6 & 45 & 10 \\
    \textit{Planck} PR3 & 100 & 28 \\
  \end{tabular}
  \end{ruledtabular}
  \begin{tablenotes}
    \item [1] The SPT-3G D1 depth is equal to $\sqrt{1/(\sum_f w_f)}$, where $w_f$ is the inverse square of the $T$ white noise level in one frequency band reported in Section~\ref{sec:full_depth_coadds}.
    The \textit{Planck} depth is calculated in the same way as the SPT-3G D1 depth, and the $T$ white noise levels used in the calculation are the 100, 143, and 217\;GHz numbers reported in \citet[Table~4]{planck18-1}.
    The ACT DR6 depth is directly taken from \citet{naess25} and is based on the maps  in the 98, 150, and 220\;GHz bands.
  \end{tablenotes}
\end{threeparttable}
\end{table}

  Two additional figures of coadds can be found in Appendix~\ref{app:coadds}.
  One figure is a version Figure~\ref{fig:coadds_spt_only_signal_tqu_strip} that shows the full field, and the other figure is a version of Figure~\ref{fig:coadds_spt_act_planck_comparison_tqu_filtered} that compares large angular-scale CMB anisotropies by using the original ACT and \textit{Planck} maps and replacing the SPT-3G D1 mid-$\ell$ \texttt{full} coadd with the low-$\ell$ counterpart.

\subsection{\label{sec:simulations}Simulations}
  To estimate the filter transfer function associated with the mid-$\ell$ mapmaking, we ran the filter-and-bin pipeline on simulated timestreams.
  We call this process a ``mock observation.''
  We describe the general steps of a mock observation and specifics of the mid-$\ell$ mock observations.

\subsubsection{\label{sec:simulations_general}General Steps}
  The input of a mock observation is a set of simulated $T/Q/U$ sky maps, from which we generate mock timestreams.
  We combine the pixel values of the simulated maps and the real pointing information of the detectors during one subfield observation to create a mock timestream for each detector and scan.
  Because a pointing sample generally does not coincide with any pixel center, an interpolation method is used to estimate the $T/Q/U$ values of the simulated sky at an arbitrary pointing sample.

  After we generate mock timestreams, we filter and bin them in the same way as we did to the corresponding real timestreams from which the pointing information was used to generate the mock timestreams.
  We apply the same filters to a mock timestream as were used for the corresponding real timestream and bin the filtered mock timestream with the weight calculated from the corresponding real timestream.
  Then, we add the output maps from multiple mock observations using the coadding process of Equation~\ref{eqn:coadd} to create mock coadds.

\subsubsection{\label{sec:simulations_specifics}Specifics}
  The input simulated $T/Q/U$ maps used for G25, C25, and O26 are HEALPix maps with \nside\ = 8192 and contain a combination of Gaussian realizations of the CMB and foreground simulations and have information up to $\ell$ = 16\,000.
  The unlensed CMB skies were created using input power spectra generated with the Code for Anisotropies in the Microwave Background (CAMB\footnote{\url{http://camb.info}} \citep{lewis11b}) from the best-fit 2018 \textit{Planck} cosmology.\footnote{
  The input file is \texttt{planck2018\_base\_plikHM\_TTTEEE\_lowl\_lowE\_\\lensing\_params\.ini}}
  The CMB skies were then lensed using {\sc LensPix}~\citep{lewis05}.
  The foreground skies were created from power spectra measured from the \texttt{Agora} simulations \citep{omori24}, with the point source masking threshold matched to that used in the mid-$\ell$ mapmaking.
  After the CMB and foreground skies were added, they were smoothed by the measured telescope beam (Section~\ref{sec:beam}).
  We created 500 realizations of simulated skies with these components.
  For the analysis in O26, other custom sets of simulated $T/Q/U$ sky maps were created, including 250 realizations of skies for which the lensing potential realization was identical but the CMB realization different.

  When generating the mock timestreams for every sky realization, we used the detector pointing information from the same 5\% of the total number of observations.
  In previous SPT analyses, we often ran mock observations of all the individual observations, but with the larger detector count of SPT-3G that becomes computationally difficult.
  Internal studies showed that the filter transfer function derived from mock observations converge at the subpercent level with 5\% of the total number of observations simulated, in part because of the highly redundant and repetitive nature of SPT observing.

  To estimate the $T/Q/U$ values of a simulated sky at an arbitrary pointing sample, we used a bilinear interpolation method that is based on the values of the four pixels that are closest to a given pointing sample.
  This interpolation in effect smooths the simulated maps, and the smoothing is equivalent to an attenuation of the power spectra of the simulated maps by approximately the fourth power of the pixel window function associated with the maps: $C_{\ell}^{\mathrm{interp}} \approx C_{\ell}^{\mathrm{input}} P_{\ell}^4$, where $C_{\ell}^{\mathrm{input}}$ is the power spectrum of a simulated $T$, $Q$, or $U$ sky, $P_{\ell}$ is the pixel window function, and $C_{\ell}^{\mathrm{interp}}$ is the power spectrum of the interpolated sky.\footnote{
  The functional form of this attenuation can be understood using a 1D toy model.
  From a smoothly varying function, to create a piecewise function whose values at a set of grid points are equal to the values of the smooth function but whose value at any other point is the linear interpolation between the values at the two closest grid points, three operations can be applied to the smooth function.
  The first operation is to multiply the smooth function by a (sampling) comb function, the second operation is to convolve the resultant function with a top hat window whose width is equal to the spacing of the comb, and the third operation is to repeat the convolution.
  The last two operations create the fourth power of the pixel window function.}
  Because the smoothing effect caused by the interpolation does not exist in a real observation, we corrected for the factor $P_{\ell}^4$ when using the mock-observation output maps to estimate the filter transfer function.

  When filtering the mock timestreams, even though we did not add emissive sources and galaxy clusters to the simulated maps, we still masked the samples of the mock timestreams that correspond to the real locations of the masked objects to simulate effects of the masking.
  We also ran an additional batch of mock observations (on 110 out of the 500 realizations of the simulated skies) in which we did not mask the mock timestream samples to further investigate the effects.
  As was the case with binning the real timestreams, we binned the mock timestreams using two pixelization schemes: HEALPix \nside\ 8192 and flat-sky $0^\prime.5625$ square pixels.

  We make the 500 realizations of input simulated sky maps and the associated mock-observation output coadds publicly accessible.
  We also include the output coadds obtained from the additional batch of mock observations run on the 110 realizations without the point source masking.
  The access information can be found in Appendix~\ref{app:data_availability}.

\section{\label{sec:beam_and_filter_transfer_function}Beam and Filter Transfer Function}
  The coadded $T/Q/U$ maps described in Section~\ref{sec:coadds} are biased with respect to the true temperature and polarization anisotropies of the millimeter-wave sky.
  Two important biases are related to the point-spread function of the telescope (often called the ``beam'') and the timestream filtering.
  Each bias can be expressed as a function of $\ell$ by which the power spectra of the true anisotropies are multiplied.
  We describe each bias only briefly in this section (Sections~\ref{sec:beam} and \ref{sec:filter_transfer_function}), and more details can be found in C25 (Sections~IV\,A and IV\,B), Huang et al. (in preparation), and Hivon et al. (in preparation).

  We make the biases as functions of $\ell$ publicly accessible along with the coadds.
  The access information can be found in Appendix~\ref{app:data_availability}.

\subsection{\label{sec:beam}Beam}
  For the beam, the function that we make publicly accessible is the spherical harmonic- or $\ell$-space beam, $B_\ell$, which is the square root of the ratio of the power spectrum of the diffraction-smoothed sky to that of the true sky.
  The procedure and results of the SPT-3G $B_{\ell}$ estimation are described in detail in C25 (Section~IV\,B) and Huang et al. (in preparation).
  Here we summarize the procedure.
  We also highlight the effort to understand how $B_{\ell}$ may differ between temperature and polarization signals.

  Following the strategy used with data from SPT-SZ (e.g., \citep{keisler11}), we estimated the SPT-3G beam for each frequency band independently using a combination of two types of astrophysical sources: planets observed through dedicated observations and bright AGN in the SPT-3G Main field.
  In 2019 and 2020, we conducted nine observations of Saturn, in which the coverage was dense enough that every detector fully observed the planet.
  These observations would be sufficient for estimating the beam, except that power from Saturn loads the detectors enough to change their responsiveness and time constants, resulting in nonlinear and slowed response to optical power.
  To correct for this, we made Saturn maps using only the approaching parts of the timestreams of each detector and stitched these planet maps with maps of bright AGN in the SPT-3G Main field.
  The final $B_\ell$ for each frequency band was estimated from 2D Fourier transforms of the stitched maps, and the uncertainty on $B_{\ell}$ was estimated from cross-spectra under variations of several systematic parameters, including which Saturn observation and bright AGN were used.

  In most previous SPT power spectrum analyses including polarization, we assumed that the polarization beam is identical to the total intensity or temperature beam, but we revisited this assumption in recent works.
  In G25, we recognized that the assumption is not physically well-motivated because beam sidelobes are unlikely to retain 100\% coherent polarization.
  In G25 and C25, we constructed a beam model with a main beam that is derived from physical optics modeling and sidelobes that are the residual between the main beam and the measured temperature beam, and we included the depolarization of the sidelobes in the fit of the power spectra to cosmology.
  This modification resulted in better interfrequency agreement in the measurements of the $TT/TE/EE$ spectra reported in C25 and better agreement in the constraints on cosmological parameters estimated from the spectra separately.
  However, in a more recent work, \citet{dehaan26}, we used polarized point sources to measure the polarization beam and found minimal sidelobe depolarization.
  This suggests that the model of sidelobe depolarization used in G25 and C25 may be too simple to capture the behavior over the full range of $\ell$ from CMB to point-source scales.

\subsection{\label{sec:filter_transfer_function}Filter Transfer Function}
  For the timestream filtering, the function that we make publicly accessible is the filter transfer function, $F_\ell$, which is the fraction of the power of the true anisotropies at each $\ell$ that still remains after the timestream filtering.
  The procedure and results of the $F_\ell$ estimation for the mid-$\ell$ filtering are described in detail in C25 (Section~IV\,A) and Hivon et al. (in preparation).
  Here we briefly summarize the procedure and provide an approximate analytical expression of the transfer function.

  As prescribed in \citet[Section~3.6]{hivon02}, the $F_\ell$ associated with given timestream filtering can be estimated by comparing the power spectra of the output maps produced by mock observations simulating the filtering with the power spectra of  the input maps.
  We used the 500 sets of input and output maps described in Section~\ref{sec:simulations} and additional 2000 sets of maps produced by a simplified but faster mock-observation pipeline (the \texttt{Quickmock} pipeline described in C25) for the estimation.

  Given the SPT-3G scan strategy and the mid-$\ell$ filtering, the 2D transfer function, which is the fraction of power retained in each $(\ell_x,\,\ell_y)$ bin after the filtering, has a simple form in the Sanson-Flamsteed (SFL) projection \citep{calabretta02}, and azimuthally averaging the 2D function results in an approximate analytical expression for $F_\ell$.\footnote{
  $\ell_x$ and $\ell_y$ are the Fourier conjugates to the map $x$ and $y$ coordinates, respectively.}
  In this projection, map rows are at constant declination, and map columns are at constant $\Delta\alpha\,\mathrm{\cos}\,\delta$, where $\delta$ is declination, and $\Delta\alpha$ is the offset in right ascension from the map center.
  The sinusoid-subtraction part of the high-pass filter results in a Heaviside function $\mathcal{H}$ at $\ell_{x,\,c,\,\mathrm{HP}} = 300$, and the low-pass filter results in $\mathrm{exp}\,[-2{(\ell_x / \ell_{x,\,c,\,\mathrm{LP}})}^{6}]$ with $\ell_{x,\,c,\,\mathrm{LP}}$ = 13\,000.
  Therefore, the 2D transfer function is approximately a pure function of $\ell_x$.\footnote{
  The effect of the polynomial subtraction is not a Heaviside function but is much smaller than the effect of the sinusoid subtraction.
  In addition, the effect of the masking is not considered here.}
  Azimuthally averaging the 2D transfer function, we obtain the following when $\ell \ll \ell_{x,\,c,\,\mathrm{LP}}$:
\begin{equation}
  F_\ell =
    \mathcal{H}(\ell - \ell_{x,\,c,\,\mathrm{HP}})\;
    \frac{2}{\pi} \arccos \left ( \frac{\ell_{x,\,c,\,\mathrm{HP}}}{\ell} \right ).
\end{equation}
  This function increases from 0 at $\ell=300$, reaches 0.6 at $\ell=500$, and 0.9 at $\ell=2000$.

  In other flat-sky projections and in spherical harmonic space, the 2D transfer function is less easy to model because the filtering is generally a function of position in the map.\footnote{
  In spherical harmonic space, by \textit{2D transfer function} we mean the fraction of power retained in each $(\ell,m)$ mode.}
  For example, in spherical harmonic space, if the sphere is oriented in the equatorial coordinate system, the timestream filters act on constant-declination HEALPix rings with declination-dependent high-pass and low-pass cutoff values in $m$.

\section{\label{sec:map_calibration}Map Calibration}
  As briefly described in Section~\ref{sec:coadding_procedure}, after producing the \texttt{full}, \texttt{half}, \texttt{one-thirtieth}, and \texttt{signflip-noise} coadds for each subfield and frequency band, we calibrated and cleaned the subfield coadds before combining them to produce the full-field coadds.
  We used the \texttt{full} and \texttt{half} coadds to estimate the calibration and cleaning coefficients and used the same coefficients for all four types of coadds.
  The calibration and cleaning were done to reduce known biases in the coadds.\footnote{
  In fact, G25 used a version of the full-field coadds that did not have these steps applied but included the biases into the forward model (G25, Appendix~B).
  C25 used the full-field coadds that had these steps applied and then marginalized over the uncertainty on some of the calibration and cleaning parameters in the cosmological parameter fitting.}

  The known biases are caused by the fact that we do not have accurate values of detector properties for every detector, and we use nominal or ideal values when binning timestreams in mapmaking.
  The detector properties in question are the gain, polarization angle, and polarization efficiency of a detector.
  We expand Equation~\ref{eqn:detector_signal_model_simple} by including the gain and polarization efficiency as follows (again after \citep{jones07}):
\begin{equation}
\label{eqn:detector_signal_model_complex}
  \begin{split}
    I_\alpha(\hat{n}) =& \frac{g_\alpha}{2}\,
      \{T(\hat{n}) + \frac{\gamma_{\alpha}}{2-\gamma_{\alpha}}\,\times \\
    & [Q(\hat{n})\,\cos\,2\psi_\alpha +
                         U(\hat{n})\,\sin\,2\psi_\alpha] \},
  \end{split}
\end{equation}
where $g_\alpha$ and $\gamma_{\alpha}$ are the gain and polarization efficiency of detector $\alpha$, respectively.
  The gain is an attenuation factor connecting the intensity of incoming radiation from the sky to the intensity actually absorbed by the detector.
  The polarization efficiency ranges from 0 (insensitive to polarized radiation) to 1 (sensitive to radiation along only one polarization direction).

  Because accurately measuring the gain, polarization angle, and polarization efficiency of every detector is not feasible, we let the inaccuracies in the values of individual detectors' properties used for binning their timestreams propagate through the mapmaking and correct for the collective effects of the inaccuracies over the full detector array at the level of the coadds through a series of map calibration and cleaning steps.
  For the mid-$\ell$ maps, we applied four steps.
  The first two steps are related to the gain (Sections~\ref{sec:gain_calibration} and \ref{sec:monopole_t_to_p_leakage_removal}), the third step is related to the polarization angle (Section~\ref{sec:polarization_angle_calibration}), and the fourth step is related to the polarization efficiency and angle (Section~\ref{sec:polarization_amplitude_calibration}).

\subsection{\label{sec:gain_calibration}Gain Calibration}
  In the first step, we corrected for an effect of the nonzero average of the miscalibrations of the detector gains over all the detectors for each frequency band.
  For each coadd, we multiplied the $T/Q/U$ maps by the same calibration factor.\footnote{
  Following Section~\ref{sec:coadding_procedure}, we continue to call a group of three weight-removed coadded maps ($\bar{T}(\hat{n})$, $\bar{Q}(\hat{n})$, and $\bar{U}(\hat{n})$) in one frequency band produced by Equation~\ref{eqn:coadd} a coadd.}

  The miscalibration of the gain of a detector (and thus its timestreams) was caused by inaccuracies in our estimates of the flux densities of the \hii{} regions measured by that detector.
  Both \hii{} regions used to calibrate the SPT-3G D1 timestreams (RCW38 and Mat5a) were also observed regularly in the SPT-SZ survey, and maps of the \hii{} regions were produced from SPT-SZ data using an absolute calibration of SPT-SZ timestreams that is based on \textit{Planck} maps \citep{hou18,mocanu19}.
  If the SPT-3G frequency bands and beams were identical to those of SPT-SZ or if the spectra of the \hii{} regions were identical to the derivative of a 2.7 K blackbody spectrum, we would expect the ``out-of-the-box'' SPT-3G calibration with the SPT-SZ \hii{} region maps to be accurate at subpercent levels.\footnote{
  The \textit{Planck} calibration in total intensity, which is based on the annual modulation of the CMB dipole from the Earth's motion around the Sun, has an estimated uncertainty of 0.07\% (in amplitude) in the 143\;GHz band \citep{planck15-8}, and the SPT-SZ calibration using \textit{Planck} maps and interfrequency comparisons has estimated uncertainties of 0.26\%, 0.17\%, and 0.48\% in the three frequency bands \citep{mocanu19}.}
  In reality, those conditions are not met, and we expect the timestream calibration that is based on each \hii{} region to be inaccurate at the 5--10\% level.
  This causes a common multiplicative bias in all the coadded $T/Q/U$ maps for each subfield and frequency band.

  We estimated the calibration factors needed to correct for the multiplicative biases by cross-correlating maps.
  For each subfield, we first cross-correlated SPT-3G 150\;GHz maps with the part of a \textit{Planck} 143\;GHz map covering that subfield to estimate the 150\;GHz calibration factor.
  Then, we cross-correlated SPT-3G 95 and 220\;GHz maps with 150\;GHz ones to estimate the 95 and 220\;GHz calibration factors.
  We now describe details of the estimation of these calibration factors.

  We estimated the 150\;GHz calibration factor, $g_{\mathrm{cal},\,150}$, by finding a best-fit constant for a binned cross-spectrum ratio that is flat in $\ell$ as follows:
\begin{align}
  R_{b,\,150}^G &= P_{b\ell} \left(
    \frac{C^{TT}_{\ell,\,\mathrm{150}{\times}\mathrm{143}}}
         {C^{TT}_{\ell,\,\mathrm{150}{\times}\mathrm{150}}}\,
    \frac{B_{\ell,\,\mathrm{150}}}{B_{\ell,\,\mathrm{143}}}\,
    \frac{P_{\ell,\,\mathrm{150}}}{P_{\ell,\,\mathrm{143}}} \right),
    \label{eqn:gcal_150_rb} \\
  \chi_{150}^2(x) &= {(R_{b,\,150}^G - x)}^T\,
                     {\left( \Sigma_b^G \right)}^{-1}\,
                     (R_{b,\,150}^G - x),
    \label{eqn:gcal_150_chi2} \\
  g_{\mathrm{cal},\,150} &= \arg \min_{x}\, \chi_{150}^2(x),
    \label{eqn:gcal_150_final}
\end{align}
where $P_{b\ell}$ is a binning operator, which acts on the product of the three ratios of functions of $\ell$ shown within the parentheses in Equation~\ref{eqn:gcal_150_rb}, $R_{b,\,150}^G$ is the resultant binned ratio, which is flat in $\ell$, $\Sigma_b^G$ is the covariance matrix associated with $R_{b,\,150}^G$, $x$ is a constant whose difference from $R_{b,\,150}^G$ was used to construct a $\chi^2$, and $g_{\mathrm{cal},\,150}$ is the value of $x$ that minimizes the $\chi^2$.
  We now describe these quantities.

  The leftmost ratio within the parentheses is the ratio of two $TT$ spectra and central to the estimation of $g_{\mathrm{cal},\,150}$.
  In this ratio, in effect we divide a \textit{Planck} 143\;GHz $T$ map by an SPT-3G 150\;GHz $T$ map in $\ell$ space to find an $\ell$-independent calibration factor that corrects the SPT-3G calibration so that it matches \textit{Planck}'s.
  The numerator is the cross-spectrum between the SPT-3G 150\;GHz \texttt{full} $T$ map and a filtered version of the \textit{Planck} 143\;GHz full-mission $T$ map, and the denominator is the cross-spectrum between the two SPT-3G 150\;GHz $T$ maps from the pair of \texttt{half} coadds.
  We created the filtered \textit{Planck} map by mock-observing the original \textit{Planck} map.
  This way, all four maps involved in the ratio have the same SPT-3G filter transfer function, so $g_{\mathrm{cal},\,150}$ is not biased by the filtering.
  Because the two \texttt{half} $T$ maps used in the denominator have uncorrelated noise, using the cross-correlation of the two maps instead of the autocorrelation of the \texttt{full} $T$ maps avoids biasing $g_{\mathrm{cal},\,150}$ by the autospectrum of the noise in the \texttt{full} $T$ map.

  We used \texttt{PolSpice}\footnote{\url{http://www2.iap.fr/users/hivon/software/PolSpice/}} \cite{szapudi01, chon04} to calculate each $TT$ spectrum (and all the spectra that appear elsewhere in Section~\ref{sec:map_calibration}).
  The apodization mask that we supplied to the program has tapers near the borders of each subfield and covers the galaxy clusters and emissive sources masked in the timestream high-pass filter.
  (Tapers also exist around the holes covering the objects.)
  When cross-correlating SPT-3G and \textit{Planck} maps, we used the masking radii six times larger than the ones used in the high-pass filter because of the larger \textit{Planck} beam.

  The ratio of the two $TT$ spectra is not flat across $\ell$ because the SPT-3G and \textit{Planck} maps are smoothed by different beams and pixel window functions, and we corrected for the differences using the other two ratios within the parentheses.
  $B_{\ell,\,\mathrm{150}}$ and $B_{\ell,\,\mathrm{143}}$ are the SPT-3G and \textit{Planck} beams, respectively.
  $P_{\ell,\,\mathrm{150}}$ is the pixel window function associated with the SPT-3G map, and $P_{\ell,\,\mathrm{143}}$ is the product of the pixel window function associated with the unfiltered \textit{Planck} map and the pixel window function introduced by the bilinear interpolation used in the mock observations.

  After obtaining a flat cross-spectrum ratio with the corrections, we binned the ratio in the $\ell$ range between 600 and 1200 with the uniform bin width 30 and chose  the constant that minimizes the $\chi^2$ shown in Equation~\ref{eqn:gcal_150_chi2} as $g_{\mathrm{cal},\,150}$.
  We estimated the covariance matrix used for the $\chi^2$ calculations by using 20 simulated cross-spectrum ratios.
  To create a set of four simulated maps to calculate a simulated cross-spectrum ratio, we generated a signal-only $T$ sky as the starting point.
  We then smoothed it with the \textit{Planck} beam, mock-observed the smoothed sky, and added a \textit{Planck} FFP10 noise map to the mock-observed sky to create a simulated \textit{Planck} map.
  We also smoothed the signal-only $T$ sky with the SPT-3G beam, mock-observed the smoothed sky, and combined it with one full-depth and two half-depth \texttt{signflip-noise} $T$ maps to create three simulated SPT-3G maps.

  We estimated the 95 and 220\;GHz calibration factors, $g_{\mathrm{cal},\,95}$ and $g_{\mathrm{cal},\,220}$, using the same method as the one shown in Equations~\ref{eqn:gcal_150_rb}--\ref{eqn:gcal_150_final} except that the product of the ratios of functions of $\ell$ has additional terms related to cross-correlating a 95 or 220\;GHz $T$ map with a 150\;GHz one as follows:
\begin{equation}
  \begin{split}
    R_{b,\,\nu}^G = P_{b\ell}\, \Biggl(
      \frac{C^{TT}_{\ell,\,\mathrm{150}{\times}\mathrm{143}}}
           {C^{TT}_{\ell,\,\mathrm{150}{\times}\mathrm{150}}}\, &
      \frac{B_{\ell,\,\mathrm{150}}}{B_{\ell,\,\mathrm{143}}}\,
      \frac{P_{\ell,\,\mathrm{150}}}{P_{\ell,\,\mathrm{143}}}\, \times \\
      & \frac{C^{TT}_{\ell,\,\mathrm{150}{\times}\mathrm{150}}}
             {C^{TT}_{\ell,\,\mathrm{150}{\times}\nu}}\,
        \frac{B_{\ell,\,\nu}}{B_{\ell,\,\mathrm{150}}} \Biggr),
  \end{split}
\end{equation}
where ${C^{TT}_{\ell,\,\mathrm{150}{\times}\nu}}$ is the cross-spectrum between a 150\;GHz \texttt{half} $T$ map and the 95 or 220\;GHz \texttt{half} $T$ map from the other half of the dataset,\footnote{
  Cross-correlating a 150\;GHz \texttt{full} $T$ map with the \texttt{full} $T$ map from another band would create a noise bias because of correlated noise between the frequency bands.}
  $B_{\ell,\,\nu}$ is the 95 or 220\;GHz beam, and the other terms are the same as the ones appearing in Equation~\ref{eqn:gcal_150_rb}.
  The additional terms represent the ratio of the 95 or 220\;GHz calibration to the 150\;GHz calibration.
  Because SPT-3G maps are deeper than \textit{Planck} maps, comparing the 95 and 220\;GHz maps with the 150\;GHz map internally within SPT-3G produces more precise calibration factors for 95 and 220\;GHz than calibrating these bands to the corresponding \textit{Planck} bands directly.

  One potential concern with comparing different frequency bands is the effect of foreground emission.
  If the foreground power is equal to some fraction $\eta$ of the CMB anisotropy power at a given $\ell$ in the 150\;GHz autospectrum, and the power differs between 150\;GHz and one of the other bands by a factor $\rho$, we would expect $C^{TT}_{\ell,\,\mathrm{150}{\times}\mathrm{150}}/C^{TT}_{\ell,\,\mathrm{150}{\times}\nu}$ to be biased by a factor $\eta \sqrt{\rho}$.
  In the $\ell$ range used for the calibration, we expect that the biases are not significant, which is confirmed by simulations using \texttt{Agora}.\footnote{
  In the $\ell$ range between 600 and 1200, the maximum expected bias on the calibration correction is $\lesssim 0.5\%$ in the 220\;GHz band and much lower in the 95\;GHz band.}
  These biases are smaller than the statistical uncertainties on the calibration factors.

  Table~\ref{tab:gain_calibration} shows the $g_{\mathrm{cal}}$ and its uncertainty for each frequency band and subfield.
  All the $g_{\mathrm{cal}}$ values are within 10\% of unity, which is consistent with our expectation of the \hii{} region calibration.
  The uncertainty on each $g_{\mathrm{cal}}$ is the 68\% confidence interval obtained from the process of minimizing the $\chi^2$ to determine that $g_{\mathrm{cal}}$, and the uncertainty reflects the noise fluctuations in the SPT-3G and \textit{Planck} maps used to calculate the cross-spectra.
  We used each $g_{\mathrm{cal}}$ to calibrate all the subfield \texttt{full}, \texttt{half}, \texttt{one-thirtieth}, and \texttt{signflip-noise} coadds for the relevant frequency band as follows:
\begin{equation}
\label{eqn:gain_calibration}
  \begin{pmatrix}
    \bar{T}(\hat{n}) \\ \bar{Q}(\hat{n}) \\ \bar{U}(\hat{n}) \\
  \end{pmatrix}_1
  =
  \begin{pmatrix}
     g_{\mathrm{cal}} & 0 & 0 \\
     0 & g_{\mathrm{cal}} & 0 \\
     0 & 0 & g_{\mathrm{cal}} \\
  \end{pmatrix}
  \begin{pmatrix}
     \bar{T}(\hat{n}) \\ \bar{Q}(\hat{n}) \\ \bar{U}(\hat{n}) \\
  \end{pmatrix}_0,
\end{equation}
where the subscripts 0 and 1 denote the coadds before and after the gain calibration step, respectively.

\begin{table}
  \caption{\label{tab:gain_calibration}%
  The gain calibration factor, $g_\mathrm{cal}$, for each subfield and frequency band.}
  \begin{ruledtabular}
  \begin{tabular}{c|ccc}
        & 95\;GHz & 150\;GHz & 220\;GHz \\
    \hline
    \texttt{el0} & 1.058 $\pm$ 0.004 & 1.013 $\pm$ 0.004 & 0.986 $\pm$ 0.009 \\
    \texttt{el1} & 1.068 $\pm$ 0.003 & 1.033 $\pm$ 0.003 & 0.999 $\pm$ 0.008 \\
    \texttt{el2} & 1.071 $\pm$ 0.003 & 1.001 $\pm$ 0.004 & 1.002 $\pm$ 0.007 \\
    \texttt{el3} & 1.081 $\pm$ 0.007 & 1.014 $\pm$ 0.008 & 1.007 $\pm$ 0.009 \\
  \end{tabular}
  \end{ruledtabular}
\end{table}

  Finally, we clarify the differences between the notations used in this work and those used in G25 and C25.
  The three calibration factors $T_{\mathrm{cal,external}}^{150}$, $T_{\mathrm{cal,internal}}^{95}$, and $T_{\mathrm{cal,internal}}^{220}$ described in G25 (Appendix~B) are the same as the calibration factors $g_{\mathrm{cal},\,150}$, $g_{\mathrm{cal},\,95}$, and $g_{\mathrm{cal},\,220}$ described in this work.
  On the other hand, the three nuisance parameters $A_\mathrm{cal}^\mathrm{ext}$, $A_\mathrm{cal}^\mathrm{rel,95}$, and $A_\mathrm{cal}^\mathrm{rel,220}$ in C25 are equivalent to the inverse of $g_{\mathrm{cal},\,150}$, $g_{\mathrm{cal},\,95}/g_{\mathrm{cal},\,150}$, and $g_{\mathrm{cal},\,220}/g_{\mathrm{cal},\,150}$ because it is the power spectra of theoretical models instead of real data that are multiplied by the nuisance parameters and because the two nuisance parameters with the superscript ``rel'' represent the relative calibration of 95 and 220\;GHz with respect to 150\;GHz.
  In C25, when choosing the width of the prior for $A_\mathrm{cal}^\mathrm{ext}$ to constrain cosmology using the full set of $TT/TE/EE$ spectra, we added two numbers in quadrature to obtain the chosen width 0.0036.
  One number is the (inverse-quadrature) combined uncertainty from the four subfield values in the 150\;GHz band shown in Table~\ref{tab:gain_calibration} (0.0026), and the other number is an uncertainty on the \textit{Planck} calibration (0.0025).

\subsection{\label{sec:monopole_t_to_p_leakage_removal}Monopole \texorpdfstring{$T$}{T}-to-\texorpdfstring{$P$}{P} Leakage Removal}
  In the second step, we corrected for an effect of the nonzero average of the differential gain miscalibrations among detectors sensitive to different polarization directions for each frequency band.
  For each coadd, we subtracted copies of the $T$ map (multiplied by coefficients much smaller than unity) from the $Q/U$ maps.

  If individual detector gains are misestimated by different amounts, the $T$ information does not properly cancel in the estimate of the $Q/U$ maps, so the $T$ information ``leaks'' into the $Q/U$ maps.
  We call this type of leakage the ``monopole $T$-to-$P$ leakage'' because it couples to $T$ itself rather than some orders of derivatives of $T$ (see, e.g., \citep{hu03} and Section~IV\,B\,2 of C25 for descriptions of higher-order leakage).
  The monopole $T$-to-$P$ leakage adds a scaled copy of $T$ to the true $Q/U$ in actually measured $Q/U$ as follows:
\begin{align}
  Q_{\mathrm{measured}} &= \epsilon_Q T + Q_{\mathrm{true}}
    \label{eqn:t_to_q_intro}, \\
  U_{\mathrm{measured}} &= \epsilon_U T + U_{\mathrm{true}}
    \label{eqn:t_to_u_intro},
\end{align}
where $\epsilon_Q$ and $\epsilon_U$ are the leakage coefficients.

  For random gain miscalibrations for individual detectors (and equal weights among detectors), the induced monopole $T$-to-$P$ leakage is of order $\sigma_\mathrm{G} / \sqrt{N}$, where $\sigma_\mathrm{G}$ is the typical (differential) gain miscalibration, and $N$ is the number of detectors contributing to the maps.
  For SPT-3G, with approximately four thousand detectors per frequency band and gain miscalibrations at the level of several percent, the effect from random gain miscalibrations is negligible.

  What can cause non-negligible leakage is gain miscalibrations correlated with polarization angles.
  While fabrication effects could cause this, the most straightforward mechanism is nonzero polarization of the \hii{} regions, which we use as calibration sources assuming they are unpolarized.
  In Z25, for each pair of detectors that are sensitive to orthogonal polarization directions and are in the same pixel of the full detector array, the relative gain between the two detectors was adjusted so that their responses to the elevation nod (Section~\ref{sec:calibration_observations}) preceding a subfield observation were the same in units of CMB fluctuation temperature.
  The quiescent emission from the atmosphere at SPT-3G frequencies is mostly unpolarized, and it was found that this gain correction produced maps with no detectable level of the monopole $T$-to-$P$ leakage, reinforcing the notion that source polarization is the primary cause.
  While eliminating the leakage is desirable, having this response matching as a mapmaking step necessitates cutting $\sim$2000 detectors across the three frequency bands that do not form complete pairs.
  For the analyses planned to be done on the mid-$\ell$ maps, that level of data volume loss counteracts the advantage of the leakage suppression.

  To estimate the coefficients to multiply the $T$ maps by before subtracting from the $Q/U$ maps, we used the following model that is based on Equations~\ref{eqn:t_to_q_intro} and \ref{eqn:t_to_u_intro}:
\begin{align}
  C^{TQ}_{\ell,\,\mathrm{h1}{\times}\mathrm{h2}}
    & = \epsilon^{TT}_Q C^{TT}_{\ell,\,\mathrm{h1}{\times}\mathrm{h2}} +
        \epsilon^{TQ}_Q C^{TQ}_{\ell,\,\mathrm{mock}}
    \label{eqn:t_to_q_model}, \\
  C^{TU}_{\ell,\,\mathrm{h1}{\times}\mathrm{h2}}
    \label{eqn:t_to_u_model}
    & = \epsilon^{TT}_U C^{TT}_{\ell,\,\mathrm{h1}{\times}\mathrm{h2}},
\end{align}
where $C^{TQ}_{\ell,\,\mathrm{h1}{\times}\mathrm{h2}}$ and $C^{TU}_{\ell,\,\mathrm{h1}{\times}\mathrm{h2}}$ are the cross-spectra between a \texttt{half} $T$ map with the \texttt{half} $Q$ and $U$ maps from the other half of the dataset, $C^{TT}_{\ell,\,\mathrm{h1}{\times}\mathrm{h2}}$ is the cross-spectrum between the two \texttt{half} $T$ maps, $\epsilon^{TT}_Q$ and $\epsilon^{TT}_U$ are the leakage coefficients, $C^{TQ}_{\ell,\,\mathrm{mock}}$ is an estimate of the $TQ$ cross-spectrum from the CMB, which we obtained by averaging the $TQ$ cross-spectra of mock-observation output maps over multiple realizations of the sky, and $\epsilon^{TQ}_Q$ is a coefficient associated with $C^{TQ}_{\ell,\,\mathrm{mock}}$.
  The \texttt{half} maps used here had the gain calibration applied.
  The parameters of interest are the monopole leakage coefficients $\epsilon^{TT}_Q$ and $\epsilon^{TT}_U$, and $\epsilon^{TQ}_Q$ is a nuisance parameter (with the expectation value of unity).

\begin{figure}
  \includegraphics[scale=1.00]{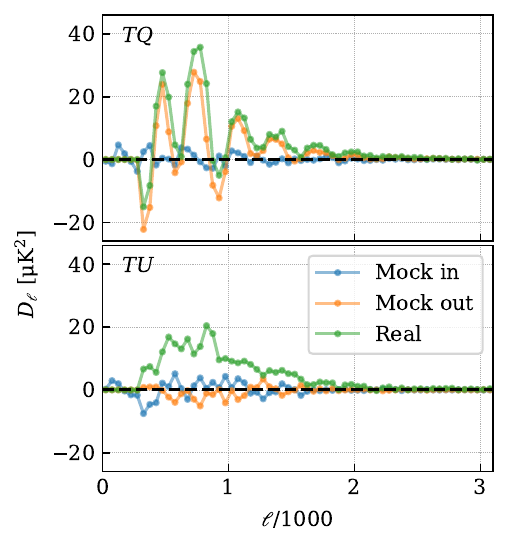}
  \caption{\label{fig:calibration_cleaning_monopole_tp}%
  Example binned $TQ/TU$ spectra from simulated maps (mock observation input and output maps) and real-data maps.}
\end{figure}

  As described in \citet[Section~IV\,G\,2]{dutcher21}, the expected 2D $TQ/TU$ spectra of the CMB are nonzero because of the CMB $TE$ spectrum, but the 1D $TQ/TU$ spectra (averaged over $m$) are expected to be zero for isotropic weighting and no filtering of the different $m$ modes.
  For the mid-$\ell$ timestream filtering, we still expect the $TU$ spectrum to be zero but not for the $TQ$ spectrum.
  This is illustrated in Figure~\ref{fig:calibration_cleaning_monopole_tp}, which shows that the $TU$ spectrum is consistent with zero for both a simulated sky and the mock-observed sky, but the $TQ$ spectrum becomes nonzero after the mock observations (timestream filtering).
  Compared with the mock-observation output $TQ/TU$ spectra, the spectra from the real data show excess power, which indicates $T$-to-$P$ leakage.
  Given the nonzero $TQ$ spectrum from the CMB, we included the expectation spectrum, $C^{TQ}_{\ell,\,\mathrm{mock}}$, in the $TQ$ part of the model.

  From the model in Equations~\ref{eqn:t_to_q_model} and \ref{eqn:t_to_u_model}, we constructed a $\chi^2$ function from a binned version of each equation and found the values of the three $\epsilon$ parameters that minimize the sum of the two $\chi^2$ functions.
  The $\chi^2$ function associated with Equation~\ref{eqn:t_to_q_model} is the following:
\begin{equation}
  \begin{split}
  \label{eqn:t_to_q_chi2}
    &\chi_{TQ}^2 (\epsilon^{TT}_Q, \epsilon^{TQ}_Q)
      = \sum_b \frac{1}{\sigma_{b,\,TQ}^2}\, \times \\
        &{\left( C^{TQ}_{b,\,\mathrm{h1}{\times}\mathrm{h2}}
               - \epsilon^{TT}_Q C^{TT}_{b,\,\mathrm{h1}{\times}\mathrm{h2}}
               - \epsilon^{TQ}_Q C^{TQ}_{b,\,\mathrm{mock}} \right)}^2,
  \end{split}
\end{equation}
where
\begin{equation}
\label{eqn:t_to_q_sigma}
  \sigma_{b,\,TQ}^2 = 
    \frac{{\left(C^{TQ}_{b,\,\mathrm{f}\times\mathrm{f}}\right)}^2+
           C^{TT}_{b,\,\mathrm{f}\times\mathrm{f}}\,
           C^{QQ}_{b,\,\mathrm{f}\times\mathrm{f}}}
         {(2\ell_b+1)f_{\mathrm{sky}}\Delta\ell}.
\end{equation}
  The terms in the numerator of the summand in Equation~\ref{eqn:t_to_q_chi2} are a binned version of the terms in Equation~\ref{eqn:t_to_q_model}.
  We binned the spectra in the $\ell$ range between 500 and 2000 with the uniform bin width 15.
  For $\sigma_{b,\,TQ}^2$, the terms in the numerator of Equation~\ref{eqn:t_to_q_sigma} are the $TQ$, $TT$, and $QQ$ spectra calculated from the $T/Q/U$ maps of the \texttt{full} coadd, $f_\mathrm{sky}$ is the sky fraction of a subfield, and $\Delta\ell$ is the bin width.
  This expression for the variance is based on \citet[Equation~28]{tristram05}.
  Similarly, the $\chi^2$ function associated with Equation~\ref{eqn:t_to_u_model} is the following:
\begin{equation}
  \begin{split}
  \label{eqn:t_to_u_chi2}
    \chi_{TU}^2 &(\epsilon^{TU}_U)
      = \sum_b \frac{1}{\sigma_{b,\,TU}^2}\, \times \\
        &{\left( C^{TU}_{b,\,\mathrm{h1}{\times}\mathrm{h2}}
           - \epsilon^{TU}_U C^{TT}_{b,\,\mathrm{h1}{\times}\mathrm{h2}} \right)}^2,
  \end{split}
\end{equation}
where
\begin{equation}
\label{eqn:t_to_u_sigma}
  \sigma_{b,\,TU}^2 = 
    \frac{{\left(C^{TU}_{b,\,\mathrm{f}\times\mathrm{f}}\right)}^2+
           C^{TT}_{b,\,\mathrm{f}\times\mathrm{f}}\,
           C^{UU}_{b,\,\mathrm{f}\times\mathrm{f}}}
         {(2\ell_b+1)f_{\mathrm{sky}}\Delta\ell}.
\end{equation}
  The $\chi^2$ minimization process used to determine the leakage coefficients was in fact combined with a similar process used to determine a bias related to detector polarization angles described in Section~\ref{sec:polarization_angle_calibration} for computational convenience, and practical aspects of the combined minimization processes are described in that section.

  Table~\ref{tab:monopole_t_to_p_leakage_removal} shows the values of $\epsilon^{TQ}_Q$ and $\epsilon^{TU}_U$ that minimize the sum of Equations~\ref{eqn:t_to_q_chi2} and \ref{eqn:t_to_u_chi2} and the uncertainties on the values for each frequency band, subfield, and Stokes parameter.
  The uncertainties are the 68\% confidence intervals obtained from the $\chi^2$ minimization process.
  For each Stokes parameter and frequency band, the four leakage coefficients tend to form two groups: the values for \texttt{el0} and \texttt{el1} are close to each other and form one group, and the values for \texttt{el2} and \texttt{el3} form the other group.
  This grouping is consistent with the scenario in which the monopole $T$-to-$P$ leakage arises from partial polarization of the \hii{} regions used for the detector gain estimation because \texttt{el0} and \texttt{el1} use RCW38, while \texttt{el2} and \texttt{el3} use Mat5a.
  We used each set of $\epsilon^{TT}_Q$ and $\epsilon^{TT}_U$ to clean all the subfield \texttt{full}, \texttt{half}, \texttt{one-thirtieth}, and \texttt{signflip-noise} coadds for the relevant frequency band as follows:
\begin{equation}
\label{eqn:monopole_t_to_p_leakage_removal}
  \begin{pmatrix}
    \bar{T}(\hat{n}) \\ \bar{Q}(\hat{n}) \\ \bar{U}(\hat{n}) \\
  \end{pmatrix}_2
  =
  \begin{pmatrix}
     1 & 0 & 0 \\
     -\epsilon^{TT}_Q & 1 & 0 \\
     -\epsilon^{TT}_U & 0 & 1 \\
  \end{pmatrix}
  \begin{pmatrix}
     \bar{T}(\hat{n}) \\ \bar{Q}(\hat{n}) \\ \bar{U}(\hat{n}) \\
  \end{pmatrix}_1,
\end{equation}
where the subscripts 1 and 2 denote the coadds that have the gain calibration applied (Equation~\ref{eqn:gain_calibration}) and the coadds that further have the monopole $T$-to-$P$ leakage removed, respectively.

\begin{table}
  \caption{\label{tab:monopole_t_to_p_leakage_removal}%
  The monopole $T$-to-$P$ leakage coefficients, $\epsilon^{TT}_Q$ and $\epsilon^{TT}_U$, expressed as percentage for each subfield and frequency band.
  The rows labeled by $Q$ and $U$ list values of $\epsilon^{TT}_Q$ and $\epsilon^{TT}_U$, respectively.
  The rightmost column shows the \hii{} region used to calibrate the timestreams for each subfield, where ``R'' and ``M'' represent RCW38 and Mat5a, respectively, to indicated the grouping of the leakage coefficients by the \hii{} regions.}
  \begin{ruledtabular}
  \begin{tabular}{cc|ccc|l}
      &   & 95\;GHz & 150\;GHz & 220\;GHz & Cal \\
    \hline
    \multirow{4}{0.1em}{$Q$}
      & \texttt{el0} & 0.25 $\pm$ 0.06 & 0.26 $\pm$ 0.06 & 0.26 $\pm$ 0.10 & \multirow{2}{0.1em}{R} \\
      & \texttt{el1} & 0.31 $\pm$ 0.07 & 0.30 $\pm$ 0.06 & 0.38 $\pm$ 0.10 & \\
      & \texttt{el2} & 0.76 $\pm$ 0.07 & 0.94 $\pm$ 0.06 & 1.86 $\pm$ 0.09 & \multirow{2}{0.1em}{M}\\
      & \texttt{el3} & 0.65 $\pm$ 0.08 & 0.88 $\pm$ 0.08 & 1.81 $\pm$ 0.11 & \\
    \hline
    \multirow{4}{0.1em}{$U$}
      & \texttt{el0} & 0.54 $\pm$ 0.05 & 0.72 $\pm$ 0.04 & 0.81 $\pm$ 0.08 & \multirow{2}{0.1em}{R} \\
      & \texttt{el1} & 0.60 $\pm$ 0.05 & 0.70 $\pm$ 0.04 & 0.65 $\pm$ 0.09 & \\
      & \texttt{el2} & 0.89 $\pm$ 0.05 & 1.30 $\pm$ 0.05 & 1.11 $\pm$ 0.09 & \multirow{2}{0.1em}{M} \\
      & \texttt{el3} & 0.87 $\pm$ 0.06 & 1.18 $\pm$ 0.06 & 1.32 $\pm$ 0.10 & \\
  \end{tabular}
  \end{ruledtabular}
\end{table}

\subsection{\label{sec:polarization_angle_calibration}Polarization Angle Calibration}
  In the third step, we corrected for an effect of the nonzero average of the miscalibrations of the detector polarization angles over all the detectors for each frequency band.
  For each coadd, we applied a $2{\times}2$ rotation matrix to the $Q/U$ maps.

  As shown in, e.g., \citep{hu03} and \citep{keating13}, a nonzero average of the miscalibrations of the polarization angles over all the detectors contributing to a set of $Q/U$ maps causes a mixing of $Q/U$ anisotropies in the form of a rotation as follows:
\begin{align}
  Q_{\mathrm{measured}} &=
    Q_{\mathrm{true}} \cos\,2\Delta\psi + U_{\mathrm{true}} \sin\,2\Delta\psi,
    \label{eqn:mixing_q} \\
  U_{\mathrm{measured}} &=
    -Q_{\mathrm{true}} \sin\,2\Delta\psi + U_{\mathrm{true}} \cos\,2\Delta\psi,
    \label{eqn:mixing_u}
\end{align}
where $\Delta\psi$ is the average miscalibration angle.
  This mixing in turn causes a mixing of $E$- and $B$-mode polarization anisotropies and causes nonzero $TB/EB$ spectra, which should be zero in the absence of cosmological parity-violating effects.

  Although we know to a high degree of accuracy the physical alignment of the antenna connected to each detector with respect to the silicon wafer on which the detector is fabricated, there are two effects that could cause miscalibrated polarization angles.
  First, it is possible that we misestimated the alignment between each wafer---or the full detector array---and the locally defined polarization coordinate system, either through a mechanical misalignment or an incomplete understanding of the polarization properties of the optical system.

  The second effect is related to the fact that the sinuous antenna design used for SPT-3G is known to have a property whereby the effective polarization angle of the antenna rotates relative to the physical alignment of the antenna in a frequency-dependent manner.
  More information on this ``polarization wobble'' effect can be found in S22 (Section~4.1) and \citep{edwards12}.
  The direction of the polarization wobble can be reversed by flipping the helicity of the antenna, and each SPT-3G detector wafer was designed with an equal number of ``left-handed'' and ``right-handed'' detectors, which should on average cancel the polarization wobble contribution to the average angle miscalibration, but it is also possible that incomplete understanding of this effect, or an unequal number of left- and right-handed detectors surviving the data-quality cuts, could cause a residual contribution.

  To estimate the angle by which we need to rotate the $Q/U$ maps to correct for the mixing, we followed the method proposed in \citep{keating13} and found the angle that causes the amplitudes of the $EB/TB$ spectra to be closest to zero by minimizing a $\chi^2$ function.
  The variable of the $\chi^2$ function is $\Delta\psi_{\mathrm{cal}}$, which is an angle applied to measured $Q/U$ maps and equivalent to $-\Delta\psi$ in Equations~\ref{eqn:mixing_q} and \ref{eqn:mixing_u}.
  For a given value of $\Delta\psi_{\mathrm{cal}}$, we obtained rotated $Q/U$ maps as follows:
\begin{align}
  Q &=
    Q'\, \cos\,2\Delta\psi_{\mathrm{cal}} - U'\, \sin\,2\Delta\psi_{\mathrm{cal}}, \\
  U &=
    Q'\, \sin\,2\Delta\psi_{\mathrm{cal}} + U'\, \cos\,2\Delta\psi_{\mathrm{cal}},
\end{align}
where $Q'/U'$ on the right-hand sides are the maps from the \texttt{full} and \texttt{half} coadds that had the first two steps of calibration and cleaning steps applied, and $Q/U$ on the left-hand sides are the corresponding rotated maps.
  We used binned spectra calculated from the rotated maps (in the $\ell$ range between 500 and 2000 with the uniform bin width 15) to calculate the $\chi^2$ for the given value of $\Delta\psi_{\mathrm{cal}}$ as follows:
\begin{equation}
\label{eqn:eb_tb_chi2}
  \chi^2 = \sum_b \left[
    \frac{{\left(C^{EB}_{b,\,\mathrm{h1}{\times}\mathrm{h2}}\right)}^2}
         {\sigma_{b,\,EB}^2}\, +
    \frac{{\left(C^{TB}_{b,\,\mathrm{h1}{\times}\mathrm{h2}}\right)}^2}
         {\sigma_{b,\,TB}^2} \right],
\end{equation}
where
\begin{align}
  \sigma_{b,\,EB}^2 = 
    \frac{{\left(C^{EB}_{b,\,\mathrm{f}\times\mathrm{f}}\right)}^2+
           C^{EE}_{b,\,\mathrm{f}\times\mathrm{f}}\,
           C^{BB}_{b,\,\mathrm{f}\times\mathrm{f}}}
         {(2\ell_b+1)f_{\mathrm{sky}}\Delta\ell},
  \label{eqn:eb_sigma} \\
  \sigma_{b,\,TB}^2 = 
    \frac{{\left(C^{TB}_{b,\,\mathrm{f}\times\mathrm{f}}\right)}^2+
           C^{TT}_{b,\,\mathrm{f}\times\mathrm{f}}\,
           C^{BB}_{b,\,\mathrm{f}\times\mathrm{f}}}
         {(2\ell_b+1)f_{\mathrm{sky}}\Delta\ell}.
  \label{eqn:tb_sigma}
\end{align}
  The notations used in Equations~\ref{eqn:eb_sigma} and \ref{eqn:tb_sigma} are similar to those used in Equations~\ref{eqn:t_to_q_sigma} and \ref{eqn:t_to_u_sigma}.

  The process of minimizing the $\chi^2$ function in Equation~\ref{eqn:eb_tb_chi2} was in fact combined with the process of minimizing the functions in Equations~\ref{eqn:t_to_q_chi2} and \ref{eqn:t_to_u_chi2} for computational convenience.
  We used one MCMC (Markov Chain Monte Carlo) run to find the minimum of the sum of the three $\chi^2$ functions in the 4D parameter space ($\epsilon^{TT}_Q$, $\epsilon^{TT}_U$, $\epsilon^{TQ}_Q$, and $\Delta\psi_{\mathrm{cal}}$) through an emulator.
  To build the emulator, we generated 100 points in the parameter space through Latin hypercube sampling.
  For each point, we used the corresponding $\epsilon^{TT}_Q$ and $\epsilon^{TT}_U$ values to calculate the $\chi^2$ values in Equations~\ref{eqn:t_to_q_chi2} and \ref{eqn:t_to_u_chi2}.
  We also subtracted the corresponding scaled copies of the gain calibration-applied \texttt{full} and \texttt{half} $T$ maps from the $Q/U$ maps, rotated the resultant $Q/U$ maps by the corresponding angle, and calculated the $\chi^2$ in Equation~\ref{eqn:eb_tb_chi2}.
  Then, we supplied the total $\chi^2$ and the associated spectra for each point to the software \texttt{CosmoPower} \citep{spuriomancini22} to train the emulator.
  The trained emulator provided predictions of the $EB/TB$ spectra of rotated maps at different points in the 4D parameter space.

  Table~\ref{tab:polarization_angle_calibration} shows the value of $\Delta\psi_{\mathrm{cal}}$ determined from the $\chi^2$ minimization process for each frequency band and subfield.
  As in the previous two tables, the uncertainty corresponds to the 68\% confidence interval.
  These angles are defined using the IAU convention instead of the COSMO convention.
  For each frequency band, we deemed the spread of the four values not large enough to indicate significant discrepancies between the subfields and decided to use the inverse-variance weighted average of the four angles to rotate the $Q/U$ maps for all the subfields.
  The average angle, $\Delta\psi_{\mathrm{cal,\,avg}}$, is shown in the last row.
  The same operation that we applied to the \texttt{full}, \texttt{half}, \texttt{one-thirtieth}, and \texttt{signflip-noise} coadds for all the subfields is the following:
\begin{equation}
\label{eqn:polarization_angle_calibration}
  \begin{pmatrix}
     \bar{T}(\hat{n}) \\ \bar{Q}(\hat{n}) \\ \bar{U}(\hat{n}) \\
  \end{pmatrix}_3
  =
  \begin{pmatrix}
     1 & 0 & 0 \\
     0 & c & -s \\
     0 & s & c \\
  \end{pmatrix}
  \begin{pmatrix}
     \bar{T}(\hat{n}) \\ \bar{Q}(\hat{n}) \\ \bar{U}(\hat{n}) \\
  \end{pmatrix}_2,
\end{equation}
where $c = \cos\,2\Delta\psi_{\mathrm{cal,\,avg}}$, $s = \sin\,2\Delta\psi_{\mathrm{cal,\,avg}}$, and the subscripts 2 and 3 denote the coadds produced by the monopole $T$-to-$P$ leakage removal step and the coadds that further had the $Q/U$ rotation applied, respectively.

\begin{table}
  \caption{\label{tab:polarization_angle_calibration}%
  The $Q/U$ rotation angle, $\Delta\psi_{\mathrm{cal}}$, for each subfield and frequency band expressed as degrees.
  The averages over the four subfields for each frequency band, $\Delta\psi_{\mathrm{cal,\,avg}}$, are shown in the last row.
  The uncertainty on $\Delta\psi_{\mathrm{cal,\,avg}}$ is the (inverse-quadrature) combined uncertainty from the four angles.}
  \begin{ruledtabular}
  \begin{tabular}{c|rrr}
         & \multicolumn{1}{c}{95\;GHz}
         & \multicolumn{1}{c}{150\;GHz}
         & \multicolumn{1}{c}{220\;GHz} \\
    \hline
    \texttt{el0}  & 0.36 $\pm$ 0.06 & 0.29 $\pm$ 0.05 & $-$0.65 $\pm$ 0.23 \\
    \texttt{el1}  & 0.46 $\pm$ 0.06 & 0.36 $\pm$ 0.05 & $-$0.70 $\pm$ 0.24 \\
    \texttt{el2}  & 0.57 $\pm$ 0.07 & 0.50 $\pm$ 0.05 &    0.09 $\pm$ 0.27 \\
    \texttt{el3}  & 0.53 $\pm$ 0.08 & 0.45 $\pm$ 0.06 &    0.34 $\pm$ 0.30 \\
    \hline
    Avg. & 0.46 $\pm$ 0.03 & 0.39 $\pm$ 0.03 & $-$0.32 $\pm$ 0.13 \\
  \end{tabular}
  \end{ruledtabular}
\end{table}

  A nonzero average polarization angle miscalibration is primarily an issue for measurements of the $B$-mode polarization anisotropy, as the $E$-mode polarization anisotropy is orders of magnitude stronger, so leakage of the $B$-mode signal into $E$ modes is far less of an issue than the reverse.\footnote{
  The miscalibration also reduces the $E$-mode amplitude, but this effect goes as the square of the miscalibration (for small miscalibrations) and is in any case corrected for in the polarization amplitude recalibration step described in the next section.}
  Because C25 did not estimate the $BB$ spectrum, the third calibration and cleaning step defined here is not strictly necessary for that analysis.
  The lensing analyses using these maps do, however, require clean $B$-mode maps, and we decided to use the cleaner, rotated maps for the analysis in C25 as well.

  A connection between the monopole $T$-to-$P$ leakage removal step and the $Q/U$ rotation step is worth noting.
  We performed these two steps serially, with the assumption that the two steps commute (i.e., the order in which they are performed does not matter).
  From the $\chi^2$ minimization process, we found little correlation between the leakage coefficients and the rotation angle, which supports this assumption.
  However, if the rotation angle were much larger, we would need to modify the model for the leakage in Equations~\ref{eqn:t_to_q_model} and \ref{eqn:t_to_u_model} so that $C^{TU}_{\ell,\,\mathrm{h1}{\times}\mathrm{h2}}$ contains a term proportional to $C^{TQ}_{\ell,\,\mathrm{mock}}$ to reflect the $Q/U$ mixing.

\subsection{\label{sec:polarization_amplitude_calibration}Polarization Amplitude Calibration}
  In the final step, we corrected for an effect of the nonzero average of the miscalibrations of the detector polarization efficiencies and the rms variation of the miscalibrations of the detector polarization angles for each frequency band.
  For each coadd, we multiplied the $Q/U$ maps by the same calibration factor.

  During the mapmaking, we assumed that $\gamma_\alpha$ = 1 (Equation~\ref{eqn:detector_signal_model_complex}) for every detector when binning their timestreams for simplicity, which means that no detector has cross-polarization coupling, but this is not the case in reality.
  Pre-deployment lab measurements of select detectors indicated few-percent coupling, and the assumption of the perfect polarization partially caused the resultant $Q/U$ maps to have lower levels of CMB polarization anisotropy than the true sky.

  If the average polarization angle miscalibration is zero, the $Q/U$ mixing described in Section~\ref{sec:polarization_angle_calibration} is absent, but the amplitudes of $Q/U$ are still suppressed relative to the true sky by a factor related to the rms variation in the miscalibrations of the polarization angles.
  This scenario applies to the combination of the design choice of having the same number of left- and right-handed detectors and the analysis choice of treating the timestreams from both types of detectors in the same way (S22, Sections~4.1 and 7.6).
  A left-handed detector and a right-handed one are the mirror images of each other, and their polarization angles are offset from a nominal value by the same magnitude with the opposite signs.
  During the mapmaking, we used the nominal value for both types of detectors, so the offsets between the angles of the two types of detectors contributed to an rms variation.

  We estimated the polarization amplitude calibration factor for each frequency band and subfield in a way similar to the estimation of the gain calibration factor.
  First, we cross-correlated SPT-3G 150\;GHz $Q/U$ maps with \textit{Planck} $Q/U$ maps, and then we cross-correlated SPT-3G 95 and 220\;GHz maps with the 150\;GHz ones.
  We now describe details of the estimation.

  We estimated the 150\;GHz calibration factors, $q_{\mathrm{cal},\,150}$ and $u_{\mathrm{cal},\,150}$, by finding an inverse-covariance weighted average of a binned cross-spectrum ratio.
  The estimation of $q_{\mathrm{cal},\,150}$ proceeded as follows:
\begin{align}
  R_{b,\,150}^Q &= P_{b\ell} \left(
    \frac{C^{QQ}_{\ell,\,\mathrm{150}{\times}\mathrm{143}}}
         {C^{QQ}_{\ell,\,\mathrm{150}{\times}\mathrm{150}}}\,
    \frac{B_{\ell,\,\mathrm{150}}}{B_{\ell,\,\mathrm{143}}}\,
    \frac{P_{\ell,\,\mathrm{150}}}{P_{\ell,\,\mathrm{143}}} \right),
  \label{eqn:qcal_150_rb} \\
  q_{\mathrm{cal},\,150} &= 
    \frac{{\mathbf{1}}^T\, {\left(\Sigma_{b,\,150}^Q\right)}^{-1}\, R_{b,\,150}^Q}
         {{\mathbf{1}}^T\, {\left(\Sigma_{b,\,150}^Q\right)}^{-1}\, \mathbf{1}}
  \label{eqn:qcal_150},
\end{align}
where $P_{b\ell}$ is a binning operator (for the $\ell$ range between 400 and 1300 with the uniform bin width 100), $C^{QQ}_{\ell,\,\mathrm{150}{\times}\mathrm{143}}$ is the cross-spectrum between the SPT-3G 150\;GHz \texttt{full} $Q$ map produced after the $Q/U$ rotation step and the \textit{Planck} 143\;GHz full-mission $Q$ map, $C^{QQ}_{\ell,\,\mathrm{150}{\times}\mathrm{150}}$ is the cross-spectrum between the pair of SPT-3G 150\;GHz \texttt{half} $Q$ maps (also after the $Q/U$ rotation step), the $B_{\ell}$ and $P_{\ell}$ terms are the same as the ones in Equation~\ref{eqn:gcal_150_rb}, $\Sigma_{b,\,150}^Q$ is the covariance matrix associated with the binned cross-spectrum ratio, and $\mathbf{1}$ is a vector of ones.
  We obtained $\Sigma_{b,\,150}^Q$ using the $Q$ version of the 20 simulated cross-spectrum ratios described in Section~\ref{sec:gain_calibration}.
  Then, we used the corresponding $U$ maps in the same procedure to estimate $u_{\mathrm{cal},\,150}$.

  We estimated the 95 and 220\;GHz calibration factors, $q_{\mathrm{cal},\,\nu}$ and $u_{\mathrm{cal},\,\nu}$, where $\nu$ = 95 or 220, by adding additional ratios to Equation~\ref{eqn:qcal_150_rb} as follows:
\begin{align}
  R_{b,\,\nu}^Q &= P_{b\ell}\, \Biggl(
    \frac{C^{QQ}_{\ell,\,\mathrm{150}{\times}\mathrm{150}}}
         {C^{QQ}_{\ell,\,\mathrm{150}{\times}\nu}}\,
    \frac{B_{\ell,\,\nu}}{B_{\ell,\,\mathrm{150}}} \Biggr),
  \label{eqn:qcal_nu_rb} \\
  q_{\mathrm{cal},\,\nu} &= q_{\mathrm{cal},\,150} \times
    \frac{{\mathbf{1}}^T\, {\left(\Sigma_{b,\,\nu}^Q\right)}^{-1}\, R_{b,\,\nu}^Q}
         {{\mathbf{1}}^T\, {\left(\Sigma_{b,\,\nu}^Q\right)}^{-1}\, \mathbf{1}},
\end{align}
where $C^{QQ}_{\ell,\,\mathrm{150}{\times}\nu}$ is the cross-spectrum between a 150\;GHz \texttt{half} map and a 95 or 220\;GHz \texttt{half} map from the other half of the dataset.
  Because $Q/U$ maps are much less contaminated by foregrounds than $T$ maps and because the SPT $Q/U$ maps are deeper than the \textit{Planck} ones, we extended the upper limit of the $\ell$ range used for the binning to 1800 (with the uniform bin width increased to 200).

  Table~\ref{tab:polarization_amplitude_calibration} shows the $q_{\mathrm{cal}}$ and $u_{\mathrm{cal}}$ factors and their uncertainties for each frequency band and subfield.
  The uncertainty on each calibration factor is the standard error of 20 simulated numbers, each of which was obtained by supplying one of the 20 simulated cross-spectrum ratios to the relevant equations used to calculate the actual calibration factor.
  The uncertainty reflects the noise fluctuations in the SPT-3G and \textit{Planck} maps used to calculate the cross-spectra.

  For each frequency band, we deemed the spread of the eight values not large enough to indicate significant discrepancies between the subfields and Stokes parameters, and we decided to use the inverse-variance weighted average of the eight values to calibrate the $Q/U$ maps for all the subfields.
  These average values, which we call $p_{\mathrm{cal,\,avg,\,\nu}}$, where $\nu$ = 95, 150, and 220, are shown in the second-to-last row of the table.

  After we obtained $p_{\mathrm{cal,\,avg,\,\nu}}$, we further multiplied it by another calibration factor, $\sqrt{0.966}$.
  This factor is based on \citet[Section~3.3.4, Equation~45]{planck18-5}.
  It is a correction that improves the calibration of the \textit{Planck} 143\;GHz $Q/U$ maps, but it was applied at the likelihood level rather than to the publicly released maps.
  Since we used the \textit{Planck} maps that do not include this factor, we applied this factor manually.
  Although this factor is cosmology-dependent, we used a wide prior (uniform prior between 0.8 and 1.2) for the nuisance parameters related to the polarization amplitude calibration factors in C25 when constraining cosmology using the full set of $TT/TE/EE$ spectra.
  We use $p_{\mathrm{cal,\,fin}}$ to denote the product of $p_{\mathrm{cal,\,avg}}$ and $\sqrt{0.966}$, which is shown in the last row of the table.

\begin{table}
  \caption{\label{tab:polarization_amplitude_calibration}%
  Various polarization amplitude calibration factors.
  The first eight rows show $q_\mathrm{cal}$ and $u_\mathrm{cal}$, which are the calibration factors for each subfield and frequency band.
  The second-to-last row shows $p_\mathrm{cal,\,avg}$, which is the weighted average calibration factor over the four subfields and two Stokes parameters for each frequency band.
  The uncertainty on $p_\mathrm{cal,\,avg}$ is the (inverse-quadrature) combined uncertainty from the eight calibration factors.
  The last row shows $p_\mathrm{cal,\,fin}$, the final calibration factor for each frequency band, and $p_\mathrm{cal,\,fin}$ is simply the product of $p_\mathrm{cal,\,avg}$ and $\sqrt{0.966}$.}
  \begin{ruledtabular}
  \begin{tabular}{cc|ccc}
      &   & 95\;GHz & 150\;GHz & 220\;GHz \\
    \hline
    \multirow{4}{0.1em}{$Q$}
      & \texttt{el0} & 1.096 $\pm$ 0.031 & 1.106 $\pm$ 0.031 & 1.233 $\pm$ 0.036 \\
      & \texttt{el1} & 1.037 $\pm$ 0.017 & 1.031 $\pm$ 0.017 & 1.106 $\pm$ 0.020 \\
      & \texttt{el2} & 1.065 $\pm$ 0.034 & 1.065 $\pm$ 0.034 & 1.146 $\pm$ 0.037 \\
      & \texttt{el3} & 1.090 $\pm$ 0.029 & 1.090 $\pm$ 0.029 & 1.201 $\pm$ 0.034 \\
    \hline
    \multirow{4}{0.1em}{$U$}
      & \texttt{el0} & 1.093 $\pm$ 0.020 & 1.116 $\pm$ 0.020 & 1.222 $\pm$ 0.023 \\
      & \texttt{el1} & 1.052 $\pm$ 0.016 & 1.077 $\pm$ 0.017 & 1.181 $\pm$ 0.019 \\
      & \texttt{el2} & 1.114 $\pm$ 0.026 & 1.128 $\pm$ 0.026 & 1.260 $\pm$ 0.030 \\
      & \texttt{el3} & 1.089 $\pm$ 0.026 & 1.106 $\pm$ 0.026 & 1.224 $\pm$ 0.030 \\
    \hline
    \multicolumn{2}{c|}{Avg.}
      & 1.073 $\pm$ 0.008 & 1.083 $\pm$ 0.008 & 1.189 $\pm$ 0.009 \\
    \multicolumn{2}{c|}{Final}
      & 1.054 $\pm$ 0.008 & 1.064 $\pm$ 0.008 & 1.169 $\pm$ 0.009 \\
  \end{tabular}
  \end{ruledtabular}
\end{table}

  It is $p_{\mathrm{cal,\,fin}}$ that we eventually used to calibrate the \texttt{full}, \texttt{half}, \texttt{one-thirtieth}, and \texttt{signflip-noise} coadds through the following operation:
\begin{equation}
\label{eqn:polarization_amplitude_calibration}
  \begin{pmatrix}
     \bar{T}(\hat{n}) \\ \bar{Q}(\hat{n}) \\ \bar{U}(\hat{n}) \\
  \end{pmatrix}_4
  =
  \begin{pmatrix}
     1 & 0 & 0 \\
     0 & p_\mathrm{cal,\,fin} & 0 \\
     0 & 0 &  p_\mathrm{cal,\,fin} \\
  \end{pmatrix}
  \begin{pmatrix}
     \bar{T}(\hat{n}) \\ \bar{Q}(\hat{n}) \\ \bar{U}(\hat{n}) \\
  \end{pmatrix}_3,
\end{equation}
where the subscripts 3 and 4 denote the coadds produced by the $Q/U$ rotation step and the coadds that further had the polarization amplitude calibration factor applied, respectively.

  As in Section~\ref{sec:gain_calibration}, we clarify the differences between the notations used in this work and those used in G25 and C25.
  The three calibration factors $P_{\mathrm{cal,external}}^{150}$, $P_{\mathrm{cal,internal}}^{95}$, and $P_{\mathrm{cal,internal}}^{220}$ described in G25 (Appendix~B) are the same as the calibration factors $p_{\mathrm{cal,\,fin},\,150}$, $p_{\mathrm{cal,\,fin},\,95}$, and $p_{\mathrm{cal,\,fin},\,220}$ described in this work.
  On the other hand, the three nuisance parameters $E_\mathrm{cal}^\mathrm{ext}$, $E_\mathrm{cal}^\mathrm{rel,95}$, and $E_\mathrm{cal}^\mathrm{rel,220}$ in C25 are equivalent to the inverse of $p_{\mathrm{cal,\,fin},\,150}$, $p_{\mathrm{cal,\,fin},\,95}/p_{\mathrm{cal,\,fin},\,150}$, and $p_{\mathrm{cal,\,fin},\,220}/p_{\mathrm{cal,\,fin},\,150}$.
  In C25, when choosing the width of the prior for $E_\mathrm{cal}^\mathrm{ext}$ to constrain cosmology using only the $EE$ spectra, we added two numbers in quadrature to obtain the chosen width 0.0095.
  One number is the uncertainty on $p_{\mathrm{cal,\,fin},\,150}$ (0.008), and the other number is an uncertainty on the \textit{Planck} calibration (0.005).

  We applied the four steps of calibration and cleaning sequentially to the subfield coadds and then combined them to produce the full-field coadds, which formed inputs of the analysis in C25.
  As described earlier, the analysis in G25 used the pre-calibration and cleaning full-field coadds and included the four steps in the forward model.
  As shown in  G25 (Figure~7), the calibration and cleaning parameters obtained from the forward model are consistent with the corresponding parameters obtained from the steps described in this work.

\section{\label{sec:residual_calibration_biases}Residual Calibration Biases}
  In this section, we described any (known) biases that are not removed by the calibration and cleaning steps described in Section~\ref{sec:map_calibration}.
  We identify two known residual biases: declination-dependent gain miscalibration and higher-order $T$-to-$P$ leakage.
  The declination-dependent gain miscalibration has a negligible impact on the recovered power spectra (Section~\ref{sec:residual_cal_bias_gain_dec_dependence}) and higher-order $T$-to-$P$ leakage is mitigated at the spectrum level (Section~\ref{sec:residual_cal_bias_higher_order_t_to_p}).

\subsection{\label{sec:residual_cal_bias_gain_dec_dependence}Declination-dependent Gain Miscalibration}
  In the gain calibration step (Section~\ref{sec:gain_calibration}), we reduced the overall gain miscalibration across a subfield by multiplying all the pixel values of the subfield $T/Q/U$ maps by a single number, but this step did not correct for any spatially dependent gain miscalibration within a subfield.
  We expect the $T/Q/U$ maps to have declination-dependent gain miscalibrations because the gain of a detector generally increases over the course of a subfield observation as the telescope moves upward in elevation (toward higher declination).

  The increase of the gain of a detector is caused by a decrease in the power received by the detector from atmospheric emission as described in Section~\ref{sec:field_observations}.
  Although splitting the SPT-3G Main field into the four subfields and observing one subfield at a time significantly helps with reducing this type of gain change, it is not completely removed.

  We estimated the gain change within a single subfield observation using the calibrator stares described in Section~\ref{sec:calibration_observations} that are interspersed with the subfield observations.
  A calibrator stare is conducted before each subfield observation, at the bottom elevation of the subfield, and after each subfield observation, at the top elevation of the subfield.
  By comparing the amplitudes of the response of each detector to the calibrator before and after the subfield observation, we can estimate how much the gain of each detector changed over the course of the observation.

  Combining measurements of these amplitude changes from typical calibrator stares with typical weights of the individual detectors used to bin their timestreams, we calculated the weighted average of the fractional gain change over the full detector array to be approximately 6\% in the 150\;GHz coadds across \texttt{el0}.
  The changes across the other three subfields or for the other two frequency bands are smaller.

\begin{figure}
  \includegraphics[scale=1.00]{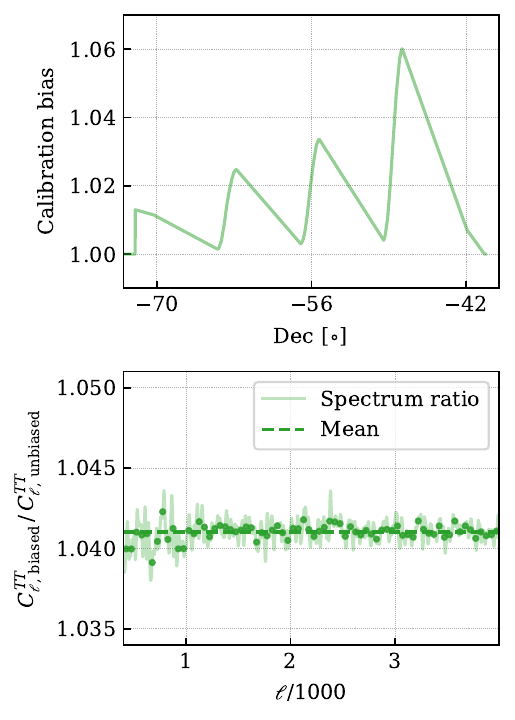}
  \caption{\label{fig:residual_cal_bias_gain_dec_dependence}%
  The simulated declination-dependent calibration bias and its effect on a power spectrum.
  The top panel shows the sawtooth pattern of calibration bias that we constructed in the toy model, and the bottom panel shows the ratio of the spectrum of the biased map to that of the unbiased one.
  The markers in the bottom panel represent a binned version of the ratio spectrum with the uniform bin width 50, which is the width used for the analysis in C25.
  The dashed line indicates the mean of the ratio spectrum.}
\end{figure}

  Using a toy model simulation that incorporates the actual measurements of the fractional gain changes and assumes a linear increase of a gain over the course of a subfield observation, we simulated the effect of this declination-dependent gain miscalibration on power spectra and concluded that it is negligible.
  In this simulation, we constructed a sawtooth pattern of calibration bias as a function of declination using the average fractional gain change over all the 150\;GHz detectors for each subfield.
  We arbitrarily chose to assume accurate gains (sawtooth function equal to unity) at the low-elevation/high-declination end of each subfield, which corresponds to the beginning of an observation.
  The sawtooth pattern of calibration bias is shown in the top panel of Figure~\ref{fig:residual_cal_bias_gain_dec_dependence}.
  We then multiplied an output $T$ map from a mock observation by this declination-dependent function and calculated the ratio of the spectrum of the modified output $T$ map to the spectrum of the unmodified map.
  This ratio is shown in the bottom panel of Figure~\ref{fig:residual_cal_bias_gain_dec_dependence}.
  Any mean offset from unity in this ratio would be removed in the gain calibration step, and the remaining fluctuations would be only at the level of 0.1\%.
  Given this result, we decided not to correct for the residual declination-dependent gain.

\subsection{\label{sec:residual_cal_bias_higher_order_t_to_p}Higher-order \texorpdfstring{$T$}{T}-to-\texorpdfstring{$P$}{P} leakage}
  Although we removed the monopole $T$-to-$P$ leakage in the second step of calibration and cleaning, we did not treat any higher-order leakage.
  Dipole $T$-to-$P$ leakage can occur when two detectors that are sensitive to orthogonal polarization directions and assumed to point in the same location in the sky actually have a pointing offset, and this type of leakage contaminates $Q/U$ maps with the first-order derivative of a $T$ map.
  Quadrupole leakage can occur when the two detectors' beams have different ellipticity, and this type of leakage contaminates $Q/U$ maps with the second-order derivative of a $T$ map.
  When there is correlation between the detector polarization angle and the differential pointing or ellipticity, the corresponding leakage does not diminish as the number of detectors increases.
  If $Q/U$ maps contain both types of leakage, the $TQ$ and $TU$ spectra contain terms that are proportional to $\ell C_{\ell}^{TT}$ and ${\ell}^2 C_{\ell}^{TT}$.

\begin{figure}
  \includegraphics[scale=1.00]{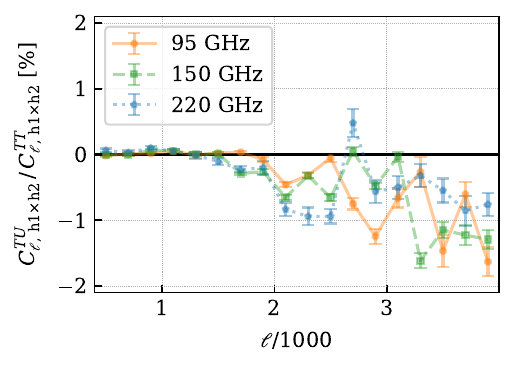}
  \caption{\label{fig:residual_cal_bias_higher_order_t_to_p}%
  The cross-spectrum ratio $TU/TT$ calculated from the \texttt{half} coadds produced at the end of the four calibration and cleaning steps for each frequency band.
  The ratio is binned with the uniform bin width 200, and the error bar for each bin is the standard error of the data points within each bin.}
\end{figure}

  After removing the monopole $T$-to-$P$ leakage, we calculated the $TQ$ and $TU$ spectra again using the \texttt{half} coadds and noticed that the spectra were not simply fluctuations around zero but had nonzero correlation between $T$ and $Q/U$ that increases as $\ell$ increases.
  This indicates that the $Q/U$ maps contain higher-order $T$-to-$P$ leakage.
  For example, in Figure~\ref{fig:residual_cal_bias_higher_order_t_to_p} we show the ratio $C^{TU}_{\ell,\,\mathrm{h1}{\times}{h2}}\,/\,C^{TT}_{\ell,\,\mathrm{h1}{\times}{h2}}$ calculated from the full-field \texttt{full} coadds produced at the end of the four steps of calibration and cleaning.

  Instead of applying additional cleaning steps to reduce higher-order $T$-to-$P$ leakage, we chose to treat the effects of the leakage at the level of power spectra calculated from the coadds.
  Details of this treatment can be found in C25 (Section~IV\,B\,2).

\section{\label{sec:map_level_null_tests}Map-level Null Tests}
  In addition to applying the four calibration and cleaning steps described in Section~\ref{sec:map_calibration} to the four types of coadds (\texttt{full}, \texttt{half}, \texttt{one-thirtieth}, and \texttt{signflip-noise}) produced from the mapmaking pipeline to reduce the known biases, we also conducted a suite of tests, which we call map-level null tests, using the fifth type of coadds (\texttt{pre-null}) to search for potential systematic errors in the \texttt{full}, \texttt{half}, and \texttt{one-thirtieth} coadds.
  In this section, we describe the procedure of the map-level null tests (Section~\ref{sec:null_procedure}), initial results from the tests (Section~\ref{sec:null_initial}), a systematic error that we detected from the initial results and had to mitigate (Section~\ref{sec:null_contamination}), and the final results (Section~\ref{sec:null_final}).

\subsection{\label{sec:null_procedure}Procedure}
  In a map-level null test, we split the full dataset into two parts and check their consistency.
  For each frequency band, we create difference maps between the two parts.
  Any common astrophysical signals are largely removed in these maps, but a potential systematic error that affects one part more than the other can be present.
  We check whether power spectra of the maps are consistent with noise fluctuations within expectations or contain excess power.

\subsubsection{\label{sec:null_procedure_splits}Data Splits}
  We split SPT-3G D1 into two parts in six different ways that are based on different anticipated systematic effects.
  We gave these six splits the following names: sun, moon, azimuth, year, scan, and wafer splits.

  We used the sun and moon splits to check whether the coadds are affected by radiation from those sources.
  For each split, the observations taken when the source was above the horizon form one part, and the observations taken when the source was below the horizon form the other part.
  While we expect the combination of our field location and our observing strategy to minimize contamination from the Sun and Moon, we usually use null tests to check this expectation.
  Details of the observing strategy can be found in Appendix~\ref{app:null_sun_avoidance}.

  We used the azimuth split to check whether the coadds are affected by thermal radiation from objects near the horizon such as nearby buildings.
  Three nearby buildings are all approximately in the direction corresponding to the azimuth angle at $153^\circ$, and we used this angle to split the full dataset.
  For each subfield observation, we assigned a representative azimuth angle to it, which is the midpoint of the azimuth range covered by all the scans of that observation.
  Then, we used all the observations whose representative azimuth angles are within ${\pm}90^\circ$ of $153^\circ$ to form one part and all the observations whose representative azimuth angles are within the other half of the circle to form the other part.

  We used the year split to check whether anything was significantly different between the data from the 2019 and 2020 austral winter observing seasons.
  Each part of the split comprises the observations taken in one of the two years.
  We did not hypothesize a particular potential systematic error associated with either year, and this split was meant to check any significant difference between the two years from any cause.

  We used the scan and wafer splits to check whether there are differences between the data from the two telescope scan directions (left-going and right-going scans) and differences between the data from two sets of detector wafers, respectively.
  For the wafer split, we divided the full detector array, which has ten detector wafers, into two groups of five detector wafers.
  We ordered the ten detector wafers by the average fractional gain change over the course of a subfield observation (Section~\ref{sec:residual_cal_bias_gain_dec_dependence}) across all the 150 GHz detectors on a detector wafer, and we set the point of split after the fifth detector wafer.
  The group that has higher fractional gain changes also happens to show a stronger spectral line near 1.42\;Hz in its timestreams.
  This is the frequency at which the helium gas is compressed and expanded in one of the two Cryomech PT-415 cryocoolers in the cryogenic system.
  Therefore, the wafer split allows us to search for systematic errors associated with at least two effects: the elevation-dependence of a detector gain and the spectral line.

  Unlike the other splits, the scan and wafer splits were applied within each subfield  observation.
  We state in Section~\ref{sec:filter_bin_details} that we used the filter-and-bin pipeline to produce weighted $T/Q/U$ maps and a weight map matrix for each subfield observation, but in fact we divided the filtered timestreams for each subfield observation into four subsets and produced weighted $T/Q/U$ maps and a weight map matrix for each subset (for each scan direction and each group of detector wafers).
  Then, to produce the \texttt{full}, \texttt{half}, \texttt{one-thirtieth}, and \texttt{signflip-noise} coadds, we added the four sets of weighted $T/Q/U$ maps and weight matrices for each subfield observation.
  To produce the \texttt{pre-null} coadds for the scan (wafer) split, for each subfield observation we added the weighted $T/Q/U$ maps and weight matrices from the two groups of detector wafers for each scan direction (the two scan directions for each group of detector wafers), and we collated the coadded maps and matrices from that scan direction (group of detector wafers) across all the subfield observations to form one part of the split.

\subsubsection{\label{sec:null_procedure_spectra}Null Spectra}
  After defining the six splits, we created difference maps between the two parts for each split.
  First, we created 25 bundles of equal depth within each part (the \texttt{pre-null} coadds).
  Then, we subtracted each bundle in one part of the split from a bundle in the other part of the split to create 25 difference bundles (hereafter null bundles).
  We created these bundles to cross-correlate them to reveal potential systematic errors common to the bundles while avoiding the noise bias.
  This method is similar to how we estimated the $TT/TE/EE$ spectra of the CMB in C25, where we cross-correlated the 30 bundles (the \texttt{one-thirtieth} coadds) described in Section~\ref{sec:coadds}.

  We primarily focused on the $TT/TE/EE$ spectra of the null bundles and treated the $BB/TB/EB$ spectra as out of the scope of this work.
  This is because the two-point correlation functions that we mainly intended to measure from the mid-$\ell$ maps were the $TT/TE/EE$ spectra.
  Using the 25 null bundles, we calculated the cross-bundle $TT/TE/EE$ spectra between all possible pairs formed from the bundles and averaged the cross-spectra as follows:
\begin{equation}
\label{eqn:null_spectrum}
  \begin{split}
    C_{\ell,\,\mathrm{avg}}^{X Y} = \frac{2}{25 \cdot 24}
      \sum_{i=1}^{24} \sum_{j=i+1}^{25}
      C_{\ell,\,i{\times}j}^{X Y},
  \end{split}
\end{equation}
where $i$ and $j$ denote null bundle numbers, $C_{\ell,\,i{\times}j}^{X Y}$ is a spectrum calculated by cross-correlating bundles $i$ and $j$ using \texttt{PolSpice}, and $X, Y \in \{T,E\}$.
  The apodization mask that we supplied to \texttt{PolSpice} has tapers near the borders of the SPT-3G Main field and covers the galaxy clusters and emissive sources masked in the timestream high-pass filter.
  (Tapers also exist around the holes covering the objects.)
  We call $C_{\ell,\,\mathrm{avg}}^{X Y}$ a null spectrum.
  From all the splits and frequency bands, we obtained 54 null spectra (6 splits $\times$ 3 frequency bands $\times$ 3 combinations of $T$ and $E$).

  To judge whether a potential systematic error is detected in a null spectrum, we calculated a $\chi^2$ value and the associated $p$-value from a binned version of the spectrum and compared the $p$-value with a predetermined threshold, which we call $p_\mathrm{sys}$ hereafter.
  We binned each null spectrum in the $\ell$ range between 400 and 4000 with the uniform bin width 50, which is the same as the binning scheme used in C25 to measure the $TT/TE/EE$ spectra of the CMB.
  We set $p_\mathrm{sys}$ to 0.05/54 (0.00093), the denominator being the Bonferroni correction \citep[page 569]{shaffer95} for the total number of null spectra.
  A $p$-value lower than $p_\mathrm{sys}$ indicates a potential systematic error that should be further investigated.

  The $\chi^2$ value for a binned null spectrum was calculated as follows:
\begin{equation}
\label{eqn:null_chi2}
  \chi^2 = \sum_b\,
    \frac{\left(C_{b,\,\mathrm{avg}}^{X Y} - C_{b,\,\mathrm{exp}}^{X Y}\right)^2}
         {\sigma_b^2}
\end{equation}
where $b$ represents a bin, $C_{b,\,\mathrm{avg}}^{X Y}$ is the binned version of Equation~\ref{eqn:null_spectrum}, $C_{b,\,\mathrm{exp}}^{X Y}$ is what we expect $C_{b,\,\mathrm{avg}}^{X Y}$ to be in the absence of potential systematic errors, and $\sigma_b^2$ is an estimate of the variance on $C_{b,\,\mathrm{avg}}^{X Y}$.
  We now describe $C_{b,\,\mathrm{exp}}^{X Y}$ and $\sigma_b^2$ in detail.

\subsubsection{\label{sec:null_procedure_expectations}Expectation Spectra}
  We used different $C_{b,\,\mathrm{exp}}^{X Y}$ for different splits.
  For the sun, moon, azimuth, and year splits, we expected astrophysical signals to be    removed in the null bundles, so we assumed $C_{b,\,\mathrm{exp}}^{X Y}$ to be zero.
  For the scan and wafer splits, we did not expect astrophysical signals to be completely removed and used simulations to model $C_{b,\,\mathrm{exp}}^{X Y}$.

  The null bundles from the scan split contain residual astrophysical signals because the bundles within the left-going (right-going) part of the split contain a shifted version of the signals from the sky to the left (right).
  This shift is caused by two effects.
  One effect is that the detectors have finite time constants and respond to changing signals from the sky with a delay as the telescope scans, and we did not correct for this effect in the timestreams during the mapmaking.
  The other effect is that the data acquisition system has an offset between the time stamp assigned to a detector timestream sample and the time stamp assigned to the corresponding telescope pointing timestream sample.
  The direction of this offset is such that the right ascension assigned to a detector timestream sample collected at a given moment is in fact the right ascension at which the detector was pointed at a slightly earlier moment.

  Coincidentally, the effects of the time offset and the time constants act in opposite directions and have similar magnitudes, and their effects nearly cancel, particularly in the 220\;GHz detectors.\footnote{
  To first order in $2 \pi f \tau$, where $\tau$ is the detector time constant, the effect of time constants is also a simple shift.}
  Nevertheless, there is a net residual shift of astrophysical signals along the scan direction.
  More information on these two effects can be found in \citet[Section~3.2]{archipley26}.
  In the next generation of SPT-3G maps, including the ones used in that work, we use timestreams that are corrected for both the detector time constants and the time offset.

  To estimate the expected nonzero null spectra associated with the time constants and time offset, we simulated the effects using mock-observation output maps.
  Starting from a set of mock-observed $T/Q/U$ maps, we decomposed each map into constant-declination HEALPix rings, created two shifted versions of each ring (one shifted to the left and the other shifted to the right), subtracted one version from the other, substituted the values of the difference ring for the values of the original ring, and calculated the $TT/TE/EE$ spectra of the maps assembled from the difference rings.

  When creating a shifted version of the values in a HEALPix ring, we used a weighted average of the time constants over the full detector array for each frequency band (6.9, 6.5, and 4.3\;ms in the 95, 150, and 220\;GHz bands, respectively) and the best estimate of the time offset (4.6\;ms) to modify the values in the ring.
  We calculated the weighted average for each frequency band by first estimating an average time constant of each detector over the two observing seasons and then combining the average time constants of the individual detectors with the weights of the detectors used in the mapmaking for a typical subfield observation.
  The average time constant of an individual detector was estimated from results of the many calibrator sweeps (Section~\ref{sec:calibration_observations}) over the two observing seasons.

  The null bundles from the wafer split contain residual astrophysical signals because the timestreams from the two groups of detector wafers were high-pass filtered with slightly different $\ell_{x,\,c}$ during the mapmaking.
  As described in Section~\ref{sec:high_pass_filter_cutoff}, the cutoff frequency of the high-pass filter used for each scan is a function of the telescope elevation such that the cutoff frequency is equal to $\ell_{x,\,c}$ = 300.
  Here the telescope elevation is defined to be the elevation at which a detector located near the center of the full detector array is pointed.
  We use this definition to calculate the cutoff frequency for each scan and use that frequency for all the detectors.
  As a result, that cutoff frequency is in turn equal to slightly different $\ell_{x,\,c}$ for detectors at different locations of the full detector array.
  This resulted in the two groups of detector wafers having slightly different filter transfer functions because they have different average elevation, which caused the imperfect cancellation of astrophysical signals in the null bundles.
  We estimated the expected nonzero null spectra by mock-observing a simulated sky using the pointing information of the two groups of detector wafers separately and calculating the $TT/TE/EE$ spectra of the difference maps.

  Although the null bundles from the scan and wafer splits have nonzero expectation spectra, these spectra do not represent systematic errors.
  For the scan split, the effects of the time constants and time offset make the beam elongated along the scan direction, but the effects are incorporated in our beam measurement because the measurement was made from maps that have the same effects.
  For the wafer split, the average of the filter transfer functions from the two groups of wafers is what is needed to make unbiased measurements of the $TT/TE/EE$ spectra of the CMB, and the mock observations described in Section~\ref{sec:simulations_specifics} yield the average transfer function because the pointing information from all the detector wafers was used.

\subsubsection{\label{sec:null_procedure_pvalues}Null Spectrum Variance and $p$-values}
  Having described the term $C_{b,\,\mathrm{exp}}^{X Y}$ in Equation~\ref{eqn:null_chi2}, we now describe the term $\sigma_b^2$.
  We tried two values for this term.
  In the first case, we used the variance of the mean of the 300 cross-spectra calculated from the pairs of the null bundles.
  We use $\sigma_{b,\,n}^2$ to denote the variance used in this case.
  The $\chi^2$ calculated with $\sigma_{b,\,n}^2$ indicates how significant the deviation of a null spectrum from the expectation is with respect to noise fluctuations in the coadds.
  In the second case, we added additional variance to $\sigma_{b,\,n}^2$.
  This additional variance is a small fraction (0.01\%) of the $TT$, $TE$, or $EE$ sample variance (the uncertainty on the estimate of the true, underlying power spectrum caused by measuring a finite number of modes from a limited sky area).
  We use $\sigma_{b,\,n+s}^2$ to denote the variance used in this case for later convenience.

  At relatively low $\ell$, we measure individual spherical harmonic modes of the temperature and $E$-mode polarization anisotropies within the SPT-3G Main field with high signal-to-noise ratios, and the total variance of our measurements of the $TT/TE/EE$ spectra of the CMB is not dominated by noise fluctuations in the coadds but by the sample variance.
  The idea of adding a small fraction of the sample variance when calculating a $\chi^2$ value is to test whether a potential systematic error detected above the noise variance is important to the measurements of the $TT/TE/EE$ spectra of the CMB and the associated constraints on the cosmological parameters.
  A potential systematic error that is still much smaller than the sample variance should cause negligible biases to the measurements and constraints.
  In \citet[Section~IV\,A]{balkenhol23}, we checked that a systematic error at the level of 1\% of the sample variance caused negligible shifts to constraints on the cosmological parameters.
  Given that the noise spectra from the 2019--2020 dataset are approximately a factor of ten lower than those from the 2018 dataset, we chose a smaller fraction of the sample variance, 0.01\%, this time.

  Given the two $p$-values calculated from each null spectrum, there were three outcomes.
  First, if the $p$-value associated with the $\chi^2$ value calculated using only the noise variance ($\sigma_{b,\,n}^2$) was above $p_\mathrm{sys}$, we concluded that we did not detect a potential systematic error.
  Second, even if that $p$-value was below $p_\mathrm{sys}$, as long as the other $p$-value, the one associated with the $\chi^2$ value calculated using the inflated variance ($\sigma_{b,\,n+s}^2$), was above $p_\mathrm{sys}$, we concluded that we detected a potential systematic error at a negligible level and did not prioritize investigations into the causes.
  Third, if the $p$-value obtained with the inflated variance was still below the $p_\mathrm{sys}$, that required further investigations.
  In this work, we checked the $p$-values only individually and did not perform the uniformity tests of distributions of $p$-values used in \citet[Section~V\,A]{dutcher21}.
  This is because, while those tests are based on the assumption that the $p$-values from the null spectra are independent, the addition of the small fraction of the sample variance adds correlation among $p$-values.

\subsection{\label{sec:null_initial}Initial Results}
\begin{figure}
  \includegraphics[scale=1.00]{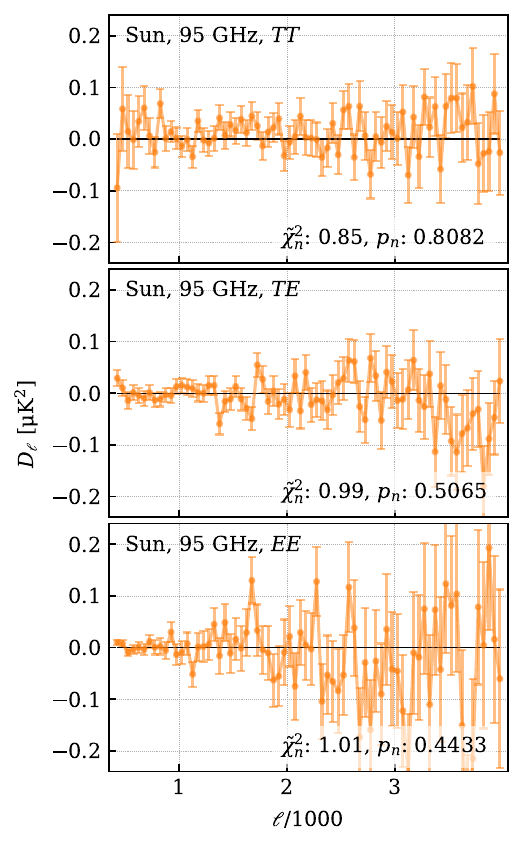}
  \caption{\label{fig:null_spectra_initial_outcome1}%
  Three 95\;GHz null spectra from the sun split as examples of the spectra that had good $p$-values.
  The null spectra are shown in the $D_{\ell}$ convention, where $D_{\ell} = [\ell(\ell+1)/(2\pi)] C_{\ell}$.
  The error bars accompanying each null spectrum represent $\sigma_{b,\,n}$.
  In each panel, the top left corner indicates the type of the null spectrum.
  The bottom right corner shows the reduced $\chi^2$ ($\tilde{\chi}^2$) value calculated from the null spectrum and the associated $p$-value.
  The expected null spectra for the sun split are zero.}
\end{figure}

  All the 18 null spectra from the sun and moon splits, 15 out of the 18 null spectra from the azimuth and year splits, and 6 out of the 18 null spectra from the scan and wafer splits had the first outcome (no detection of potential systematic errors).
  Figure \ref{fig:null_spectra_initial_outcome1} shows three null spectra from the sun split as examples.

\begin{figure}
  \includegraphics[scale=1.00]{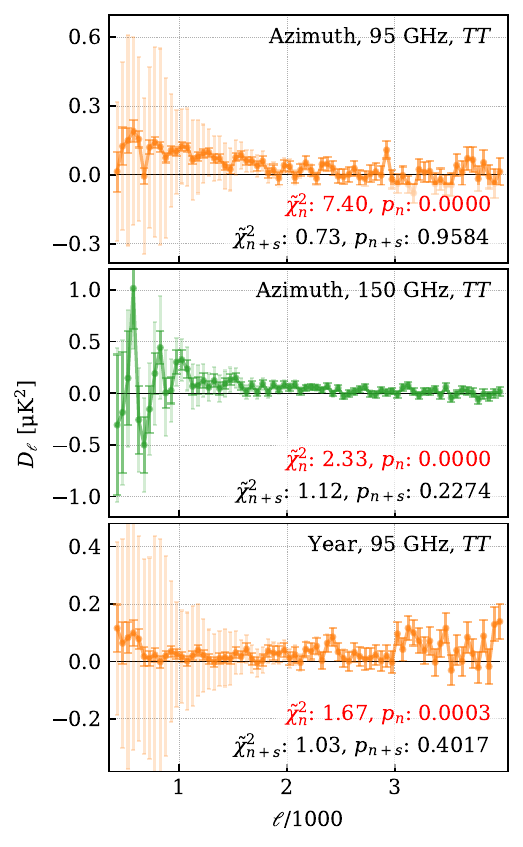}
  \caption{\label{fig:null_spectra_initial_outcome2}%
  The three null spectra that revealed small potential systematic errors.
  This figure shows the same types of quantities as Figure~\ref{fig:null_spectra_initial_outcome1} except that two sets of error bars ($\sigma_{b,\,n}$ in a darker shade and $\sigma_{b,\,n+s}$ in a lighter shade) are shown and that two sets of $\tilde{\chi}^2$ values and $p$-values are printed.
  The first set is shown in the first row near the bottom right corner and was calculated using $\sigma_{b,\,n}^2$.
  The second set is shown in the second row and was calculated using $\sigma_{b,\,n+s}^2$.
  The expected null spectra for the azimuth and year splits are again simply zero.}
\end{figure}

  The remaining three null spectra from the azimuth and year splits had the second outcome (a detection of a potential systematic error but still much smaller than the sample variance).
  They are the 95 and 150\;GHz $TT$ null spectra from the azimuth split and the 95\;GHz $TT$ null spectrum from the year split.
  Figure \ref{fig:null_spectra_initial_outcome2} shows each null spectrum with two sets of error bars.
  The smaller (darker color) and the larger (shallower color) error bars represent $\sigma_{b,\,n}$ and $\sigma_{b,\,n+s}$, respectively.
  Also shown are the reduced $\chi^2$ ($\tilde{\chi}^2$) and $p$-value associated with each set of error bars.
  For each null spectrum, the $p$-value calculated with $\sigma_{b,\,n}^2$ is essentially zero, indicating a detection of a potential systematic error.
  With the larger variance, $\sigma_{b,\,n+s}^2$, however, the $p$-value increases to well above $p_\mathrm{sys}$.

\begin{figure}
  \includegraphics[scale=1.00]{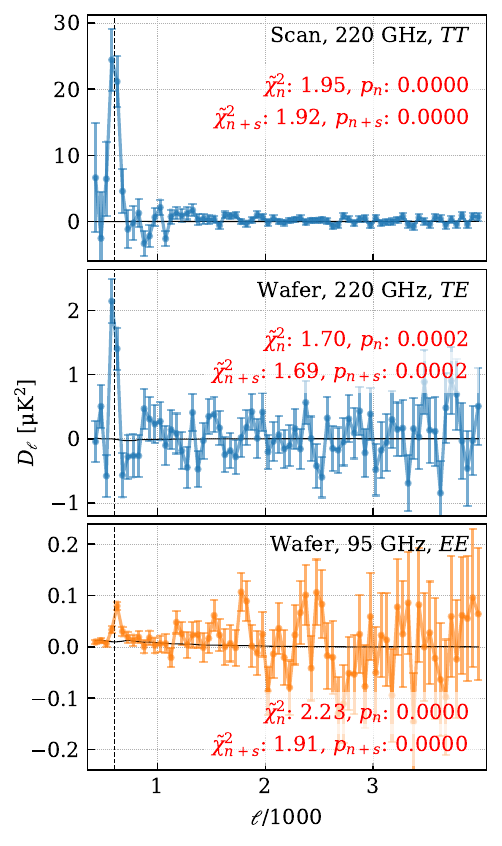}
  \caption{\label{fig:null_spectra_initial_outcome3}%
  Three examples of the null spectra from the scan and wafer splits that revealed large, narrowband contamination at $\ell$ close to 600.
  The vertical black dashed lines are drawn at $\ell$ = 600.
  This figure shows the same types of quantities as Figure~\ref{fig:null_spectra_initial_outcome2}.
  Although $\sigma_{b,\,n+s}$ is also shown, it is indistinguishable from $\sigma_{b,\,n}$.
  The expected null spectra for the scan and wafer splits are nonzero and shown as black curves, but they have much smaller amplitudes compared with the y-axis ranges.}
\end{figure}

  Finally, seven null spectra from the scan and wafer splits revealed large, narrowband contamination at $\ell$ close to 600 and had the third outcome.
  Figure~\ref{fig:null_spectra_initial_outcome3} shows examples of the contamination.
  In these cases, using $\sigma_{b,\,n+s}^2$ still yielded low $p$-values.
  The remaining five null spectra from the scan and wafer splits also revealed the contamination but at lower levels, and those spectra technically had the second outcome because using $\sigma_{b,\,n+s}^2$ barely yielded high enough $p$-values.

\subsection{\label{sec:null_contamination}Narrowband Contamination}
  The narrowband contamination in the null spectra appears to originate from narrowband contamination in timestreams.
  In the $\ell,m$ space, this contamination is localized not only at $\ell$ close to 600 but also at $m$ close to 400.
  The localization in $m$ suggests that timestreams contain contamination localized at frequencies close to 1.1 Hz.\footnote{
  Because the speed of a telescope scan is 1 deg/s in right ascension, a 1\;Hz oscillation in a timestream corresponds to an oscillation in $\phi$ with $m$ = 360.}

\begin{figure}
  \includegraphics[scale=1.00]{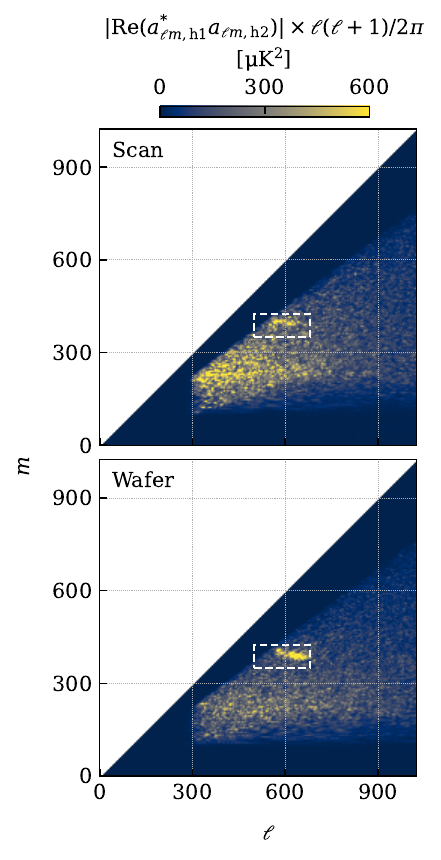}
  \caption{\label{fig:null_spectra_initial_contamination_alm}%
  A 2D version of the 220\;GHz $TT$ null spectra from the scan (top panel) and wafer (bottom panel) splits obtained by cross-correlating the pair of half-depth null coadds.
  The images show the quantity $\ell(\ell+1)/(2\pi) \times |\Re(a_{{\ell}m,\,\mathrm{h1}}^{\ast}\,a_{{\ell}m,\,\mathrm{h2}})|$, where $a_{{\ell}m,\,\mathrm{h1}}$ and $a_{{\ell}m,\,\mathrm{h2}}$ represent the spherical harmonic coefficients of the two $T$ null maps for each split.
  The rectangular region surrounded by the white dashed line is the region that we masked.}
\end{figure}

  Although the exact physical mechanism causing this contamination is unclear, we were able to mitigate it by masking contaminated modes in the $\ell,m$ space (setting their spherical harmonic coefficients to zero).
  The masked region is a rectangle in the $\ell,m$ space with the following boundaries: $500 \le \ell \le 680$ and $350 \le m \le 425$.
  Figure~\ref{fig:null_spectra_initial_contamination_alm} shows a 2D version of the 220\;GHz $TT$ null spectra from the scan and wafer splits and the masked region in spherical harmonic space.
  We obtained the 2D spectra for each split by cross-correlating a pair of half-depth null coadds, which we produced by adding the first 12 null bundles and the last 13 null bundles separately.

\begin{figure}
  \includegraphics[scale=1.00]{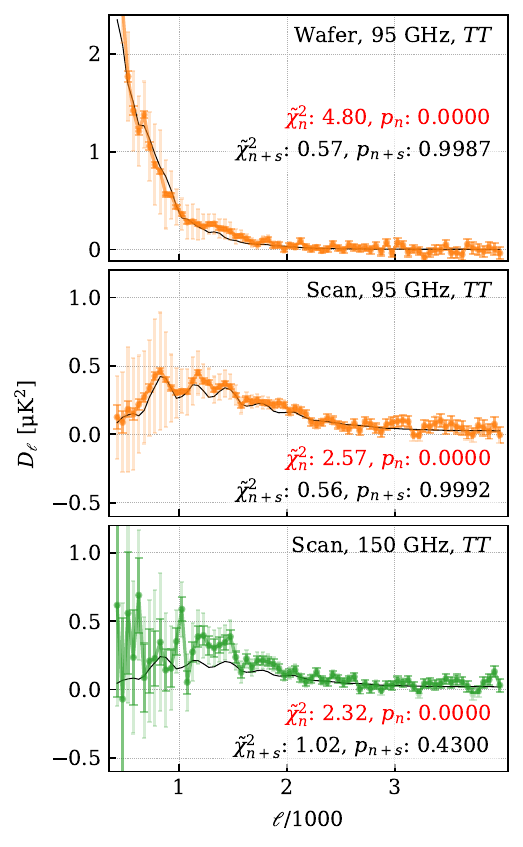}
  \caption{\label{fig:null_spectra_final_outcome2}%
  The three null spectra from the scan and wafer splits that still had low $p$-values after the harmonic-space mask was applied and when $\sigma_{b,\,n}^2$ was used but had high enough $p$-values when $\sigma_{b,\,n+s}^2$ is used.
  This figure shows the same types of quantities as Figures~\ref{fig:null_spectra_initial_outcome2} and \ref{fig:null_spectra_initial_outcome3}.
  The expected null spectra are shown as the black curves.}
\end{figure}

  With this harmonic-space mask, we removed the large peaks in the null spectra from the scan and wafer splits and obtained good $p$-values with $\sigma_{b,\,n}^2$ for all but three null spectra from these two splits.
  The three null spectra that still had low $p$-values are the 95\;GHz $TT$ null spectrum from the wafer split and the 95 and 150\;GHz $TT$ null spectra from the scan split.
  Regardless, the $p$-values increased to well above $p_\mathrm{sys}$ when we used the $\sigma_{b,\,n+s}^2$ as shown in Figure~\ref{fig:null_spectra_final_outcome2}.
  We also recalculated the null spectra from the other four splits using this harmonic-space mask and checked that using the mask did not move any $p$-values to below $p_\mathrm{sys}$.

  While we do not have a full understanding of the contamination, we list here several pieces of information that we gathered on how the contamination appears in the null bundles.
  First, this contamination appears mostly near the left and right edges of the SPT-3G Main field: null spectra from the scan and wafer splits calculated using only the central 1000 $\mathrm{deg}^2$ of the field (the full extent in declination but from $21^{\mathrm{h}}40^{\mathrm{m}}0^{\mathrm{s}}$ to $2^{\mathrm{h}}20^{\mathrm{m}}0^{\mathrm{s}}$ instead of from $20^{\mathrm{h}}40^{\mathrm{m}}0^{\mathrm{s}}$ to $3^{\mathrm{h}}20^{\mathrm{m}}0^{\mathrm{s}}$ in right ascension) do not show the peaks seen in the null spectra calculated from the full area of the field.
  Second, the phase of the 1.1\;Hz oscillation at either edge of the field appears to be the same regardless of declination: when a full-depth $T$ null map for either split is projected in the Cartesian projection and filtered with a bandpass filter to highlight the contaminated region of spherical harmonic space, the troughs and peaks of the 1.1\;Hz plane wave occur in the same map columns ($x$ location) independent of map row ($y$ location).
  Third, the contamination appears to be localized not only within the range of right ascension near the left or right edge of the field but also within the range of declination that covers \texttt{el0} and \texttt{el1}.
  The contamination appears only in \texttt{el0} in the null bundles from the scan split and only in \texttt{el0} and \texttt{el1} for the wafer split.
  Fourth, the frequency of the contamination appears to decrease slightly as the telescope elevation increases.

\begin{figure}
  \includegraphics[scale=1.00]{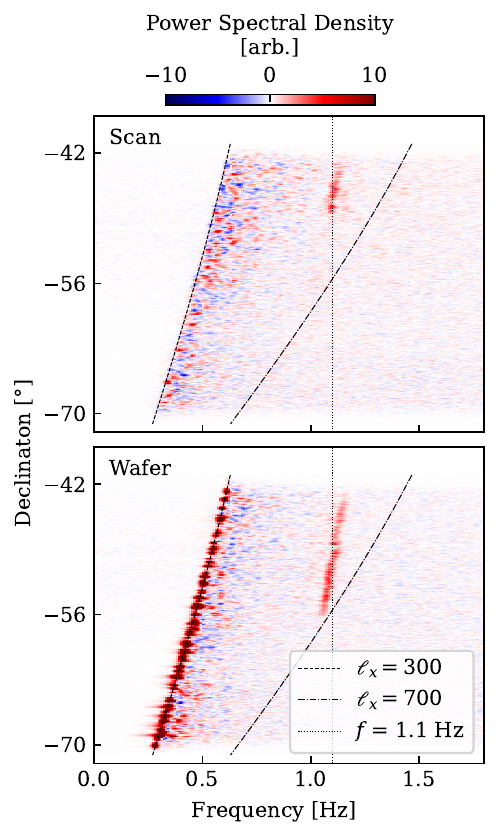}
  \caption{\label{fig:null_spectra_initial_contamination_theta_hert}%
  Power spectral densities of 95\;GHz null bundles from the scan (top panel) and wafer (bottom panel) splits obtained by cross-correlating a pair of half-depth coadds of the null bundles on a ring-by-ring basis using the constant-declination HEALPix rings.
  In each panel, each row shows the power spectral density for one constant-declination ring as a function of frequency.
  The red regions intersecting the vertical dotted line, which represents 1.1\;Hz, are visual representations of the contamination discovered in the null spectra in this declination-frequency space.
  Two contours that have constant $\ell_x$ are also shown.}
\end{figure}

  The third and fourth points are illustrated in Figure~\ref{fig:null_spectra_initial_contamination_theta_hert}, which shows power spectral densities of 95\;GHz null bundles for each split in a combined Fourier-real-space projection.
  The $x$-axis of each panel is the effective frequency of pixel values on a constant-declination HEALPix ring projected back into timestreams using the telescope scan rate at that declination; the $y$-axis of each panel is the ring declination.
  The color scale indicates the amplitude of the (squared) one-dimensional FFT of the timestream created from the ring pixel values.
  The figure shows that the frequency of the $\sim$1.1\;Hz oscillation drifts to slightly lower values as the declination changes and that the contamination disappears abruptly at certain declination.

  Finally, a 1.1 Hz line was detected in the power spectral densities of both the telescope elevation and azimuth encoder readings, but that line has different phenomenology from that of the contamination in the null bundles.
  The line in the elevation and azimuth timestreams is present in all the subfields instead of only \texttt{el0} and \texttt{el1} and does not have a frequency drift.

  The findings listed earlier point to a physical source of the contamination having to do with the telescope drive system, in particular the onset of scanning.
  The localization of the contamination in right ascension and its phase stability in declination are consistent with an oscillation that is excited at the start of a scan and decays as the scan proceeds.
  For an oscillation in elevation, atmospheric emission that is modulated at $\sim$1.1 Hz would be responsible for the excess power in the null bundles.
  The declination localization is consistent with the telescope being more or less balanced at different elevation.
  This could also explain the drift in frequency with declination, though this is difficult to reconcile with the lack of drift in the oscillation frequency seen in the elevation and azimuth encoder data.

  Regardless of the true physical cause, these findings also point to potentially more efficient ways to mitigate the contamination, albeit at the possible cost of higher complexity in the analysis.
  For instance, the line could be notched in the timestreams with a filter that is only active in the right ascension and declination ranges in which the contamination is worst and that has a declination-dependent notch frequency.
  This would remove less overall bandwidth than the masking in the $\ell, m$ space, but it would cause a position-dependent transfer function.
  For the mid-$\ell$ maps, we decided not to develop such a filter and remake maps but to move forward with the power spectrum analyses by applying the harmonic-space mask instead.

\subsection{\label{sec:null_final}Final Results}
  We recalculated all the 54 null spectra and the associated $p$-values after using the harmonic-space mask described in Section~\ref{sec:null_contamination}.
  With this mask, all but five null spectra yielded $p$-values above $p_\mathrm{sys}$ with $\sigma_{b,\,n}^2$.
  The five null spectra with low $p$-values are all $TT$ null spectra; they are from the azimuth split (95 and 150\;GHz), wafer split (95\;GHz), and the scan split (95 and 150\;GHz).\footnote{
  The 95\;GHz $TT$ null spectrum from the year split had a low $p$-value slightly below $p_\mathrm{sys}$ in the initial version, but using the harmonic-space mask happened to increase value slightly to above $p_\mathrm{sys}$.}
  For those five null spectra, using $\sigma_{b,\,n+s}^2$ increased the $p$-values to above $p_\mathrm{sys}$.
  As a result, all 54 null spectra yielded high enough $p$-values with $\sigma_{b,\,n+s}^2$, and we concluded that no significant sources of potential systematic errors were detected from the null tests after using the harmonic-space mask.

  Table \ref{tab:null_spectra_ptes} shows two $p$-values for each of the 54 null spectra, one calculated using $\sigma_{b,\,n}^2$ and the other calculated using $\sigma_{b,\,n+s}^2$.
  In addition, we show all 54 null spectra with and without the harmonic-space mask (final and initial null spectra) in Appendix~\ref{app:null_spectra}.

\begin{table*}
  \caption{\label{tab:null_spectra_ptes}%
  The $p$-values of each of the 54 null spectra calculated with $\sigma_{b,\,n}^2$ (the number on the left in each cell) and with $\sigma_{b,\,n+s}^2$ (the number on the right).}
  \begin{ruledtabular}
  \begin{tabular}{cc|cccccc}
      &  & Wafer & Scan & Azimuth & Year & Moon & Sun \\
    \hline
    \multirow{3}{0.5em}{$TT$}
      &  95\;GHz &
          0.0000/0.9987 & 0.0000/0.9992 & 0.0000/0.9589 & 
          0.0010/0.4029 & 0.7456/1.0000 & 0.7905/1.0000 \\
      & 150\;GHz &
          0.0140/0.9853 & 0.0000/0.4300 & 0.0000/0.2518 & 
          0.0600/0.8184 & 0.5877/0.9990 & 0.7156/0.9870 \\
      & 220\;GHz &
          0.9148/0.9270 & 0.1037/0.1190 & 0.3249/0.3496 & 
          0.1485/0.1671 & 0.8124/0.8319 & 0.9374/0.9405 \\
    \hline
    \multirow{3}{0.5em}{$TE$}
      &  95\;GHz &
          0.0529/0.9724 & 0.1482/0.9853 & 0.1881/0.7263 & 
          0.8483/0.9982 & 0.1640/0.6739 & 0.4837/0.8939 \\
      & 150\;GHz &
          0.0418/0.3591 & 0.0050/0.1407 & 0.7522/0.9747 & 
          0.2411/0.7424 & 0.9205/0.9867 & 0.9141/0.9746 \\
      & 220\;GHz &
          0.5917/0.6000 & 0.1821/0.1857 & 0.0647/0.0671 & 
          0.3132/0.3172 & 0.9608/0.9616 & 0.3375/0.3420 \\
    \hline
    \multirow{3}{0.5em}{$EE$}
      &  95\;GHz &
          0.5155/0.5939 & 0.3169/0.3888 & 0.4728/0.5835 & 
          0.0283/0.0499 & 0.6685/0.7302 & 0.4273/0.4892 \\
      & 150\;GHz &
          0.0420/0.1049 & 0.0047/0.0177 & 0.7176/0.8169 & 
          0.1931/0.3210 & 0.6285/0.7530 & 0.1080/0.1658 \\
      & 220\;GHz &
          0.7650/0.7656 & 0.3431/0.3439 & 0.0049/0.0049 & 
          0.6052/0.6057 & 0.4272/0.4274 & 0.4786/0.4788 \\
  \end{tabular}
  \end{ruledtabular}
\end{table*}

\section{\label{sec:conclusion}Conclusion}
  We have described the production and validation of a set of maps of the millimeter-wave sky from SPT-3G, the third-generation camera on the South Pole Telescope.
  The maps are in frequency bands centered at 95, 150, and 220\;GHz.

  These maps are based on a dataset that we call SPT-3G D1, which denotes the observations that we conducted on the SPT-3G Main field, covering 4\% of the sky, during the 2019 and 2020 austral winter observing seasons.
  The SPT-3G Main field is divided into four subfields, and SPT-3G D1 comprises 3286 two-hour subfield observations.
  In each subfield observation, on average 10~697 detectors, which were divided nearly equally among the three frequency bands, passed quality checks and were used for the mapmaking.

  To convert the detector timestream samples acquired during each subfield observation into the pixel values of maps of the $T$, $Q$, and $U$ Stokes parameters, we used the filter-and-bin approach as has been traditionally done in SPT mapmaking.
  We applied a high-pass filter to the timestreams to reduce low-frequency correlated noise.
  We chose 300 as the scan-direction cutoff multipole number.
  We binned the filtered timestreams into HEALPix \nside\ 8192 and ZEA 0.5625$^\prime$ pixels.

  Having produced $T/Q/U$ maps for each subfield observation, we combined the individual observations maps for each subfield in different ways to produce data products needed for different analysis tasks, calibrated and cleaned the coadds for each frequency band and subfield to reduce known biases, and combined the calibrated/cleaned coadds across the subfields.
  The different types of coadds include full-depth coadds, half-depth coadds, and noise maps.
  The calibration and cleaning steps comprise gain calibration, monopole $T$-to-$P$ leakage removal, polarization angle calibration, and polarization amplitude calibration.
  White noise levels of the full-depth $T$ maps are 5.4, 4.4, and 16.2 $\mathrm{\mu}$K--arcmin in the 95, 150, and 220\;GHz bands, respectively, and for the $Q/U$ maps 8.4, 6.6, and 25.8 $\mathrm{\mu}$K--arcmin.

  After the calibration and cleaning, two types of known biases still remained in the coadds: declination-dependent gain miscalibration and higher-order $T$-to-$P$ leakage.
  The former is not concerning because we expect it to cause a negligible calibration bias to the $TT/TE/EE$ spectra of the CMB measured from the coadds.
  The latter is a non-negligible effect, and we corrected for it at the level of power spectra instead of maps.

  In addition to reducing the known biases through the calibration and cleaning steps, we searched for potential systematic errors in the coadds with null tests.
  We implemented tests based on six types of splits in total (sun, moon, azimuth, year, scan, and wafer) and calculated the $TT/TE/EE$ null spectra for each frequency band.
  From the scan and wafer splits, we detected a significant narrowband contamination at $\ell \sim$ 600 and mitigated it by masking contaminated modes in spherical harmonic space when calculating power spectra from the coadds.
  With that mask, 49 out of the 54 null spectra yielded $p$-values above our predetermined threshold for passing.
  Although we detected potential systematic errors in the other five null spectra, all of which are $TT$ null spectra, the amount by which each null spectrum deviates from the expectation beyond noise is well below the expected sample variance on our measurements of the $TT$ spectrum of the CMB.
  Therefore, we concluded that all the null test results were acceptable.

  These maps were used for significantly improved SPT-3G measurements of temperature and $E$-mode polarization anisotropies and gravitational lensing of the CMB, which are presented in G25, C25, and O26.
  We make the maps and supporting data products that enable post-map analyses, such as power spectrum estimation, lensing reconstruction, and cross-correlation, publicly accessible.
  The access information can be found in Appendix~\ref{app:data_availability}.
  The released maps represent the deepest wide-area, high-resolution CMB temperature and polarization data available, and they can be used for a wide range of applications, including multi-wavelength cross-correlation and stacking studies, estimates of higher-order statistics of CMB primary and secondary anisotropies, and tests of new probes that require deep maps.

\begin{acknowledgments}
  The South Pole Telescope program is supported by the National Science Foundation (NSF) through awards OPP-1852617 and OPP-2332483. Partial support is also provided by the Kavli Institute of Cosmological Physics at the University of Chicago.

  Work at Argonne National Laboratory was supported by the U.S. Department of Energy, Office of High Energy Physics, under contract DE-AC02-06CH11357.
  We gratefully acknowledge the computing resources provided on Crossover, a high-performance computing cluster operated by the Laboratory Computing Resource Center at Argonne National Laboratory.
  Work at the Fermi National Accelerator Laboratory (Fermilab), a U.S. Department of Energy, Office of Science, Office of High Energy Physics HEP User Facility, is managed by Fermi Forward Discovery Group, LLC, acting under Contract No. 89243024CSC000002.
  The SLAC group is supported in part by the Department of Energy at SLAC National Accelerator Laboratory, under contract DE-AC02-76SF00515.
  The Paris group has received funding from the European Research Council (ERC) under the European Union's Horizon 2020 research and innovation program (grant agreement No 101001897), and funding from the Centre National d'Etudes Spatiales.
  The UC Davis group acknowledges support from Michael and Ester Vaida.
  The CAPS authors are supported by the Center for AstroPhysical Surveys (CAPS) at the National Center for Supercomputing Applications (NCSA), University of Illinois Urbana-Champaign.
  The Melbourne authors acknowledge support from the Australian Research Council's Discovery Project scheme (No. DP210102386).

  This work was made possible by software packages including \texttt{NumPy} \citep{numpy}, \texttt{SciPy} \citep{scipy}, and \texttt{Matplotlib} \citep{matplotlib}.
  Some results in this work were obtained using services provided by the OSG Consortium \citep{osg07, osg09, osg06, osg15}. which is supported by the National Science Foundation awards \#2030508 and \#2323298, and some results were derived using the \texttt{healpy} and \texttt{HEALPix} packages \citep{zonca19, gorski05}.
\end{acknowledgments}

\appendix

\section{\label{app:data_availability}Data Availability}
  We make the SPT-3G D1 mid-$\ell$ HEALPix maps and ancillary products available through the Laboratory Computing Resource Center at Argonne National Laboratory.
  A webpage containing instructions for accessing the data products and example code for using them can be found at the SPT website.\footnote{\url{https://pole.uchicago.edu/public/Data\%20Releases.html}}

  The original HEALPix maps, which have \nside\ = 8192 and $\ell_{\mathrm{max}}\sim$ 16\,000, are provided.
  We provide the \texttt{full}, \texttt{half}, \texttt{one-thirtieth}, \texttt{signflip-noise}, and \texttt{pre-null} coadds (Section~\ref{sec:coadding_procedure}).

  The ancillary data products are grouped into two categories.
  One category comprises the products that were used to correct the biased power spectra calculated from the maps to obtain the band powers reported in C25.
  These products include the beams, filter transfer functions, $a_{{\ell}m}$ masks, filtering-artifact biases, and inpainting biases, and the $\ell_{\mathrm{max}}$ of these data products is 4000.
  The other category includes products that can be more widely applied to analyses of the maps and support $\ell$ up to 16\,000.
  These products include beams, filter transfer functions, and apodization masks.

  We also provide the mock-observation input and output maps (Section~\ref{sec:simulations}).
  We provide the 500 sets of standard input and output maps.
  We also provide the 110 sets of output maps produced without the masking in the high-pass filter.

\section{\label{app:mapmaking}Additional Details of the Mapmaking}
  We describe additional details related to Section~\ref{sec:mapmaking}.
  We show additional figures related to the four test mapmaking runs conducted to inform the decision on the high-pass filter cutoff (Appendix~\ref{app:high_pass_filter_cutoff}), describe major changes in the mapmaking parameters between the SPT-3G D1 mid-$\ell$ mapmaking run and the counterpart for the 2018 dataset (Appendix~\ref{app:changes_from_2018}), describe the procedure to manipulate weight map matrices when generating the \texttt{signflip-noise} coadds (Appendix~\ref{app:weight_map_matrix}), and show additional figures of the \texttt{full} coadds (Appendix~\ref{app:coadds}).

\subsection{\label{app:high_pass_filter_cutoff}High-pass Filter Cutoff}
\begin{figure*}
  \includegraphics[scale=1.00]{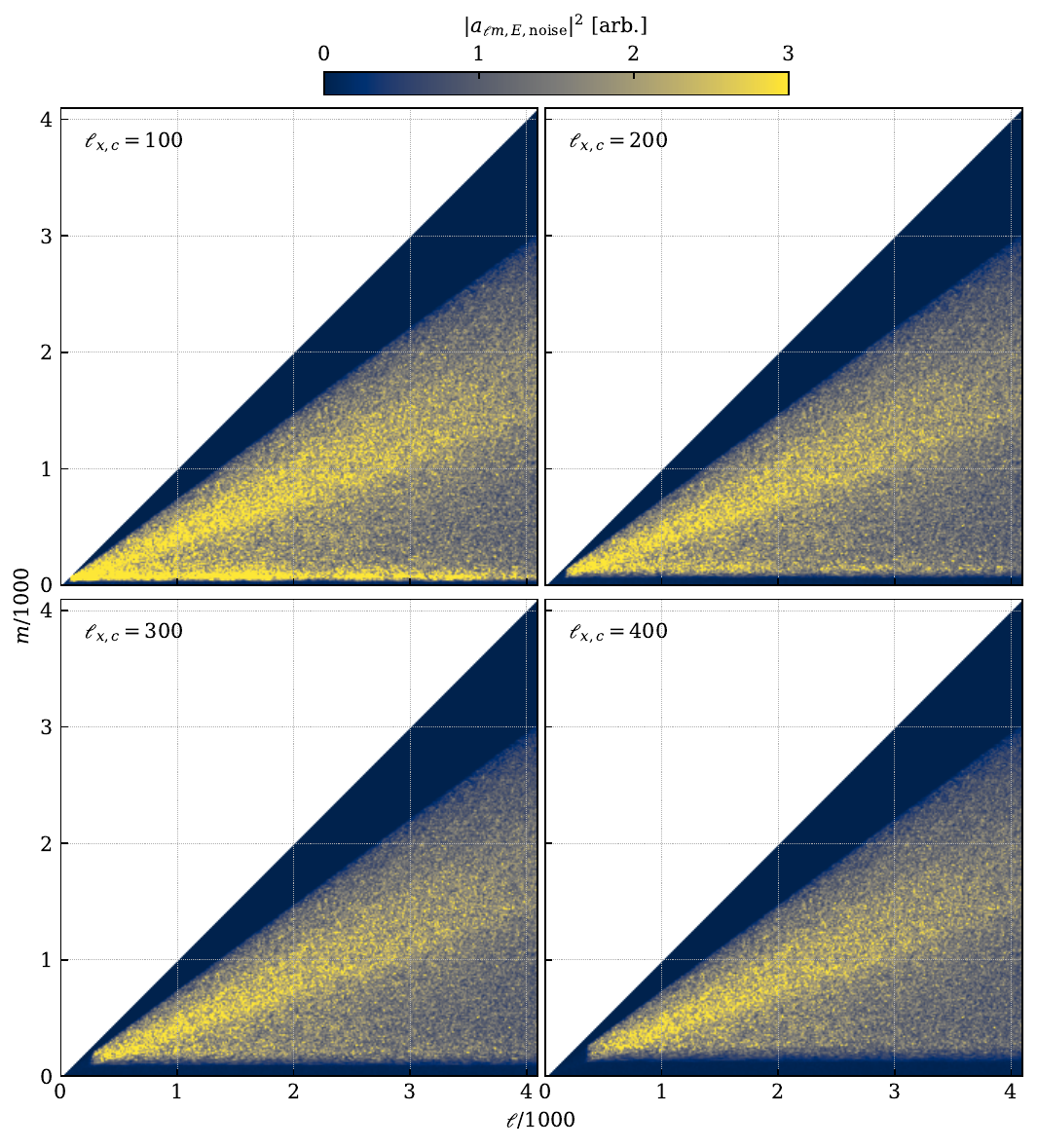}
  \caption{\label{fig:noise_ee_alm_triangle}%
  The 2D $EE$ noise spectra in the 150\;GHz band obtained from the four test mapmaking runs.
  Unlike the $TT$ version (Figure~\ref{fig:noise_tt_alm_triangle}), a linear scale is used here with a different arbitrary normalization.
  In the linear scale, it can be more clearly seen that the modes that lie between  $m\sim\ell/3$ and $m\sim\ell/2$ are noisier than the modes that lie outside the region.
  This is because the observations used for the test mapmaking runs were not balanced well across the subfields and is not a fundamental property of SPT data.}
\end{figure*}

\begin{figure}
  \includegraphics[scale=1.00]{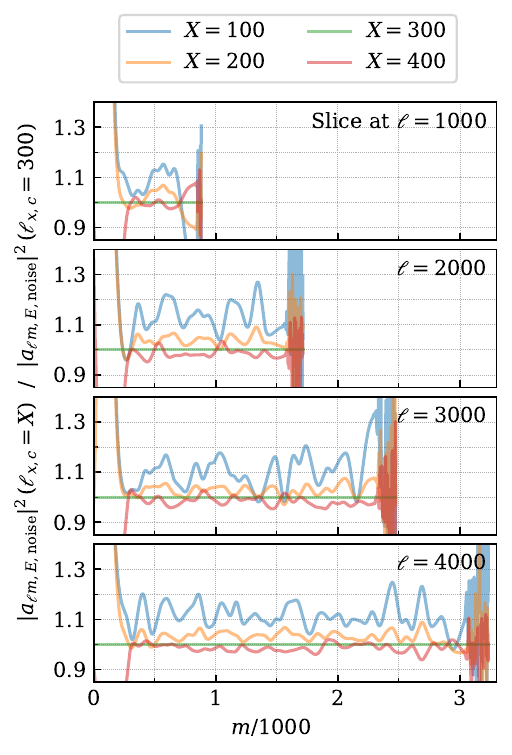}
  \caption{\label{fig:noise_ee_alm_xsec_ratio}%
  The ratio of each 2D power spectrum shown in Figure~\ref{fig:noise_ee_alm_triangle} to the power spectrum in the $\ell_{x,\,c} = 300$ case along four constant-$\ell$ slices.}
\end{figure}

  We show two additional figures, Figures~\ref{fig:noise_ee_alm_triangle} and \ref{fig:noise_ee_alm_xsec_ratio}, related to the considerations behind the decision of setting $\ell_{x,\,c}$ to 300 described in Section~\ref{sec:high_pass_filter_cutoff}.
  These figures are similar to Figures~\ref{fig:noise_tt_alm_triangle} and \ref{fig:noise_tt_alm_xsec_ratio} and show the 2D $EE$ noise spectra instead of the $TT$ ones.

  In addition, we present the results of a semi-analytical model for the leakage of atmospheric noise from low frequencies in timestreams to high $\ell$ in sky maps.
  This model will be described in detail in an upcoming publication; we summarize the main features here.
  Underpinning this model are three assumptions: 1) the frozen-screen approximation to the atmosphere (Section~\ref{sec:high_pass_filter_cutoff}); 2) the atmospheric fluctuations have a well-defined angular correlation function (as would be expected for Kolmogorov turbulence, see, e.g., \cite{coerver24}); 3) the atmospheric noise seen by the telescope in each scan is an independent realization of the underlying atmospheric noise power spectrum, uncorrelated with the atmospheric noise in any other scan.
  The combination of \#1 and \#3 is assumed to be achieved by a wind speed that is low enough that the modulation of the atmospheric signal in detector data is dominated by the telescope scan, not the motion due to wind, but high enough that, during the time the telescope takes to turn around between scans, the atmosphere has shifted relative to the background sky by an angle larger than the largest angle of interest in the analysis.

\begin{figure}
  \includegraphics[scale=1.00]{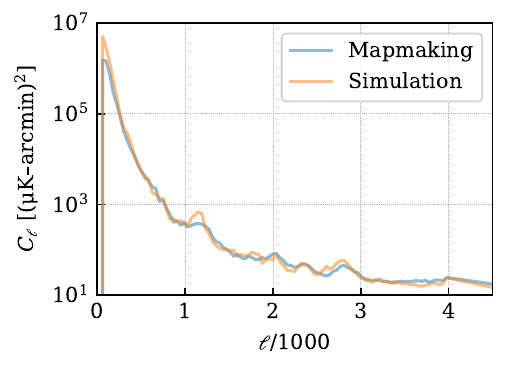}
  \caption{\label{fig:noise_tt_cl_model}%
  The comparison between the 1D $TT$ noise spectrum obtained from the mapmaking run with $\ell_{x,\,c}$ = 50 and the simulated spectrum obtained from the model.}
\end{figure}

  In this scenario, the noise from the atmosphere appears in SPT-3G data in each scan as a snapshot of the instantaneous realization of the atmosphere during that scan, superimposed on the background astronomical sky.
  In the limit of a detector array with infinite extent and infinitely fine sampling (and identically zero wind), this snapshot would be an unbiased representation of the true atmospheric screen, but, for a real detector array, the information in the cross-scan direction will be limited by both the array height and the finite sampling of cross-scan positions.
  The semi-analytic model for the noise in SPT-3G assumes that, in each scan, the camera ``observes'' a ``sky'' of noise with some underlying power spectrum in the scan direction.
  The noise is assumed to be correlated in the cross-scan direction with some known correlation function, which can be anywhere from a Dirac $\delta$-function (if we are modeling, for example uncorrelated detector or readout $1/f$ noise) to a product of the true one-dimensional atmosphere correlation function and the true cross-scan-direction distribution of detector positions (weighted inversely by the measured detector variance).
  The noise is simulated directly on a map pixel grid (made possible by the constant-declination scan strategy of SPT), and the resulting simulated map data are filtered with the map-space version of the timestream filtering described in Section~\ref{sec:mapmaking}.
  In Figure~\ref{fig:noise_tt_cl_model}), we show the prediction of the model for the 150 GHz $TT$ power spectrum and a scan-direction high-pass filter cutoff of $\ell_{x,\,c} = 50$.
  The relevant ``noise bump'' features are reproduced at reasonably high fidelity by this simple model, indicating that the source of these features is reasonably well understood.
  Adding a sub-dominant but non-zero wind component to the model is expected to improve the fidelity further.

\subsection{\label{app:changes_from_2018}Changes from the 2018 Mapmaking}
  We highlight two major changes in the filter-and-bin parameters between the SPT-3G D1 mid-$\ell$ maps and the maps produced for the 2018 analyses (\citet{dutcher21} and \citet{pan23}).

  One change is related to the high-pass filter.
  In addition to the polynomial and sinusoid subtraction, the 2018 mapmaking included common-mode subtraction by frequency band and detector wafer.
  This is a very efficient means of removing atmospheric noise from the timestreams because the SPT-3G detector beams are nearly overlapping at the mean height of water vapor in the atmosphere (\citet[Section~5.1]{coerver24}).
  It also was an effective means to reduce correlated low-frequency noise in 2018 timestreams that we attributed to vibrations in the stage supporting the full detector array, induced by telescope motion.
  The stage was replaced by a stiffer version between the 2018 and 2019 austral winter observing seasons, as described in S22 (Section~3.1).
  The common-mode filter did, however, induce a type of $T$-to-$P$ leakage, which complicated the analysis in \citet[Section~IV\,G\,2]{dutcher21}.
  Because the stage-related noise was no longer an issue, we chose to skip the common-mode filter step when producing the mid-$\ell$ maps to avoid the $T$-to-$P$ leakage.

  Another change is related to the pixelization scheme.
  The 2018 analyses---and all previous power spectrum analyses on datasets from the SPTpol and SPT-SZ cameras---used a flat-sky projection and two-dimensional fast Fourier transforms as an approximation to spherical harmonic transforms (SHTs) on the curved sky.
  While this was adequate for the smaller patches of sky treated in the SPTpol and SPT-SZ works, on the 1500-square-degree SPT-3G Main field, we found two significant non-idealities related to the flat-sky approximation.
  (Although the SPT-SZ data analyzed in \citet{story13} cover 2500 square degrees, the analysis was performed in 19 individual patches of $< 200$ square degrees each.)

  First, we found that the projection resulted in excess off-diagonal correlation between bandpowers (binned power spectra) in adjacent bins \citep[Appendix~A]{balkenhol23}.
  Second and relatedly, we found that we could no longer trust the analytical calculation of the mode-coupling matrix that was found to be highly accurate on the 500-square-degree sky patch analyzed in \citet{henning18}, and we had to use expensive simulations instead \citep[Section~IV\,C]{dutcher21}.
  For these reasons, we chose to measure the $TT/TE/EE$ spectra from SPT-3G D1 using a pixelization scheme that works well with sky curvature and using SHTs.
  As described in Section~\ref{sec:binning}, we chose HEALPix for the curved-sky pixelization and HEALPix-based tools (in particular \texttt{Polspice}) to perform fast SHTs and estimate angular power spectra (including correcting for certain masking effects).

\subsection{\label{app:weight_map_matrix}Weight Map Matrix}
  We describe the exact procedure that we used to manipulate the weight map matrices in the production of the \texttt{signflip-noise} coadds described in Section~\ref{sec:coadds}.
  In that section, we describe the two additional steps needed to produce the coadds.
  In the first step, because the individual-observation maps contain weighted $T/Q/U$ maps as defined in the beginning of Section~\ref{sec:coadds}, the subtraction was done by weighting the \texttt{full} coadds by the weight map matrix associated with each subfield observation.
  In the second step, we multiplied only the weighted $T/Q/U$ and not the weight map matrix by $+1$ or $-1$.
  In an equation form, the production of a \texttt{signflip-noise} coadd is the following:
\begin{equation}
  N = {\left(W_{\mathrm{full}}\right)}^{-1}
      \sum_j \epsilon_j (\bar{S}_j^W - W_j \bar{S}_{\mathrm{full}}),
\end{equation}
where $\bar{S}_j^W = {[\bar{T}_j^W(\hat{n})\ \bar{Q}_j^W(\hat{n})\ \bar{U}_j^W(\hat{n})]}^T$ and contains the weighted $T/Q/U$ maps from observation $j$, $W_j$ is the weight map matrix associated with that observation, $\bar{S}_{\mathrm{full}}$ is the \texttt{full} coadd, $\epsilon_j$ is $+1$ or $-1$, and $W_{\mathrm{full}}$ is the weight map matrix associated with the \texttt{full} coadd.

\subsection{\label{app:coadds}Additional Figures of Coadds}
\begin{figure*}
  \includegraphics[scale=1.0]{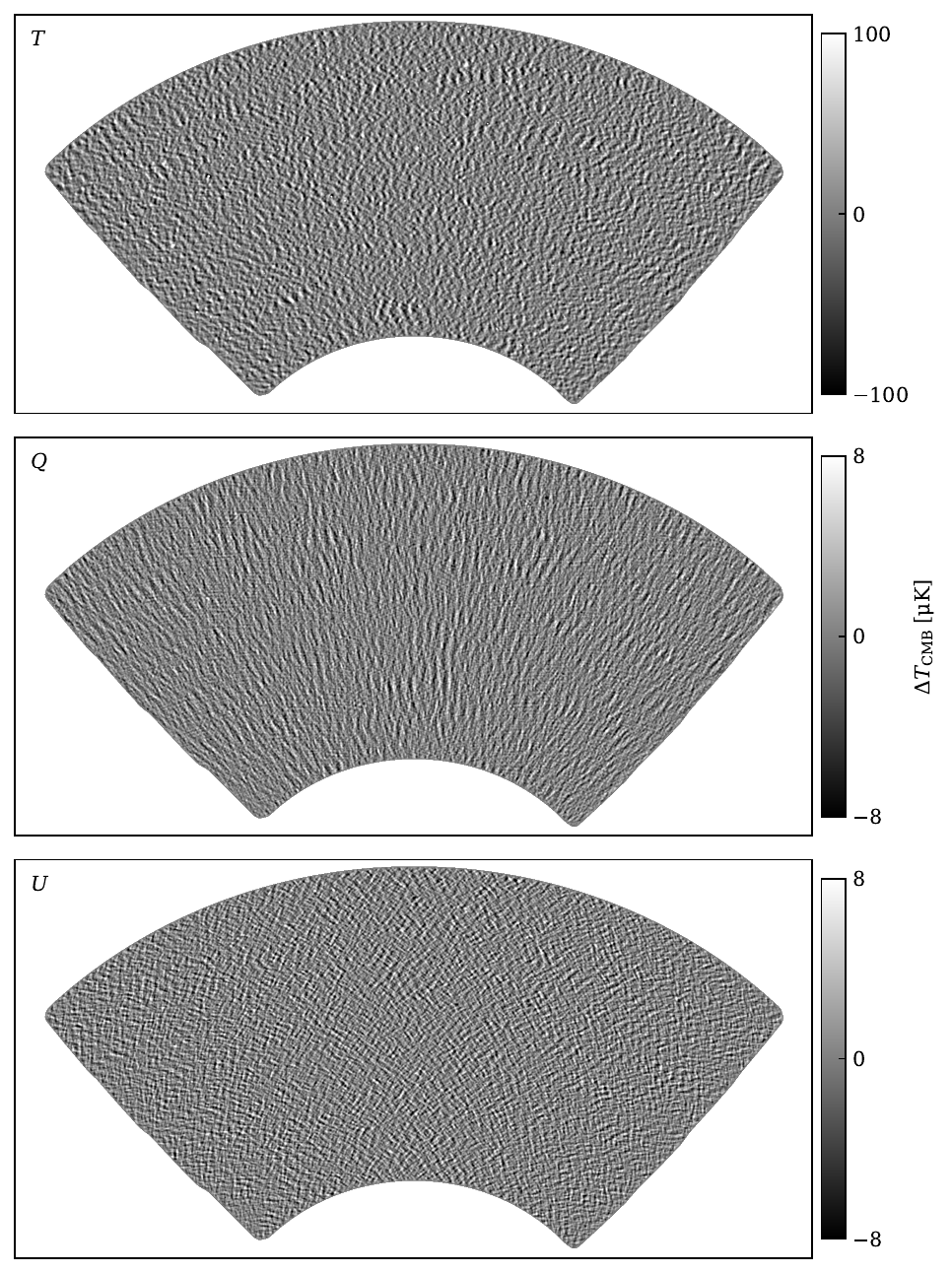}
  \caption{\label{fig:coadds_spt_only_signal_tqu_full}%
  The 150\;GHz \texttt{full} coadd over the full SPT-3G Main field.
  Each map was multiplied by an apodization mask to suppress noisy borders and then smoothed by a Gaussian beam with 6$^\prime$ full width at half maximum before the map was plotted.}
\end{figure*}

\begin{figure*}
  \includegraphics[scale=1.0]{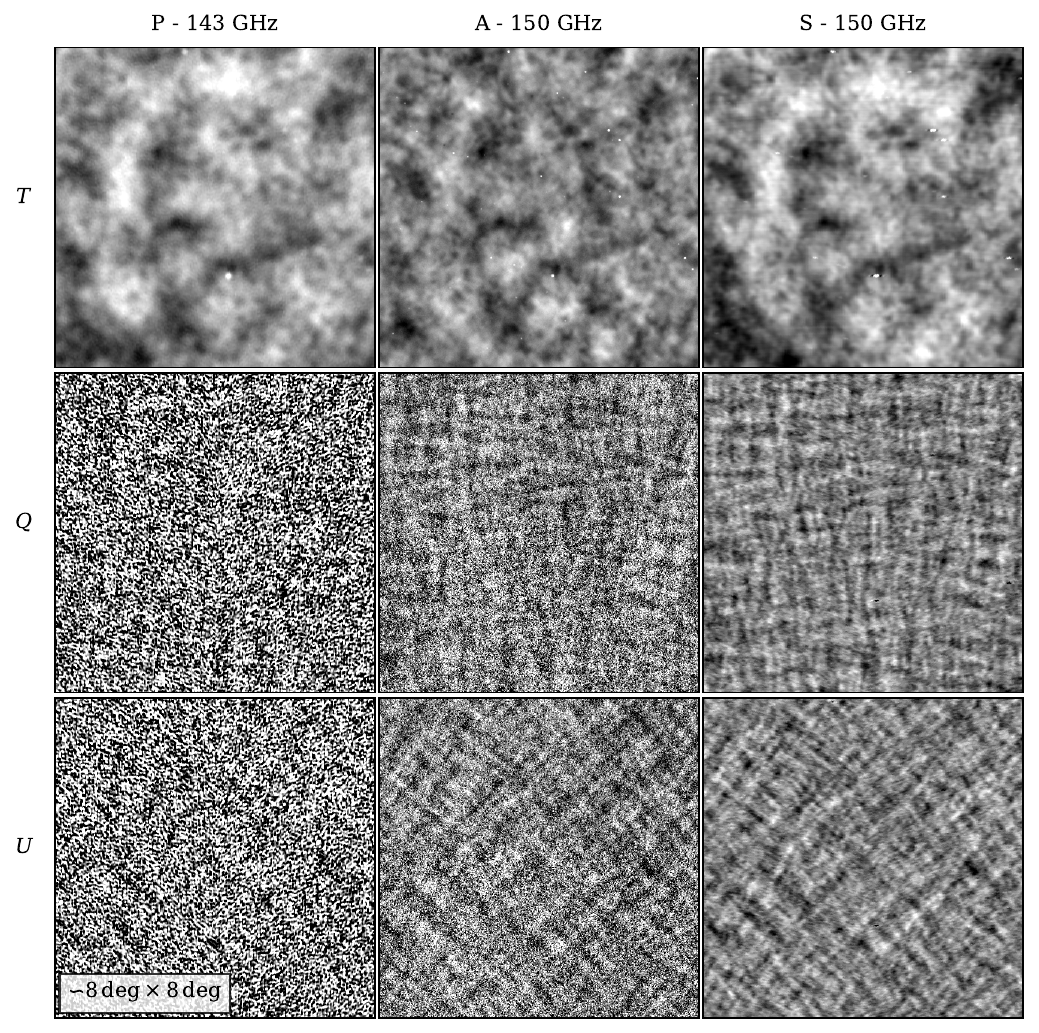}
  \caption{\label{fig:coadds_spt_act_planck_comparison_tqu_unfiltered}%
  A version of Figure~\ref{fig:coadds_spt_act_planck_comparison_tqu_filtered} that compares large angular-scale CMB anisotropies.
  The ACT and \textit{Planck} maps are not filtered.
  The SPT maps are the low-$\ell$ \texttt{full} coadd instead of the mid-$\ell$ one.
  High-S/N objects in the SPT $T$ map look elongated in the horizontal direction because the low-$\ell$ maps were produced with a timestream low-pass filter that has a much lower $\ell_{x,\,c}$ at 3000 as opposed to 13\,000 used for the mid-$\ell$ maps.}
\end{figure*}

  We show additional map figures of the \texttt{full} coadds.
  In Figure~\ref{fig:coadds_spt_only_signal_tqu_full}, we expand Figure~\ref{fig:coadds_spt_only_signal_tqu_strip} to show the $T/Q/U$ maps of the 150\;GHz \texttt{full} coadd across the full field.
  Figure~\ref{fig:coadds_spt_act_planck_comparison_tqu_unfiltered} is a version of Figure~\ref{fig:coadds_spt_act_planck_comparison_tqu_filtered} that shows the unfiltered \textit{Planck} and ACT maps and the SPT-3G D1 low-$\ell$ maps.

\section{\label{app:map_level_null_tests}Additional Details of the Null Tests}
  We describe additional details related to Section~\ref{sec:map_level_null_tests}.
  We describe the effects of the Sun on our observing strategy (Appendix~\ref{app:null_sun_avoidance}) and show additional figures of the null spectra (Appendix~\ref{app:null_spectra}).

\subsection{\label{app:null_sun_avoidance}Sun Avoidance}
  First, we describe our observing strategy 
  The observing strategy includes a sun-avoidance criterion, which partially defines an austral winter observing season.
  The criterion is approximately equivalent to a minimum distance in right ascension of $4^\mathrm{h}$ between the edge of the SPT-3G Main field and the Sun.
  When the Sun is closer to the field than this, features pointing in the direction of the Sun are visible by eye in maps made with minimal filtering.
  These features stem from a sidelobe pattern caused by diffraction off of the gaps between primary mirror panels \citep{carlstrom11}.
  We have found empirically that these features appear on or around December 1 and are undetectable before that date, hence the boundary of late November for the observing season.
  The Sun stays close enough to the field to cause these features until it sets (at the South Pole) around March 21, hence the start of the observing season shortly thereafter.
  Thus, in approximately two out of the eight months (from South Pole sunrise around September 21 to the end of the observing season in late November) we take data with the Sun above the horizon.
  We apply no active moon-avoidance strategy; the Moon is a factor of 20 dimmer than the Sun at millimeter wavelengths, and the distance of the SPT-3G Main field from the Ecliptic plane enforces an effective $\sim$15-degree minimum Moon distance.

\subsection{\label{app:null_spectra}Full Sets of Null Spectra}
\begin{figure*}
  \includegraphics[scale=0.50]{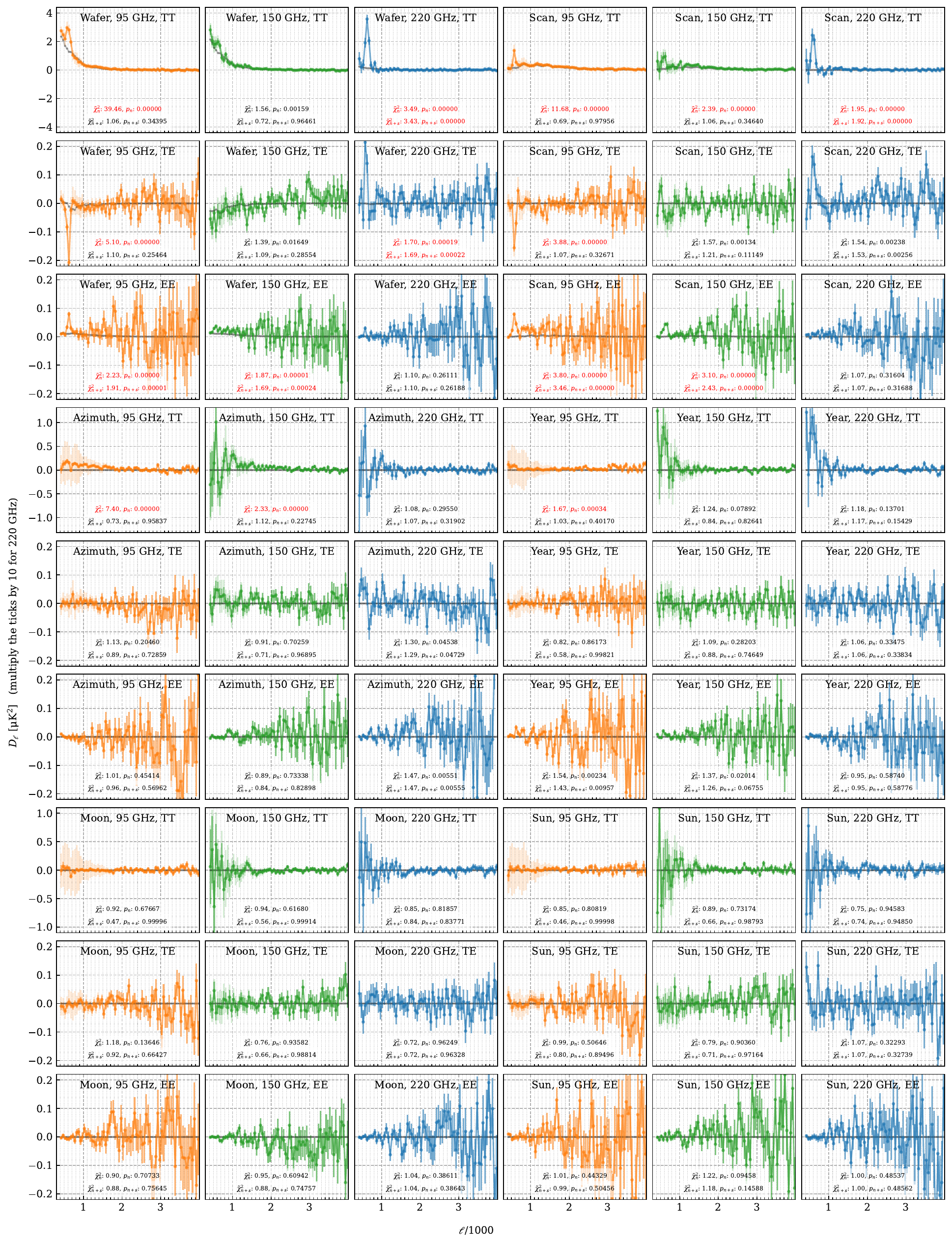}
  \caption{\label{fig:null_spectra_initial_all}%
  All the initial null spectra, which were calculated without the harmonic-space mask.}
\end{figure*}

\begin{figure*}
  \includegraphics[scale=0.50]{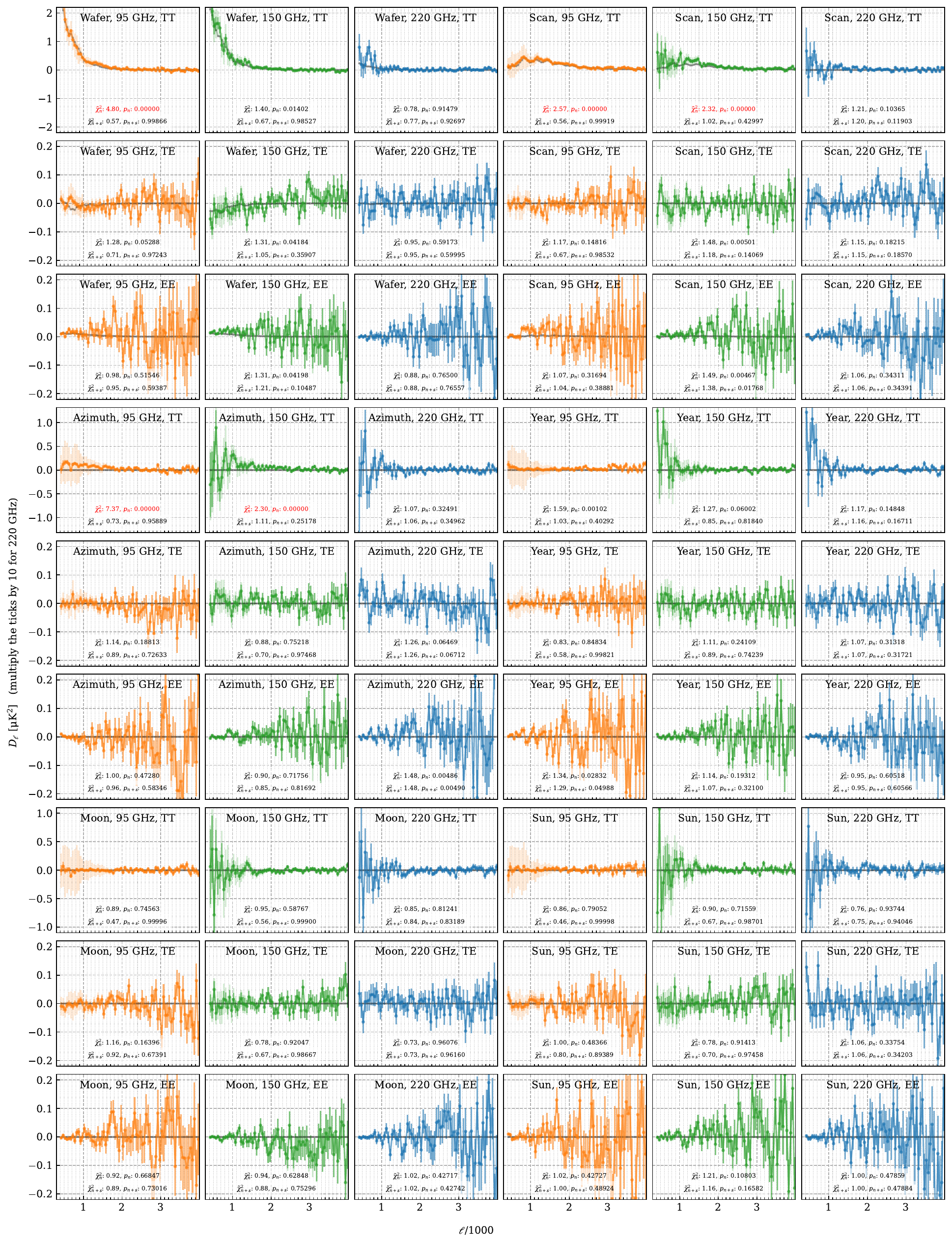}
  \caption{\label{fig:null_spectra_final_all}%
  All the final null spectra, which were calculated with the harmonic-space mask.}
\end{figure*}

  We show two additional figures related to the null spectra (Section~\ref{sec:null_procedure_spectra}).
  Each figure shows all 54 spectra.
  Figure~\ref{fig:null_spectra_initial_all} shows the initial null spectra, without the harmonic-space mask (Section~\ref{sec:null_contamination}), and Figure~\ref{fig:null_spectra_final_all} shows the final null spectra, after masking.

  Each figure contains 54 panels.
  The panels in the top, middle, and bottom three rows show the null spectra from the wafer and scan, azimuth and year, and sun and moon splits, respectively.
  Each panel shows the same types of quantities as Figure~\ref{fig:null_spectra_initial_outcome2}.
  The $p$-values shown in Figure~\ref{fig:null_spectra_final_all} are the same as the ones shown in Table~\ref{tab:null_spectra_ptes}.
  For the 220\;GHz null spectra, the tick labels need to be multiplied by 10.

\FloatBarrier

\bibliography{
  ./bibliographies/spt_before_1995,
  ./bibliographies/spt_1995_to_2000,
  ./bibliographies/spt_2000_to_2005,
  ./bibliographies/spt_2005_to_2010,
  ./bibliographies/spt_2010_to_2015,
  ./bibliographies/spt_2015_to_2020,
  ./bibliographies/spt_2020_to_2025,
  ./bibliographies/spt_2025_and_after}

\end{document}